\documentclass[final,5p,times,twocolumn]{elsarticle}

\usepackage{amssymb}
\usepackage{amsmath}

\usepackage{amsmath,amssymb,amsfonts}
\usepackage{algorithmic}
\usepackage{graphicx}
\usepackage{tikz}
\usepackage{textcomp}
\usepackage{xcolor}
\usepackage{subcaption}
\usepackage{booktabs}
\usepackage{tabularx}
\usepackage{multicol}
\usepackage{siunitx}
\usepackage{colortbl}

\usepackage[nolist]{acronym} 
\usepackage{verbatim}
\usepackage{float}
\usepackage{hyperref}
\usepackage{cleveref}
\usepackage{nicefrac}
\usepackage{tablefootnote}
\usepackage{booktabs}
\usepackage[super]{nth}
\usepackage{etoolbox}
\usepackage{threeparttable}
\usepackage{listings}
\usepackage{pgfplots}
\usepackage{tabularx}
\usepackage{csquotes}
\usepackage{multirow} 

\pgfplotsset{compat=1.18} 
\newenvironment{code}{\captionsetup{type=listing}}{}

\newcommand\circled[3]{%
	\tikz[baseline=(char.base)]{
		\node[shape=circle, fill=#1, inner sep=0pt, text width=10pt, align=center]
		(char) {\textcolor{#2}{\normalfont{\bfseries\scriptsize #3}}};
	}%
}

\newcommand\rcircle[1]{
	\circled{red}{white}{#1}%
}

\newcommand\orcircle[1]{
	\circled{orange}{white}{#1}%
}
\newcommand\pcircle[1]{
	\circled{purple}{white}{#1}%
}

\newcommand{\myurl}[1]{{{\href{#1}{#1}}}}
\newcommand{\myhref}[2]{{{\href{#1}{#2}}}}
\newcommand{\ie}[0]{\textit{i.e.}~} 
\newcommand{\eg}[0]{\textit{e.g.}~}

\usepackage[listings,skins]{tcolorbox}
\lstdefinelanguage{Toml}{
    comment = [l]{\#},
    keywords = {true, false},
    morestring = [b]{"}
}
\newtcblisting{mylisting}[1]{%
    enhanced,
    listing only,
    title={\texttt{#1}},
    attach boxed title to top right,
    listing options={language=#1}
}

\newtcbinputlisting{\myinputlisting}[2][]{%
    listing file={#2},
    enhanced,
    listing only,
    title={\scriptsize\texttt{S#1}},
    attach boxed title to top right={yshift=-\tcboxedtitleheight/2},
    listing options={language={#1}},
    coltitle=black,
    colbacktitle=gray!10
}
\journal{Computer Communications}

\begin{document}

\begin{frontmatter}



\title{Deterministic and Reliable Software-Defined Vehicles: key building blocks, challenges, and vision}




\author[overall,duarte,overall,overall]{Pedro V. Teixeira, Duarte Raposo, Rui Eduardo Lopes, Susana Sargento} 

\cortext[pedrovt]{Pedro V. Teixeira acknowledges financial support from the National Science and Technology Foundation within grant \href{https://doi.org/10.54499/2023.04459.BD}{FCT.2023.04459.BD}.}

\affiliation[overall]{organization={Department of Electronics, Telecommunications and Informatics, University of Aveiro / Instituto de Telecomunicações Aveiro},
            addressline={Campus Universitário de Santiago}, 
            city={Aveiro},
            postcode={3810-193}, 
            country={Portugal}}

\affiliation[duarte]{organization={Instituto de Telecomunicações Aveiro},
            addressline={Campus Universitário de Santiago}, 
            city={Aveiro},
            postcode={3810-193}, 
            country={Portugal}}

\begin{abstract}
As vehicle systems become increasingly complex, with more features, services, sensors, actuators, and processing units, it is important to view vehicles not just as modes of transportation moving toward full autonomy, but also as adaptive systems that respond to the needs of their occupants. Vehicular services can be developed to support these adaptations. However, the increasing complexity of vehicular service development, even with current standardizations, best practices and guidelines, are insufficient to tackle the high complexity of development, with expectations of up to 1 (U.S.) billion lines of code for a fully (level 5) autonomous vehicle.\\Within this survey, the paradigm of Deterministic Software Defined Vehicles is explored, aiming to enhance the quality and ease of developing automotive services by focusing on service-oriented architectures, virtualization techniques, and the necessary deterministic intra- and inter-vehicular communications. Considering the main open challenges for such verticals, a vision architecture towards improved services development and orchestration is presented, focusing on: a) a deterministic network configurator; b) a data layer configurator; c) a hypervisor configurator; d) the vehicle abstraction layer; and e) a software orchestrator.  
\end{abstract}



\begin{keyword}
Software-Defined Vehicles, Connected Vehicles, Vehicular Applications, Vehicular Services, Intelligent Transportation Systems, Deterministic Networks, Time Sensitive Networks, Survey



\end{keyword}

\end{frontmatter}





\begin{acronym}[AAAAAAAAA]
    \acro{3gpp}[3GPP]{Third Generation Partnership Project}
    \acro{5qi}[5QI]{5G QoS Identifier} 
    \acro{aadl}[AADL]{Architecture Analysis and Design Language}
    \acro{abs}[ABS]{Anti-lock Braking System}
    \acro{acm}[ACM]{Association for Computing Machinery}
    \acro{adas}[ADAS]{Advanced Driving Assistance System}
    \acro{anacom}[ANACOM]{Autoridade Nacional de Comunicações}
    \acro{ap}[AP]{Access Point}
    \acro{api}[API]{Application Programming Interface}
    \acro{ar}[AR]{Augmented Reality}
    \acro{asic}[ASIC]{Application-Specific Integrated Circuit}
    \acro{asil}[ASIL]{Automotive Safety Integrity Level}
    \acro{asn1}[ASN.1]{Abstract Syntax Notation One interface description language}
    \acro{atcll}[ATCLL]{Aveiro Tech City Living Lab}
    \acro{ats}[ATS]{Asynchronous Traffic Shaping}
    \acro{autosar}[AUTOSAR]{Automotive Open System Architecture}
    \acro{avb}[AVB]{Audio Video Bridging}
    \acro{bosch}[Bosch]{Bosch Car Multimédia Portugal}
    \acro{boschOvar}[Bosch Ovar]{Bosch Security Systems, S.A}
    \acro{bsm}[BSM]{Basic Safety Message} 
    \acro{cam}[CAM]{Cooperative Awareness Message}
    \acro{can}[CAN]{Controller Area Network bus}
    \acro{canfd}[CAN-FD]{CAN Flexible Data Rate}
    \acro{cav}[CAV]{Connected Autonomous Vehicle}
    \acro{cbs}[CBS]{Credit Based Shaper}
    \acro{ccu}[CCU]{Central Control Unit}
    \acro{cnc}[CNC]{Centralized Network Controller}
    \acro{cni}[CNI]{Container Network Interface}
    \acro{coap}[CoAP]{Constrained Application Protocol}
    \acro{covesa}[COVESA]{Connected Vehicle Systems Alliance}
    \acro{cpm}[CPM]{Collective Perception Message}
    \acro{cps}[CPS]{Collective Perception Service}
    \acro{cpu}[CPU]{Central Processing Unit}
    \acro{csma-ca}[CSMA-CA]{CSMA-Collision Avoidance}
    \acro{csma}[CSMA]{Carrier-Sense Multiple Access}
    \acro{cuc}[CUC]{Centralized User Configuration}
    \acro{cuda}[CUDA]{Compute Unified Device Architecture}
    \acro{cv2x}[C-V2X]{Cellular Vehicle-to-Everything}
    \acro{dcgbr}[DC-GBR]{Delay Critical GBR}
    \acro{dcu}[DCU]{Data Collection Unit}
    \acro{dds-sdp}[SDP]{Simple Discovery Protocol}
    \acro{dds}[DDS]{Data Distribution Service}
    \acro{denm}[DENM]{Decentralized Environmental Notification Message}
    \acro{detnet}[DetNet]{Deterministic Networking}
    \acro{dl}[DL]{Deep Learning}
    \acro{dos}[DoS]{Denial of Service}
    \acro{dpdk}[DPDK]{Data Plane Development Kit}
    \acro{dry}[DRY]{Don't repeat yourself}
    \acro{dsl}[DSL]{Domain-Specific Language}
    \acro{dsp}[DSP]{Digital Signal Processor}
    \acro{dsrc}[DSRC]{Dedicated Short Range Communications}
    \acro{ebpf}[eBPF]{extended Berkeley Packet Filter}
    \acro{ecu}[ECU]{Electronic Control Unit}
    \acro{ee}[E/E]{Electrical/Electronic}
    \acro{elk}[ELK]{Elasticsearch-Logstash-Kibana}
    \acro{etsi}[ETSI]{European Telecommunications Standards Institute}
    \acro{eva}[EVA]{Emergency Vehicle Alert}
    \acro{evtol}[eVTOL]{electrical Vertical Take Off and Landing}
    \acro{face}[FACE]{Future Architecture for Computing Environment}
    \acro{fdma}[FDMA]{Frequency-Division Multiple Access}
    \acro{fpga}[FPGA]{Field-Programmable Gate Array}
    \acro{frer}[FRER]{Frame Replication and Elimination for Reliability}
    \acro{gbr}[GBR]{Guaranteed Bit Rate}
    \acro{gda}[GDA]{Geographical Destination Area}
    \acro{gnss}[GNSS]{Global Navigation Satellite System}
    \acro{gprs}[GPRS]{General Packet Radio Service}
    \acro{gps}[GPS]{Global Positioning System}
    \acro{gpu}[GPU]{Graphical Processing Unit}
    \acro{grpc}[gRPC]{Google RPC}
    \acro{hmd}[HMD]{Head Mounted Display}
    \acro{hpc}[HPC]{High Performance Computer}
    \acro{http}[HTTP]{Hypertext Transfer Protocol}
    \acro{hud}[HUD]{Heads-Up Display}
    \acro{i2x}[I2X]{Infrastructure to Everything}
    \acro{ic}[IC]{Integrated Circuit}
    \acro{ica}[ICA]{Intersection Collision Announcement}
    \acro{ieee}[IEEE]{Institute of Electrical and Electronics Engineers}
    \acro{ietf}[IETF]{Internet Engineering Task Force}
    \acro{imu}[IMU]{Inertial Measurement Unit}
    \acro{io}[I/O]{Input/Output}
    \acro{iot}[IoT]{Internet of Things}
    \acro{iou}[IoU]{Intersection over Union}
    \acro{ip}[IP]{Internet Protocol}
    \acro{ipfs}[IPFS]{InterPlanetary File System}
    \acro{it}[IT-Aveiro]{Institute of Telecommunications of Aveiro}
    \acro{its-s}[ITS-S]{Intelligent Transport System Station}
    \acro{its}[ITS]{Intelligent Transport System}
    \acro{jdl}[JDL]{U.S. Joint Directors of Laboratories}
    \acro{json}[JSON]{JavaScript Object Notation}
    \acro{kpi}[KPI]{Key Performance Indicator}
    \acro{kvm}[KVM]{Kernel-based Virtual Machine}
    \acro{lidar}[Lidar]{LIght Detection And Ranging}
    \acro{lin}[LIN]{Local Interconnect Network bus}
    \acro{lora}[LoRaWAN]{Long Range Wide-Area Network}
    \acro{lpwan}[LPWAN]{Low-Power Wide-Area Network}
    \acro{lte}[LTE]{Long Term Evolution}
    \acro{ltea}[LTE-A]{Long Term Evolution Advanced}
    \acro{m2m}[M2M]{Machine to Machine}
    \acro{mac}[MAC]{Media Access Control}
    \acro{manet}[MANET]{Mobile Ad-hoc Network}
    \acro{mcm}[MCM]{Manoeuvre Coordination Message}
    \acro{mec}[MEC]{Multi-access Edge Computing}
    \acro{misra}[MISRA]{Motor Industry Software Reliability Association}
    \acro{ml}[ML]{Machine Learning}
    \acro{most}[MOST]{Media Oriented Systems Transport}
    \acro{mpls}[MPLS]{Multiprotocol Label Switching}
    \acro{mqtt}[MQTT]{Message Queuing Telemetry Transport}
    \acro{mtbf}[MTBF]{Mean Time Between Failures}
    \acro{mtu}[MTU]{Maximum Transmission Unit}
    \acro{nap}[NAP]{Network Architectures and Protocols}
    \acro{nbiot}[NB-IoT]{Narrowband IoT}
    \acro{netconf}[NETCONF]{Network Configuration Protocol}
    \acro{nfc}[NFC]{Near Field Communication}
    \acro{ngsiv2}[NGSIv2]{Next Generation Service Interfaces version 2}
    \acro{nic}[NIC]{Network Interface Controller/Card}
    \acro{noma}[NOMA]{Non-orthogonal Multiple Access}
    \acro{ntp}[NTP]{Network Time Protocol}
    \acro{oasis}[OASIS]{Organization for the Advancement of Structured Information Standards}
    \acro{oatd}[OATD]{Open Access Theses and Dissertations}
    \acro{obd2}[OBD2]{OnBoard Diagnostic v2}
    \acro{obu}[OBU]{On-Board Unit}
    \acro{oem}[OEM]{Original Equipment Manufacturer}
    \acro{ofdma}[OFDMA]{Orthogonal FDMA}
    \acro{omg}[OMG]{Object Management Group}
    \acro{oop}[OOP]{Object Oriented Programming}
    \acro{os}[OS]{Operating System}
    \acro{osal}[OSAL]{Operating system abstraction layer}
    \acro{osi}[OSI]{Open Systems Interconnection}
    \acro{ota}[OTA]{Over-the-air}
    \acro{p2i}[P2I]{Pedestrian to Infrastructure}
    \acro{p2p}[P2P]{Pedestrian to Pedestrian}
    \acro{p2v}[P2V]{Pedestrian to Vehicle}
    \acro{p2x}[P2X]{Pedestrian to Everything}
    \acro{pcp}[PCP]{Priority Code Point}
    \acro{per}[PER]{Packet Error Rate}
    \acro{posix}[POSIX]{Portable Operating System Interface}
    \acro{pps}[PPS]{Pulse Per Second}
    \acro{psm}[PSM]{Personal Safety Message}
    \acro{ptp}[PTP]{Precision Time Protocol}
    \acro{qos}[QoS]{Quality of Service}
    \acro{radar}[Radar]{RAdio Detection And Ranging}
    \acro{rap}[RAP]{Resource Allocation Protocol}
    \acro{rat}[RAT]{Radio Access Technology}
    \acro{rdma}[RDMA]{Remote Direct Memory Access}
    \acro{rest}[REST]{Representational State Transfer}
    \acro{restconf}[RESTCONF]{NETCONF through REST}
    \acro{rfc}[RFC]{Request for Comments}
    \acro{roi}[ROI]{Region of Interest}
    \acro{ros}[ROS]{Robotic Operating System}
    \acro{rpc}[RPC]{Remote Procedure Call}
    \acro{rsu}[RSU]{Road Side Unit}
    \acro{rte}[RTE]{Run Time Environment}
    \acro{rtos}[RTOS]{Real-Time Operating System}
    \acro{rtps}[RTPS]{Real-Time Publish/Subscribe Protocol}
    \acro{rtt}[RTT]{Round Trip Time}
    \acro{rtwt}[rTWT]{restricted Target Wake Time}
    \acro{sae}[SAE]{Society of Automotive Engineers}
    \acro{scs}[SCS]{Stream Classification Service}
    \acro{sdk}[SDK]{Software Development Kit}
    \acro{sdn}[SDN]{Software-Defined Network}
    \acro{sdr}[SDR]{Software-Defined Radio}
    \acro{sdv}[SDV]{Software-Defined Vehicle}
    \acro{sdvc}[SDVC]{Software-Defined Vehicular Cloud}
    \acro{sip}[SIP]{Session Initiation Protocol}
    \acro{sla}[SLA]{Service Level Agreement}
    \acro{soa}[SOA]{Service Oriented Architecture}
    \acro{soafee}[SOAFEE]{Scalable Open Architecture for Embedded Edge}
    \acro{soc}[SoC]{System-on-chip}
    \acro{someip}[SOME/IP]{Scalable service-Oriented MiddlewarE over IP}
    \acro{spice}[SPICE]{Automotive Software Process Improvement and Capability Determination}
    \acro{spof}[SPoF]{Single Point of Failure}
    \acro{srp}[SRP]{Stream Reservation Protocol}
    \acro{tas}[TAS]{Time Aware Shaper}
    \acro{tcp}[TCP]{Transmission Control Protocol}
    \acro{tdd}[TDD]{Test Driven Development}
    \acro{tdma}[TDMA]{Time-Division Multiple Access}
    \acro{tim}[TIM]{Traveler Information Message}
    \acro{tsf}[TSF]{Timing Synchronization Function}
    \acro{tsn}[TSN]{Time Sensitive Network}
    \acro{ttc}[TTC]{Time To Collision}
    \acro{ttethernet}[TTEthernet]{Time-Triggered Ethernet}
    \acro{ttl}[TTL]{Time To Live}
    \acro{ua}[DETI/UA]{Department of Electronics, Telecommunications and Informatics, University of Aveiro}
    \acro{uc}[UC]{Use Case}
    \acro{udp}[UDP]{User Datagram Protocol}
    \acro{ue}[UE]{User Equipment}
    \acro{ui}[UI]{User Interface}
    \acro{upf}[UPF]{User Plane Function}
    \acro{us}[U.S.]{United States of America}
    \acro{usb}[USB]{Universal Serial Bus}
    \acro{utp}[UTP]{Unshielded Twisted Pair cable}
    \acro{v2i}[V2I]{Vehicle to Infrastructure}
    \acro{v2p}[V2P]{Vehicle to Pedestrian}
    \acro{v2v}[V2V]{Vehicle to Vehicle}
    \acro{v2x}[V2X]{Vehicle to Everything}
    \acro{val}[VAL]{Vehicle Abstraction Layer}
    \acro{vam}[VAM]{Vulnerable Road User Awareness Message}
    \acro{vanet}[VANET]{Vehicular Ad-hoc Network}
    \acro{vcu}[VCU]{Vehicle Control Unit}
    \acro{vm}[VM]{Virtual Machine}
    \acro{vpn}[VPN]{Virtual Private Network}
    \acro{vru}[VRU]{Vulnerable Road User}
    \acro{vxcan}[VXCAN]{Virtual CAN tunnel}
    \acro{wan}[WAN]{Wide Area Network}
    \acro{wave}[WAVE]{IEEE 802.11p WAVE}
    \acro{wpan}[WPAN]{Wireless Personal Area Network}
    \acro{wsg84}[WGS84]{EPSG:4326/World Geodetic System 1984}    
    \acro{wsgi}[WSGI]{Web Server Gateway Interface}
    \acro{xdp}[XDP]{eXpress Data Path}
    \acro{xml}[XML]{eXtended Markup Language}
    \acro{yang}[YANG]{Yet Another Next Generation}
    \acro{yolo3}[YOLO]{You Only Look Once real-time object detection system}
\end{acronym}

\section{Introduction}
\label{chapter:introduction}

\begin{figure}[ht]
	\centering
	\includegraphics[width=\columnwidth]{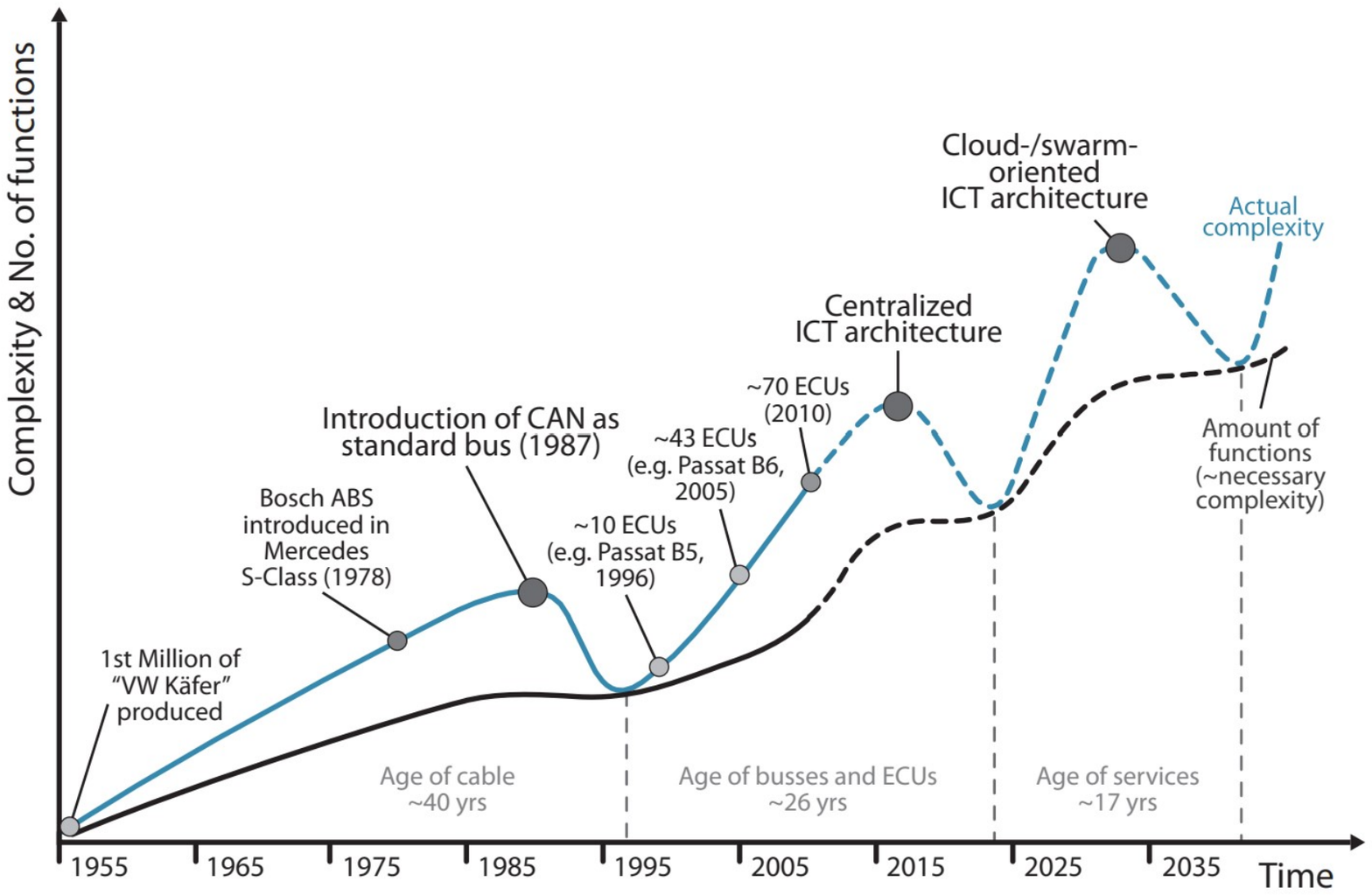}
	\caption[Evolution of complexity in vehicular architectures]{Evolution of complexity in vehicular architectures~\cite{6183198}.}
	\label{fig:background-paradigm}
\end{figure}

Vehicles require the introduction of more advanced systems to support (semi)autonomous driving and \acl{adas}s, decrease the environmental impact and improve the overall quality of life for passengers and drivers~\cite{H1stvisi11:online}. \Acp{sdv}, by considering that most features are now defined through software services and not uniquely through mechanical solutions, can support a new myriad of those advanced systems. However, current vehicular systems supporting a \ac{sdv} are extremely complex, with many \acp{ecu}. Vehicles can have at least around 100\footnote{Described in \myhref{https://www.fortunebusinessinsights.com/industry-reports/automotive-electronic-control-unit-ecu-market-101595}{www.fortunebusinessinsights.com}.} sensors, and actuators scattered throughout the vehicle. Within these \acp{ecu} there are more than 100 million lines of code~\cite{ASHJAEI2021102137}\footnote{More than in a military or commercial aircraft (F-35, 787 Dreamliner) ~\cite{2022-Problem-8}.}, with perspective of reaching 500 million to 1 billion with level 5 autonomous vehicles~\cite{2022-Problem-8,ASHJAEI2021102137}. This is part of an ever-growing tendency of increasing complexity in vehicular architectures, as depicted in~\cref{fig:background-paradigm}.

\begin{figure}[t!]
    \centering
    \includegraphics[width=\columnwidth]{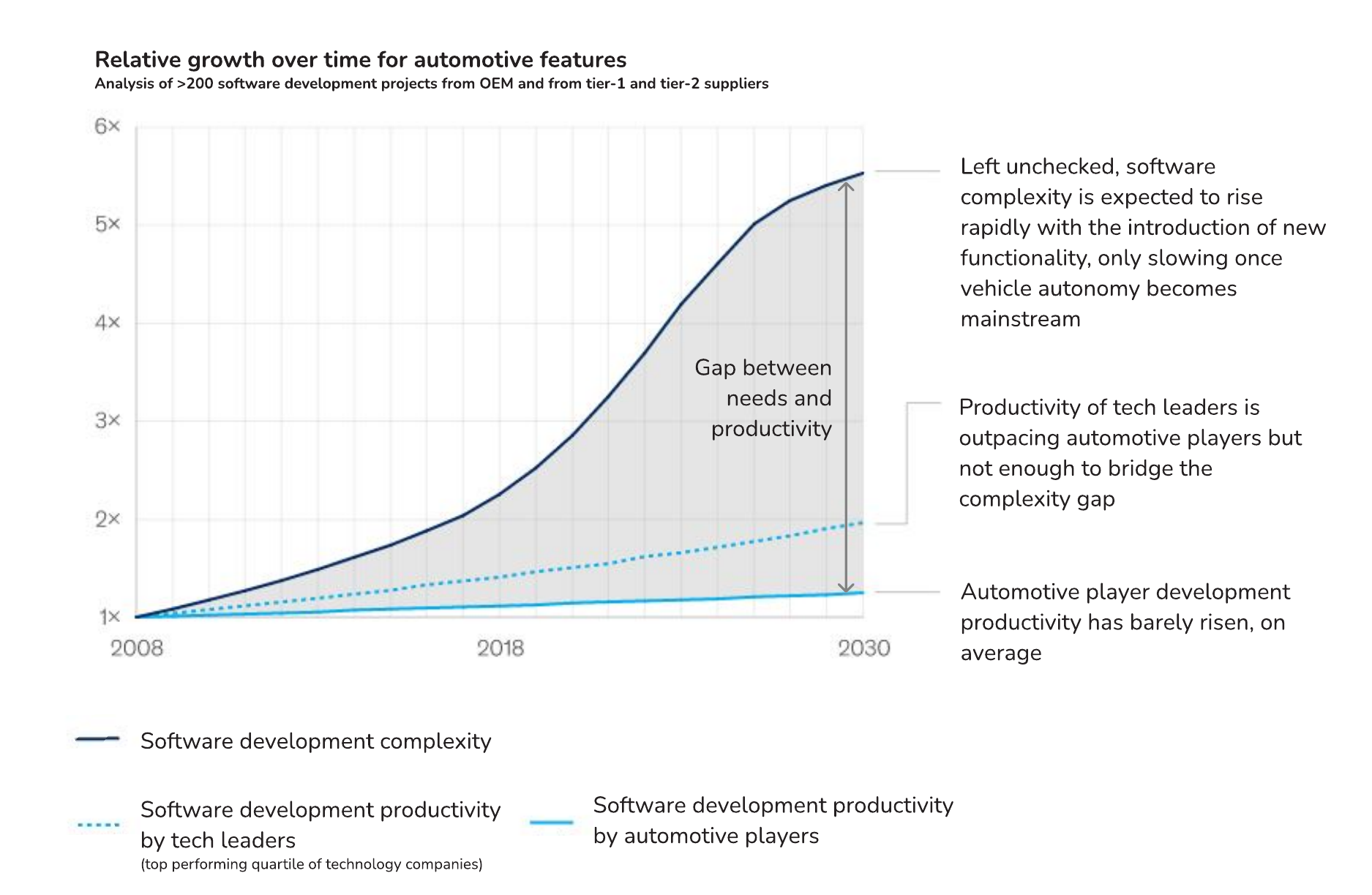}
	\caption[Average complexity of individual software projects in the automotive industry]{Average complexity of individual software projects grown by 300 percent over the past decade, creating an unsustainable gap between software complexity and the productivity of software developers~\cite{2022-Problem-8,Thecasef27:online}.}
	\label{fig:background-sdv-mckinsey}
\end{figure}

This growing code base is becoming unmanageable to vehicle manufacturers and developers of vehicular services within a \ac{sdv}. As described in~\cref{fig:background-sdv-mckinsey}, automakers will tend to leave unchecked, unverified and poorly tested software go into production vehicles. If the growth trend of the code base continues,  the results can be potentially deadly\footnote{Unverified logic and deadly effects on Tesla's autopilot per \myhref{https://www.washingtonpost.com/technology/2023/06/10/tesla-autopilot-crashes-elon-musk/}{washingtonpost.com}.}. The overall intra-vehicular systems, as developed following legacy standards and practices, are deemed too complex and static — and thus costly to maintain, update and keep safe — by vehicle manufacturers, with the reduction of the overall complexity (both in software and hardware terms, in particular wiring harness weight savings) becoming a major goal in order to reduce operational costs and decrease potential security issues from unchecked, poorly tested code~\cite{2021-Problem-5}. 

Standardization of best practices and development guidelines for intra vehicular software gained traction during the early 2000s~\cite{AUTOSARE36:online}. In addition, programming methodologies and guidelines\footnote{Such as the \acs{asil} within ISO 26262 or \acs{misra}-C, a set of guidelines for safety-related systems, such as automotive vehicles. Details at \myhref{https://www.synopsys.com/automotive/what-is-iso-26262.html}{synopsys.com}, \myhref{https://misra.org.uk/}{misra.org.uk} and \myhref{https://www.synopsys.com/software-integrity/static-analysis-tools-sast/misra.html}{synopsys.com}.} are critical and cannot be disregarded. However, they are no longer enough to tackle such complexity of development.

The context of deterministic pieces of software that carry out something within the vehicle or its environment can provide a solution to this problem. Vehicular services, since they may be critical, shall be deterministic in a broad sense in order to ensure that the result is exactly what was intended in the exact moment it was intended. In this article, we propose that \emph{the quality and easiness of the development of services for automotive can be improved using the paradigm of Deterministic Software Defined Vehicles}. The main contributions of this survey are as follows: 
%

\begin{enumerate}
    \item Demonstrate the suitability and relevance of the software defined vehicles for a sustainable development of vehicular services.
    \item Improve the state-of-the-art on Software Defined Vehicle concept and its main building blocks.
    \item Propose an integration of Time-Aware and Deterministic Networks concepts and solutions within a Software Defined Vehicle.
    \item Definition of a set of potential use cases that aim to frame the proposed integration not only in the foreseeable future, but in a long-term perspective.
    \item Survey of hardware and software alternatives for validation of the proposed integration.
\end{enumerate}

The remaining of this document is organized into the following sections. \Cref{sec:sdv} introduces the background concepts, defining vehicular services and their characteristics, and the concept of \ac{sdv} and additional information on the industrial landscape. It also shows how automakers or suppliers are facing the \ac{sdv} paradigm. Next,~\cref{sec:blocks} presents the main building blocks for a~\ac{sdv} based vehicle (and vehicular services) development, and~\cref{sec:sota-sdv-discussion} discusses on the main challenges. \Cref{sec:use-cases} follows with the main use cases and requirements that can be drawn from the use cases. \Cref{sec:architecture} considers those requirements for the proposal of an architecture and of an evaluation process to validate the work. This considers the required hardware and software for both simulation and real-world testing. Finally, conclusions are provided in~\cref{sec:conclusions}.

\section{Software Defined Vehicles}
\label{sec:sdv}
\begin{figure*}[t!]
    \centering
    \includegraphics[width=0.9\textwidth]{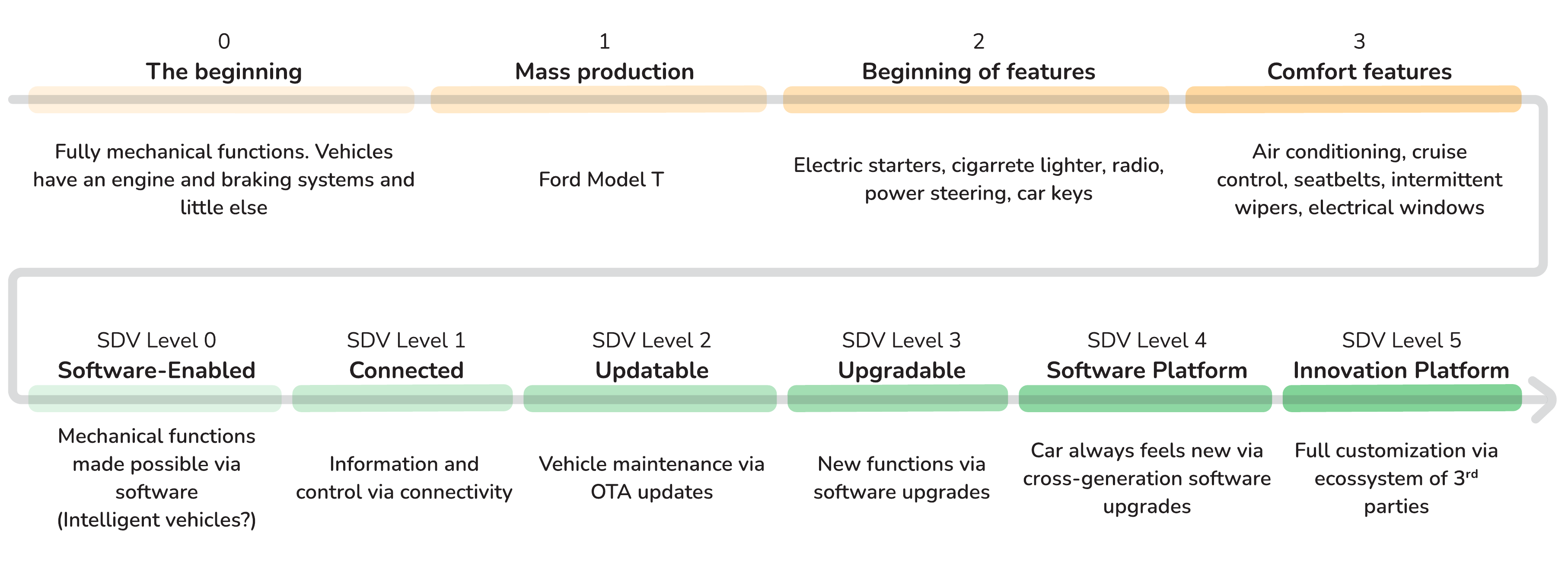} 
    \caption[Evolution of vehicular features over time, including levels of \acs*{sdv}. Partially based on~\cite{moritz,historycars}]{Evolution of vehicular features over time, including levels of \acs*{sdv}. Partially based on~\cite{moritz,historyOfCars}.}
    \label{fig:sdv-levels}
\end{figure*}

At the core of \ac{sdv} there is the concept that features within a vehicle are mostly defined through software. This follows a trend where vehicles, that started with little to no features, then evolved to features defined by mechanical components and by the 1980s started to define the features through software. \Cref{fig:sdv-levels} summarizes this evolution.

The \ac{sdv} paradigm main ideas are not new. Prior to the advent of \acp{sdv}, the automotive industry focus on software was mostly restricted to connectivity between vehicles and the user's smartphones to provide basic insights into the state of the vehicle~\cite{Zhang2018}. By 1998,~\cite{1998-Vehicle-Abstraction-Layer-40} considered intelligent vehicles as essential to autonomous transportation, designing a proposal of a vehicle capable of communicating with a centralized management control system. Also in 1998,~\cite{1998-Vehicle-Abstraction-Layer-41} proposes that smart vehicles are a junction of communication architectures, system architecture, sensors, tasks, kinematics and man-machine interface. The authors propose a communication architecture in a producer-distributor-consumer pattern with buses such as RS232 and \ac{can}, where time coherence is necessary. By 2001, a reference model architecture for intelligent vehicles~\cite{1992-Vehicle-Abstraction-Layer-39} considered the vehicle as having intelligent control systems with each layer with one or more computational nodes interconnected and connected to sensors.

Automotive vehicles need to evolve to a new concept of \ac{sdv} since in-vehicle services must be dynamic, and should be able to adapt over the vehicle life-cycle. At the same time, in-vehicle services need to interact with the vehicle’s systems, sensors, and the surrounding infrastructure within a well-defined time-frame. Therefore, systems must also be interconnected with a high-performance, high-precision, reliable set of networks and protocols with time guarantees.

\ac{sdv} aims to improve cars along their life-cycle~\cite{2023-TSN-19}, while potentially mitigating the increased unmanaged complexity of modern vehicular software, hardware, and communications. The paradigm states that vehicles manage their operations, add functionality, and enable new features primarily or entirely through software\footnote{More details at \myhref{blackberry.qnx.com/en/ultimate-guides/software-defined-vehicle}{blackberry.qnx.com}.}. Since vehicles evolve over their lifetime, with more features, solutions that may require remote control or remote updates, main features being delivered by software adds flexibility, reducing the difficulty of adding new features and allowing new business models to be considered by automakers during the lifetime of the vehicle.

\begin{figure}[t!]
	\centering
	\includegraphics[width=\columnwidth]{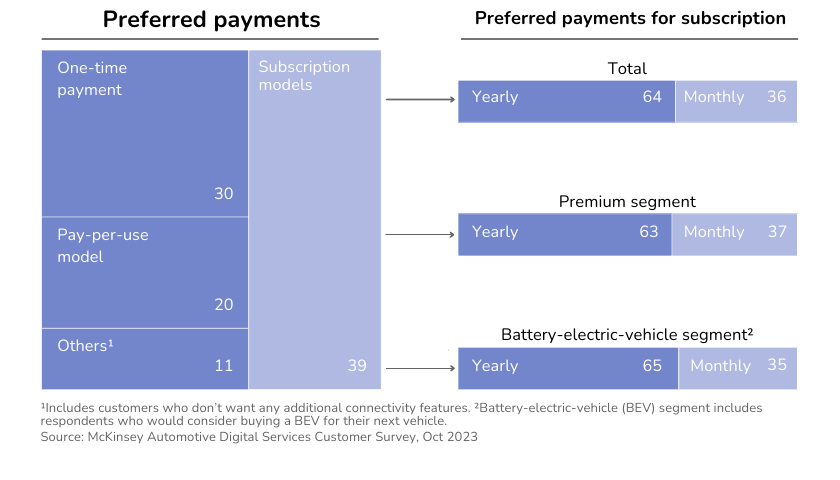}
	\caption[Preferred payments scheme to purchase/activate connectivity services, \% of respondents (n = 1649) showcasing a shift in preferred ownership payment models~\cite{mckinsey}]{Preferred payments scheme to purchase/activate connectivity services, \% of respondents (n = 1649) showcasing a shift in preferred ownership payment models~\cite{mckinsey}.}
	\label{fig:background-subscription}
\end{figure}

The	shift towards \ac{sdv} means that new features can be pushed to existing and new vehicles, which are now more defined through software than through hardware or mechanical systems as in the past~\cite{2022-Problem-8}. Not only this adds flexibility to vehicle manufacturers, but also creates new business models where owners can pay for features after buying the vehicle and during its lifetime. This follows a trend where vehicles are now seen by potential owners in a subscription payment model, instead of the one-time payments or leasing/pay-per-use models (\cref{fig:background-subscription}).

Such support not just improves the development of vehicular services, but also helps towards a decrease in hardware complexity (contributing towards a much needed decrease in the number of \acp{ecu}, cables and the cost and weight savings which follow), with more uniform software code-base (and therefore reduction of the costs and complexity of the development) and greater adaptability (essential specially if the future of the car ownership is car sharing~\cite{2020-Problem-11}). This adaptability also supports a greater capability to process higher volumes of data while fulfilling the overall performance requirements — essential in \acp{cav}.

\subsection{Industrial landscape}
\label{sec:sdv:industrial}
Vehicular manufacturers are aware of the necessary changes in thinking. However, the transition from the previous mentality of closed source, customized solutions for each model of vehicle towards a more extendable, interoperable, architecture is far from being over. The Renault, Volvo/Polestar and Volkswagen groups seems to be ahead in terms of a more interoperable architecture based on \acp{ecu} interconnected through Ethernet that run Android and well-known versions of chipsets. It is however not possible to ignore efforts from other vehicle manufacturers that may be developing equivalent solutions but have not, at the time of this survey, announced their efforts. The opacity of the vehicular industry, while expected within a competitive market, it remains and will remain a threat to any related work, including this work. To mitigate this threat, research of \textit{gray literature} is essential, targeting not only automakers but their suppliers and open source community initiatives.

\begin{figure}[t!]
	\centering
	{
\sffamily
\begin{tikzpicture}
  \begin{axis}[
    xmin=-10, xmax=10,
    ymin=-10, ymax=10,
    axis lines=center,
    axis on top=true,
    domain=-10:10,
    xticklabels={,,},
    yticklabels={,,},
    xtick style={draw=none},
    ytick style={draw=none},
    xlabel style={align=left}, xlabel=More features\\More versatile, 
    ylabel style={align=left}, ylabel=More standard\\More open,
    scale=1.0,
    label style={font=\tiny},
    ]
    
    \addplot [
        mark=*,
        only marks,
        nodes near coords, 
        every node near coord/.style={font=\small, align=left, anchor=west},
        scatter src=explicit symbolic
    ] table [meta index = 2] {
x y label
-8  5   Primitive\\vehicle\\system
-8  -5  Explosion\\of\\vehicle\\systems
0   -9  AUTOSAR\\Classic
3   -5  AUTOSAR\\Adaptive
8   -2  CARIAD\\BMW
6   3   Renault\\Volvo
8   6   Opportunity?
10   8  Tesla
    };
    
    \draw [dashed] (axis cs:0,-10) -- (axis cs:0,10);
    \draw [dashed] (axis cs:-10,0) -- (axis cs:10,0);  
    
    \draw [blue,dashed] (axis cs:8,6) circle [radius=2];

  \end{axis}
\end{tikzpicture}
}
	\caption[Positioning of main automakers and frameworks over the overall industrial landscape]{Positioning of main automakers and frameworks over the overall industrial landscape.}
	\label{fig:soa-sdv-automakers}
\end{figure}
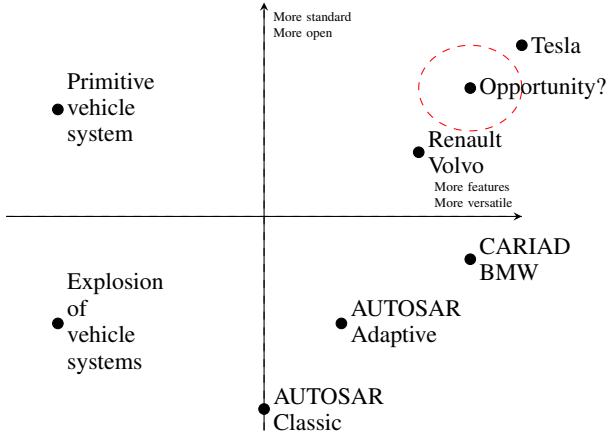

\Cref{fig:soa-sdv-automakers} summarizes the efforts of automakers on adoption the main aspects related to \ac{sdv}, with~\cref{tab:soa-sdv-automakers} providing further detail per associated areas.

\begin{table}[!htpb]
	\centering
	\caption[Overview of the manufacturers effort on \acs*{sdv}]{Overview of the manufacturers effort on \acs*{sdv}.}
	
	\resizebox{\columnwidth}{!}{%
		\begin{threeparttable}[t]
			\begin{tabular}{p{0.2\columnwidth}ccccccc}
				\hline
				\multirow{ 2}{*}{\textbf{Entity}}& \multicolumn{7}{c}{\textbf{State of integration}} \\
				& Ethernet   & \acs{tsn}  & Zonal \acp{ecu} & \acs{soa}  & \acs{sdv}  & Open  & Standard \\ \hline
				\acs*{autosar} Classic             & $\times$   & $\times$   & $\times$   & $\times$     & $\times$   & $\times$  & \checkmark \\ \hline
				\acs*{autosar} Adaptive            & \checkmark & $\times$   & $\times$   & \checkmark-  & \checkmark-  & $\times$ & \checkmark \\ \hline
				BMW                         & \checkmark & \checkmark & \checkmark & \checkmark & \checkmark & \checkmark & $\times$ \\ \hline
				Renault\tnote{a}            & \checkmark & \checkmark & \checkmark & \checkmark & \checkmark & \checkmark- & \checkmark-           \\ \hline
				Volvo                       & \checkmark & \checkmark & \checkmark & \checkmark & \checkmark & \checkmark-\tnote{c} & $\times$   \\ \hline
				Volkswagen group\tnote{b}   & \checkmark? & \checkmark? & \checkmark? & \checkmark? & \checkmark & $\times$             & $\times$   \\ \hline
				NVIDIA                      & \checkmark & \checkmark & $\times$        & \checkmark & $\times$   & $\times$             & $\times$   \\ \hline
				Tesla                       & \checkmark & \checkmark & \checkmark &  \checkmark  & \checkmark &  \checkmark & \checkmark-            \\ \hline
				Eclipse                     & \checkmark & \checkmark & \checkmark &  \checkmark  & \checkmark &  \checkmark & $\times$            \\
				\hline
			\end{tabular}
			
			\begin{tablenotes}
				\item[a] Renault software factory
				\item[b] CARIAD
				\item[c] Partial access through Volvo \acp{api}.
			\end{tablenotes}
		\end{threeparttable}%
	}
	
	\label{tab:soa-sdv-automakers}
\end{table}

Existing automakers were not agile enough within the development of their vehicular services~\cite{2020-Problem-6}, which lead to the rise of new automakers, such as Tesla, in very software-intensive tasks such as autonomous driving. The Tesla company is in fact probably the most well-known example of application of the \ac{sdv} paradigm principles. Within its vehicles, the \acp{ecu} — such as the Autopilot Control Unit, the Media Control Unit, and several Body Control Units — are interconnected through Ethernet, and are in a lower number than the usual within vehicles up to that point. This zonal architecture supports \ac{adas} systems for level 2+ autonomous driving, such as the Full Self-Driving (Supervised) that is then sold to end-users in a subscription manner\footnote{Details at \myhref{https://www.tesla.com/support/full-self-driving-subscriptions}{tesla.com/support/full-self-driving-subscriptions}.}.

\begin{figure}[!ht]
	\centering
	\includegraphics[width=0.8\columnwidth]{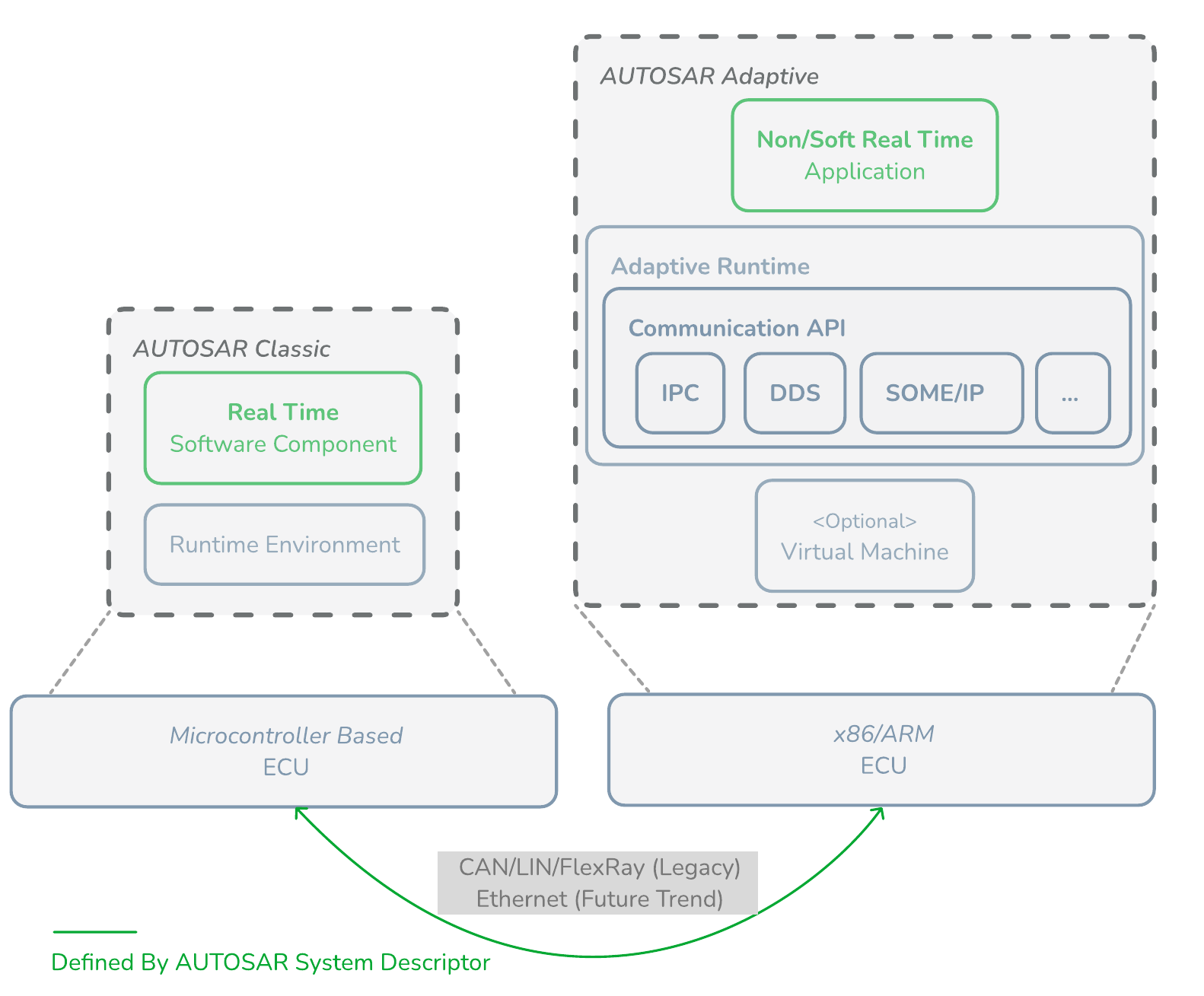}
	\caption[\acs*{autosar} Classic VS Adaptive]{\acs*{autosar} Classic VS Adaptive.}
	\label{fig:soa-autosarclassic-overview}
\end{figure}

As a standardized "standardized software framework and open E/E system architecture", the \ac{autosar}\footnote{\myhref{https://www.autosar.org/}{autosar.org}} is a \emph{de facto} standard, with the support from vehicle manufacturers and vendors. It is a highly complex ecosystem, with more than 100 modules and its documentation reaching more than 21 000 pages~\cite{AUTOSARE36:online}. The standard has two different types of platforms as depicted in~\cref{fig:soa-autosarclassic-overview}: the Classic, for traditional hard real-time systems based on micro-controllers with a static configuration, and set of pre-compiled monolithic binaries; and also a newer Adaptive platform with greater flexibility and adaptability. 
The Classic platform is aimed at safety critical systems with hard real-time requirements with a stated linkage between services and CPU cores. Development through the original Classic Platform is however, very hard-coded to the vehicle and its \ac{ee} architecture~\cite{2010-Vehicle-Abstraction-Layer-30}. The Adaptive is capable of considering heterogeneous high performance \acp{ecu} and departs from legacy buses~\cite{Bandur2021} such as \ac{can} towards a \ac{soa} communication architecture (see ~\cref{sec:blocks:soa}) with Ethernet based on \ac{someip} as bus protocol~\cite{2016-Vehicle-Abstraction-Layer-26} and is overall more dynamic but less time predictable~\cite{2021-TSN-17,2023-TSN-20}, essential within critical vehicular systems.

After the landscape changing forced by the United States through Tesla, European and Asiatic automakers such as BMW~\cite{2020-TSN-22}; the Renault group through the Renault Software Factory\footnote{More details on \myhref{ https://www.reuters.com/business/autos-transportation/renault-seeks-software-architecture-par-with-tesla-by-2026-2023-04-24/}{www.reuters.com}.}; Volvo; Kia and Hyundai\footnote{More details on \myhref{https://www.carscoops.com/2023/04/hyundai-and-kia-lead-new-consortium-to-develop-advanced-software-defined-vehicles/}{www.carscoops.com}.}; and the Volkswagen group through the CARIAD\footnote{Available at \myhref{https://cariad.technology/}{cariad.technology}.}, supported by companies such as Vector~\cite{Vector2020} and VXworks, are shifting towards a software-first development of the vehicles, based on novel hardware and communication paradigms. 

A major player within the European market, the Volkswagen group take on \ac{sdv} is based on creating and controlling the whole platform — cloud (Volkswagen Automotive Cloud VW.AC), hardware and software (VW.\ac{os})~\cite{2023-Vehicle-Abstraction-Layer-23}, affecting interoperability with other automakers and hardware providers. While in initial stages, it highlights a possible tendency towards a close proprietary \ac{sdv} paradigm. However, it was not possible to obtain documentation on \acp{api} or \acp{sdk} that should be used to developed vehicular services using the CARIAD ecosystem. 
A notable exception is the Roobadge\footnote{Product description at \myhref{https://www.volkswagen.com.au/en/roobadge.html}{https://www.volkswagen.com.au/en/ roobadge.html}.}, a solution to avoid  collisions with kangaroos — very common in Australia — by adding, through a front badge or license plate holder sensors to detect the kangaroos. This solution not only showcases a real example of the potential of \ac{sdv}, since new features are added to an existing vehicle through a new deployed service, but is also compatible with other brand's vehicles, showcasing a real \ac{sdv} example that considers brand interoperability (even if the actual implementation is closed).

Still within Europe, the Renault group considers a backbone of \acp{ecu} interconnected with Ethernet, with gateways to bridge the Ethernet connected \acp{ecu} with legacy sensors or actuators interconnected by \ac{can}~\cite{2023-TSN-19,2019-TSN-23}. These \acp{ecu} run Android and are based on Qualcomm chipsets, supporting an interoperable, open \ac{sdv} architecture\footnote{More details at \myhref{https://www.qualcomm.com/news/releases/2021/09/qualcomm-works-google-bring-premium-and-intelligent-vehicle-experiences}{www.qualcomm.com}.}. Volvo and Polestar have a similar approach, with zone controllers (more in~\cref{sec:blocks:zonal}) — high performance controllers in specific zones of the vehicle — interconnected and connected to a central master \ac{vcu} with an Ethernet backbone~\cite{2023-TSN-20}. The Zone Controllers can also act as a bridge to legacy buses, like \ac{can} and \ac{lin}. Volvo within their \ac{sdv} efforts also published OpenAPI specifications for their \acp{cav}~\cite{2023-Vehicle-Abstraction-Layer-43}. 

Other European players include Mercedes-Benz, BMW and Stellantis. Mercedes-Benz, after a transition to a zonal \ac{ee} architecture (with separations between infotainment, body and comfort, \ac{adas} and interior domains), is working on the MB.OS, a proprietary chip-to-cloud platform capable of ensuring decoupling software from hardware. This solution is already allowing \ac{ota} updates to significantly change the MB User Experience (MBUX), with future focus on standardization efforts~\cite{MercedesBenz}. 
The BMW group also aims to go from signal oriented to unified vehicle wide layered \ac{soa} architecture with providers and consumers~\cite{2020-TSN-22,Unveiled86:online}. This also requires latencies, configurations and bandwidth options encoded in the system. The Stellantis group aims at supporting sustainability and user experience use cases through service oriented solutions. \ac{soa} can provide the necessary scalability previous service development paradigms could not — namely 1 application per \ac{ecu} — while allowing a firm contract between services~\cite{IEEE_SA_d1_02}, as described in~\cref{sec:blocks:soa}.

Within the United States of America, General Motors is working towards their \ac{sdv} efforts through the uServices solution, a set of \acp{api} for developers and Ultifi, a end-to-end software platform that \enquote{will enable \ac{ota} updates and in-car subscription services}~\cite{GMnowhas36:online}.

When on the tier-2+ suppliers for automakers, it is clear they are also working towards providing automakers with new tools. Continental, with the Amazon \ac{iot} for Automotive\footnote{More details at \myhref{https://catalog.workshops.aws/awsiotforautomotive/en-US}{catalog.workshops.aws/awsiotforautomotive/}.} support, is also working towards a \enquote{a modular multi-tenant hardware and software framework that connects the vehicle to the cloud}\footnote{More details at  \myhref{https://aws.amazon.com/pt/blogs/architecture/developing-a-platform-for-software-defined-vehicles-with-continental-automotive-edge-caedge/}{aws.amazon.com}.}. 
Nvidia is entering the market with a clear AI first approach — the Nvidia DRIVE \ac{os}, which includes a embedded \ac{rtos}, a hypervisor, \ac{cuda} libraries and TensorRT, and features such as secure boot, security services, firewall, and \ac{ota} updates\footnote{Developer documentation at \myhref{https://developer.nvidia.com/drive/drive-sdk}{developer.nvidia.com/drive}}. Vector\footnote{At \myhref{https://www.vector.com/}{vector.com}} within its ecosystem propose tools to support development of vehicles with \acp{ecu} interconnected by Ethernet with TSN and of services using \ac{autosar} within \ac{soa} architectures. MotionWise~\cite{2023-Vehicle-Abstraction-Layer-33} proposes a middleware for safe orchestration of vehicle software across \acp{ecu} for automated driving platforms, with support for \ac{autosar} classic and adaptive, \ac{ros}, and \ac{dds} support. Hardware companies such as Renesas propose the AosCloud\footnote{More details at \myhref{https://www.renesas.com/us/en/software-tool/aosedge-platform}{renesas.com}.}, a solution to support \acp{oem} and service providers from seamless software installation to operation in the vehicle, and  implement various new services and functional enhancements simply. 

Regarding the open-source community, alternatives such as the Eclipse Foundation, within its \ac{sdv} working group, is also working within incubating initiatives to define: a) orchestrators and virtualization managers (Ankaios, Velocitas); b) vehicle abstraction models (Chariott); c) a development platform (Kuksa); and d) an abstraction layer for communication between nodes (eCAL) and operating system (MoEc.\ac{os}) with support for \ac{dds} as well as simulators such as SUMO and CARLA (see~\cref{sec:arch-testbed}). While in an initial development phase, these projects can be a good future basis for anyone entering the \ac{sdv}. The \ac{soafee}\footnote{\myhref{https://www.soafee.io/}{soafee.io}, more details at \myhref{https://www.arm.com/blogs/blueprint/cloud-native-automotive-development}{arm.com}.} presents a cloud-native alternative with an architecture enhanced for mixed-criticality automotive applications with corresponding open-source reference implementations to enable commercial and non-commercial offerings. Autoware, an open-source project for autonomous driving\footnote{\myhref{https://autoware.org/}{autoware.org}}, through its Open AD Kit, the first \ac{soafee} blueprint, enables cloud-native DevOps of autonomous driving solutions for \ac{sdv}\footnote{Developer documentation at \myhref{https://autowarefoundation.github.io/open-ad-kit-docs/latest/version-2.0/}{autowarefoundation.github.io}.}.
 

These individual initiatives are being target towards a cooperation. By 2024, at CES 2024, \ac{autosar}, \ac{covesa}, Eclipse \ac{sdv}, and \ac{soafee} announced a \enquote{collaboration of collaborations} to \enquote{align efforts in the \ac{sdv} ecosystem and create a joint \ac{sdv} vision clear and unified definition of what constitutes an \ac{sdv}}, which points towards a brighter future in \ac{sdv} with greater interoperability~\cite{202401SD57:online}. Innovation projects such as the HAL4SDV\footnote{\myhref{https://www.hal4sdv.eu/}{hal4sdv.eu}} also aim to contribute to a more holistic solution by including in its goals a "standardized interfaces to sensors, actors, compute resources and persistent storage".

\begin{figure*}[t!]
    \centering
    \includegraphics[width=\textwidth]{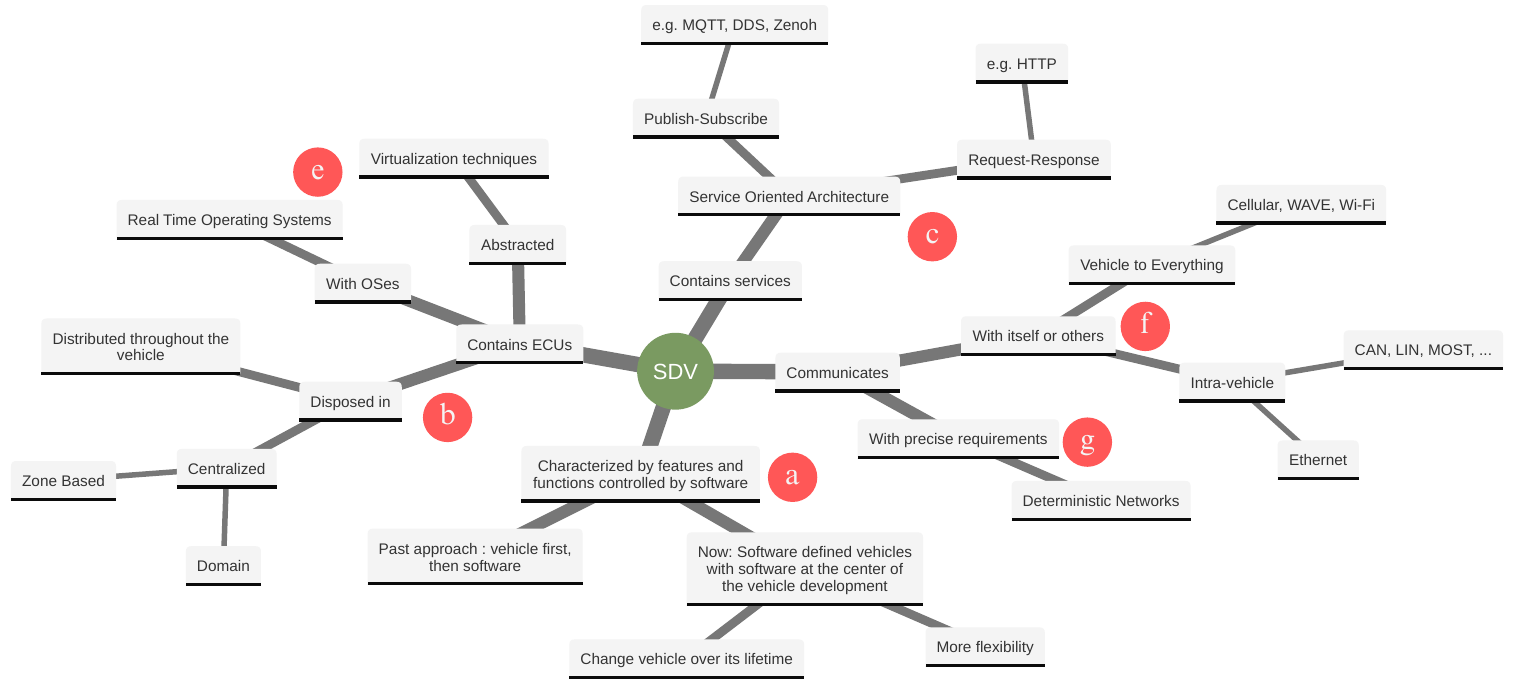} 
    \caption[Overview of the topics involved in modern vehicular development to be referred]{Overview of the topics involved in modern vehicular development to be referred, lightly based on~\cite{Liu2022}.}
    \label{fig:intro-overview}
\end{figure*}

\section{Main building blocks for Software Defined Vehicles}
\label{sec:blocks}

Developing a vehicle within the \ac{sdv} paradigm~(\rcircle{a}) requires considerations on several domains within the context provided by~\cref{fig:intro-overview}, including:

\begin{itemize}

    \item \textbf{Physical disposition of the vehicular \acp{ecu} and specifications of each \ac{ecu}}~(\rcircle{c},~\cref{sec:blocks:zonal}): this domain explores how \acp{ecu} are physically disposed throughout the vehicle, namely considering a zonal \acl*{ee} architecture.
    
    \item \textbf{\aclp*{soa}}~(\rcircle{b},~\cref{sec:blocks:soa}): this domain explores how vehicular services are developed and communicate between one another and abstractions for the vehicular service development and for transferring data between services. 

    \item \textbf{Virtualization techniques and operating systems}~(\rcircle{d},\rcircle{e},~\cref{sec:blocks:virtualization,sec:blocks:os}): this domain considers how vehicular services are deployed within each \ac{ecu}, including  the software stack within each \ac{ecu} and virtualization techniques.

    \item \textbf{Intra and inter vehicular communication architectures, protocols and physical layers}~(\rcircle{f},~\cref{sec:blocks:comms-intra,sec:blocks:comms-inter} this domain considers how vehicular \acp*{ecu} communicate with each other and with \acp*{ecu} on other vehicles (intra-vehicular and inter-vehicular communications, respectively).
    
    \item \textbf{Deterministic vehicular communications}~(\rcircle{g},~\cref{sec:blocks:deterministic}): this domain considers how to ensure vehicular communications always happen and within the necessary reliability and timing requirements.
\end{itemize}

The next subsections further detail each one of these building blocks, including their state of the art and associated main challenges. 

\subsection{Physical E/E architectures}
\label{sec:blocks:zonal}

\begin{figure*}[!ht]
	\centering
	\includegraphics[width=2\columnwidth]{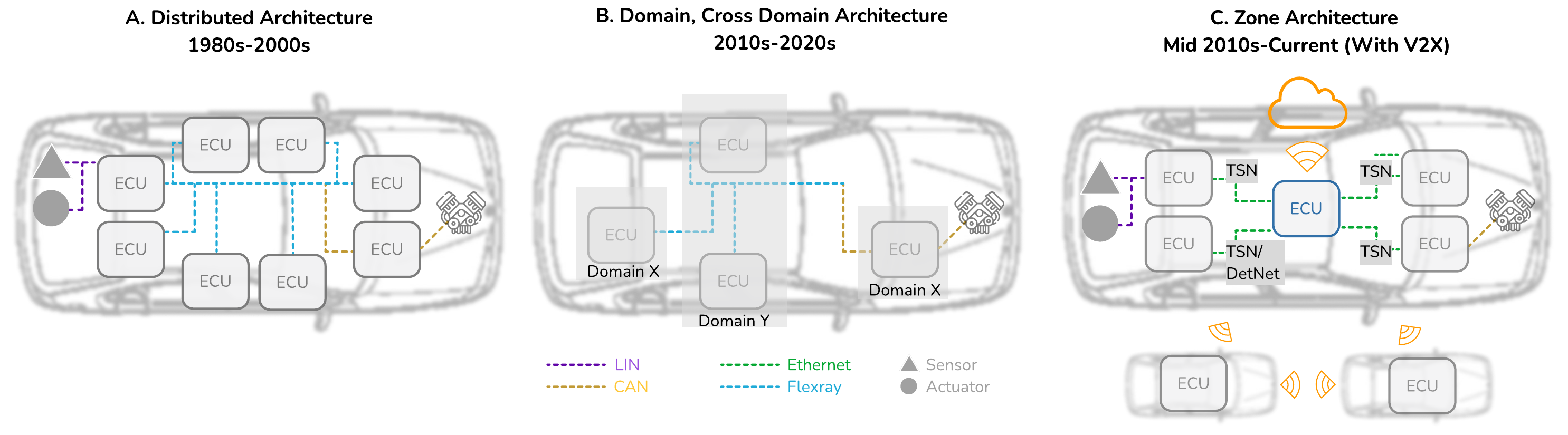}
	\caption[Legacy distributed, domain, and zonal \acs*{ee} architectures within a vehicle]{Legacy distributed, domain, and zonal \acs*{ee} architectures within a vehicle.}
	\label{fig:background-sdv-zonal_soa_overview}
\end{figure*}

To support \acp{sdv}, the \acl{ee} and computational architectures within vehicles must also evolve from the fully distributed vision, with \acp{ecu} scattered throughout the vehicle towards a more centralized solution, with zone-based architectures, \ie \acp{ecu} positioned in specific parts of the vehicle responsible for certain types of actions~\cite{2020-Problem-7}. \Cref{fig:background-sdv-zonal_soa_overview} provides an overview of this evolution.

During the 1980s (\cref{fig:background-sdv-zonal_soa_overview}a)), new features were added at the cost of adding new specialized \acp{ecu} , close to the sensors and actuators and connected through \acs*{can} or other equivalent bus. With the explosion of features in vehicles, especially luxury cars, during the 1990s, the number of ECUs increased significantly, making vehicles more complex, and thus increasing the amount of systems that need to be choreographed, developed, tested and that have to communicate, as well as the possibilities of something going wrong~\cite{HowHaveV59:online, HowAreVe60:online}. 

By the 2000s, the high complexity of intra vehicular systems became subject to \textit{de facto} standards such as \ac{autosar}~\cite{HowHaveV59:online,HowAreVe60:online} and buses such as FlexRay aimed at mitigating some of the shortcomings of \ac{can}. Since the \acp{ecu} were scattered throughout the vehicle, with little to no centralization, these legacy architectures were distributed, allowing a better decoupling, separation of concerns and higher tolerance to equipment failure. However, challenges such as the scalability of the deploying services across multiple \acp{ecu} (considering hardware variations and the absense of virtualization), communication performance, and the high costs of the associated complex, bulky, wiring harnesses have driven a trend towards a more centralized architecture, consolidating processing into fewer computing units.  

A centralized architecture can reduce costs and complexity by decreasing the number of \acp{ecu}, increasing software reusability, and enabling greater flexibility.The centralized \ac{ee} architectures supporting \acp{sdv} follow the principle of aggregation of \acp{ecu} into a small subset of more powerful \acp{hpc}. Each \ac{hpc} can potentially contain multi core \acp{cpu}, \acp{gpu}, \acp{dsp}, and potentially \acp{fpga}~\cite{HowHaveV59:online,HowAreVe60:online,TheEEarc76:online}, as well as high performance network capabilities (\eg Ethernet). Some \acp{ecu} can eventually be dedicated to computing intensive tasks, including \ac{ml} and \ac{dl} workloads supporting \ac{adas} systems~\cite{Kugele2018}. This follows previous trends within other industries, in particular the aviation industry~\cite{Bandur2021}.


Centralization within the intra-vehicular systems can be achieved at a domain, cross domain or zone/vehicle centralized way~\cite{Bandur2021}. By 2010s, domain and cross domain architectures (\cref{fig:background-sdv-zonal_soa_overview}b)) were a first step towards mitigating those issues through the aggregation of functions per domain. In a domain-oriented solution, the grouping of \acp{ecu} into \acp{hpc} is done by vehicle function since each function requires heavy interaction between \acp{ecu} (specially in complex systems such as \ac{adas}). In a cross-domain solution, functions of several domains are consolidated within more powerful \acp{ecu}. This minimizes, without removing, the high bandwidth required for communication between \acp{ecu}. However, the integration of multiple domains required by systems such as \ac{adas} still require coordination between different domains~\cite{HowHaveV59:online,HowAreVe60:online}.

Zone-oriented architectures (\cref{fig:background-sdv-zonal_soa_overview}~(b)) aimed to solve the issue of coordination between domains. Each zone of the vehicle collects and forwards information to a much more powerful central \ac{ecu}, or even offloads to the cloud. Bosch reported on achieving a 15–20\% reduction in wiring harness weight when using a zone-oriented architecture instead of a domain-oriented architecture, showcasing the potential of zonal architectures versus the previous approaches.

Future approaches should also consider other vehicles and infrastructure (\eg smart cities) as an integral part of the \ac{sdv} environment services can consider for deployment.The next phases of automotive software development should also have a deep integration with cloud and follow the best practices of software development (such as CI/CD)~\cite{2021-Vehicle-Abstraction-Layer-38} and should consider deterministic networking concepts and standards, as discussed in~\cref{sec:blocks:deterministic}. 


\subsection{Service-oriented architectures}
\label{sec:blocks:soa}

\begin{figure}[!ht]
	\centering
	\includegraphics[width=\columnwidth]{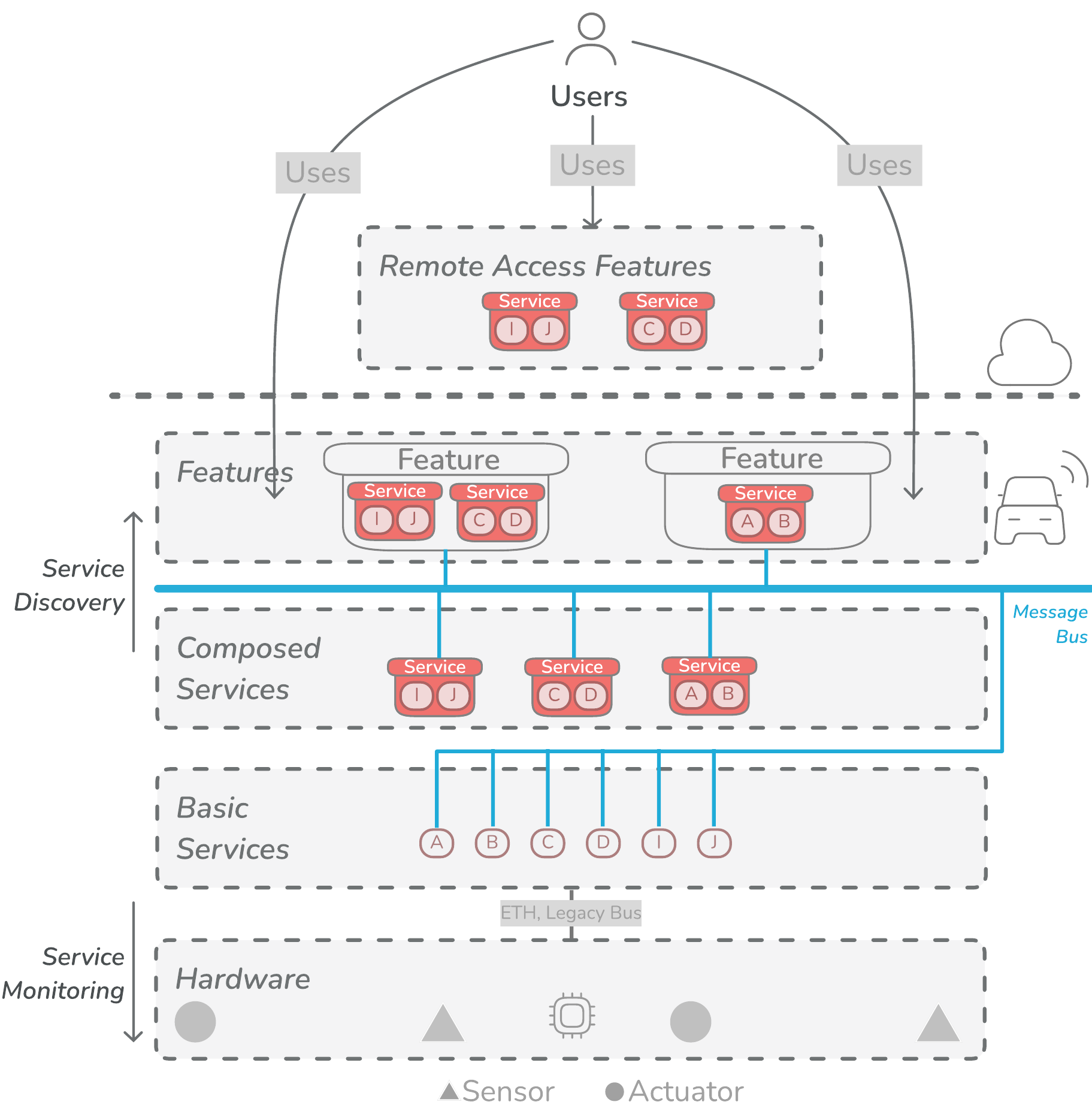}
	\caption[Service-oriented architectures in vehicles as perceived by the Renault Group automaker]{\acs*{soa} in vehicles as perceived by the Renault Group automaker (adapted). Services (in red) are the basic unit of vehicular applications and communicate with one another via one or more message bus~\cite{2023-TSN-19}.}
	\label{fig:background-sdv-soa-renault}
\end{figure}

As vehicular \ac{ee} architectures become more complex, standards and well defined patterns are required in order to model both the \acp{ecu} hardware and the software. \aclp{soa} are an example of such architectural pattern, where each component provide and consume services. \Cref{fig:background-sdv-soa-renault} presents an overview of the \acs*{soa} paradigm. Services in this context are a discrete unit of functionality, from directly interacting with sensors and actuators, up to more complex interactions between simpler services. A service contains the \enquote{\textit{mechanism to enable access to\textbf{ one or more capabilities}, where the access is provided using a \textbf{prescribed interface} and is exercised consistent with constraints and policies as specified by the \textbf{service description}}}~\cite{OASISSOA53:online}. 
 
When implemented on a vehicle, on a more reduced set of (zonal) \acp{ecu}, the \ac{soa} paradigm brings benefits such as the decoupling of hardware and software, increased modularity and separation of concerns, and contribute towards the overall decrease in hardware complexity.~\cite{richards2015microservices}. These advantages are, however, at the cost of increased complexity in maintaining contracts between services and ensuring availability and discovery of the services~\cite{richards2015microservices}. This is particularly complex within a vehicular context, with a \ac{soa} supporting at most 60 to 80 services~\cite{2020-TSN-22}, not addressing situations where there are hundreds or thousands of micro-services. 


Within \ac{soa} based architectures, data layers (message bus in \cref{fig:background-sdv-soa-renault} can abstract the physical medium of services by providing a unified way of exchanging and accessing data between services on different components/\acp{ecu}~\cite{NXPVehicleA52:online}. Information from vehicle sensors can then be fed from/to services leveraging the \acp{ecu} and their communication paradigm through data layer solutions.

Three main paradigms can be considered for implementing a data layer~\cite{2022-TSN-21}: a) a fully periodic publisher-subscriber based or b) a fully event triggered client-server based. Examples for the publish/subscribe paradigms include the \ac{dds}\footnote{At \myhref{https://www.dds-foundation.org/}{dds-foundation.org}.} middleware, \ac{someip}, or \ac{mqtt}~\cite{2023-TSN-20}. Publish/subscribe paradigms can respond to a data centric paradigm, when the data sending is event triggered, unlike solutions such as \ac{http} or \ac{coap} based on the request/response paradigm~\cite{Vector2020}. 
Abstraction layers such as uprotocol\footnote{At \myhref{https://projects.eclipse.org/projects/automotive.uprotocol}{projects.eclipse.org/projects/automotive.uprotocol}.} can be considered to decouple vehicular services from the chosen data layer.

\Cref{tab:background_iotStandards} provides a summary on some of the main possible solutions for \ac{soa} data middleware. While all these solutions can ensure delivery of data, there is a tendency to consider \ac{dds} and Zenoh the best performant solutions since they provide better flexibility and performance. According to~\cite{liang2023performance}, which compared performance of \ac{mqtt}, \ac{dds}, and Zenoh, the latter shows (for multi-machine scenarios, more realistic in zonal vehicular architectures) a double increase on the throughput of \ac{dds} and 50 times of \ac{mqtt}, while \ac{dds} usually provides the best latency and better flexibility with more \ac{qos} levels.
\begin{table*}[!ht]
\centering
\caption[Qualitative comparison of \acs*{iot} standards]{Qualitative comparison of \acs*{iot} standards (partially based on \cite{HowDoesD67:online, SASG72Ro66:online, Deployme63:online}).}
\label{tab:background_iotStandards}
\resizebox{0.9\linewidth}{!}{%
	\begin{threeparttable}[t]
		\begin{tabularx}{\linewidth}{p{0.15\linewidth}XXXXX}
		\hline
		\textbf{Standard} &
		   \textbf{\acs*{mqtt}}&\textbf{\acs*{dds}} &
		  \textbf{\acs*{someip}} &
		  \textbf{Zenoh}&
		  \textbf{\acs*{http}}  \\ \hline
		\textbf{Transport} &
		   \acs*{udp} (in \acs*{mqtt}-SN), \acs*{tcp}&\acs*{udp}, \acs*{tcp} &
		  \acs*{ip} &
		  \acs*{udp}, \acs*{tcp}, QUIC&
		  \acs*{tcp},  QUIC  \\ \hline
		\textbf{Paradigm} &
		   Pub / Sub&Pub / Sub &
		  Pub / Sub &
		  Pub / Sub
		
		Data storage&
		  Request / Reply  \\ \hline
		\textbf{Node   Discovery}&
		   No&Yes (\acs*{dds-sdp}  and alternatives) &
		  Yes &
		  Yes&
		  No  \\ \hline
		\textbf{Fault  Tolerance} &
		   Broker is  \acs*{spof}&Yes, fully distributed by default&
		  Yes &
		  Yes, can be fully distributed&
		  Client and server are \acs*{spof}  \\ \hline
		\textbf{Data  Schema} &
		   None&Required &
		  Required &
		  Only encoding is required&
		  Optional\tnote{c}  \\ \hline
		\textbf{QoS  Policies}&
		   At delivery  level&At multiple entities\tnote{a} &
		  N/A &
		  At 2 levels\tnote{b} &
		  N/A  
		 \\ \hline\end{tabularx}
		 
		 	\begin{tablenotes}
			 	\item[a] For delivery, presentation, transport, availability, cache, resource management.
			 	\item[b] For reliability and fragmentation level.
			 	\item[c] Based for example on the \texttt{Content-Type}.
		 	\end{tablenotes}
 	\end{threeparttable}%
}
\end{table*}

\subsection{Vehicular operating systems}
\label{sec:blocks:os}

\acp{ecu} within a vehicle run services that require deterministic scheduling from the \ac{os}. Failure to have a deterministic real-time scheduling can lead to services being dispatched too soon or too late to the \ac{os}, with dramatic effect on the services behavior. Therefore, \acl{rtos} are a need within vehicular computing domains to ensure services are dispatched in a determined specific order and time-frame. While many \acp{os} can support real-time constraints — including any generic Linux kernel based \ac{os} with \verb|PREEMPT_RT|\footnote{Available at \myhref{https://wiki.linuxfoundation.org/realtime/}{wiki.linuxfoundation.org/realtime}.} patches~\cite{gutiérrez2018realtime}, some solutions are heavily automotive focused, such as Android Automotive, Nvidia Drive \ac{os} and QNX.

Android Automotive\footnote{Not to be confused with Android Auto, a platform to use a smartphone as a infotainment alternative.} and its extensions such as SnapOS\footnote{Available at \myhref{https://www.snappautomotive.io/snappos}{snappautomotive.io}.} provide an \ac{os} and a platform that is based on Android — and can therefore leverage the existing application ecosystem — while supporting specific characteristics of vehicles, including the support for the creation of a vehicle abstraction layer required to abstract the vehicle's specific/proprietary components, UI support for setting vehicle's settings and \ac{ota} updates.

Nvidia Drive \ac{os}\footnote{Dev documentation at \myhref{https://developer.nvidia.com/drive/driveos}{developer.nvidia.com/drive}.} is the \enquote{reference operating system and associated software stack designed specifically for developing and deploying autonomous vehicle applications on DRIVE AGX-based hardware}. It is in fact a custom \ac{rtos} with a Nvidia Hypervisor and necessary \ac{cuda} and TensorRT libraries to fully utilize the hardware (graphical) potential.

Yocto\footnote{Available at \myhref{https://www.yoctoproject.org/}{yoctoproject.org}.} based \acp{os} are also a possibility. Yocto supports the creation of custom \ac{os} based on Linux, allowing adding custom layers to embedded systems. Yocto based Wind River Linux\footnote{At \myhref{https://www.windriver.com/products/linux}{windriver.com}.} is an example of such family of \ac{os}, providing  support for multiple board support packages in both x86 and Arm hardware. Automotive Grade Linux\footnote{At \myhref{https://docs.automotivelinux.org/}{docs.automotivelinux.org}, \myhref{https://github.com/nxp-auto-linux/auto\_yocto\_bsp}{github.com/nxp-auto-linux}.} is another example of a Yocto based \ac{os} and a industry supported initiative to develop software for the \enquote{infotainment, instrument cluster, \ac{hud}, telematics, connected car, \ac{adas}, functional safety, and autonomous driving}~\cite{Nxpcorpo93:online}.

Alternative \acp{os} include the VxWorks\footnote{Available at \myhref{https://www.windriver.com/products/vxworks}{windriver.com}.}~\cite{2022-Virtualization-44}, a \ac{rtos} for mission critical embedded systems. While usually considered in space scenarios, automotive are also supported for example through the integration with \ac{autosar} Adaptive Platform. The QNX\footnote{Available at \myhref{https://blackberry.qnx.com}{blackberry.qnx.com}.} is another alternative as a micro-kernel \ac{rtos} with extensive foothold in the automotive sector installed in more than 215 million vehicles.
Another open source initiative, albeit in initial stages (launched by January 2024), is the Eclipse ThreadX\footnote{Previously Azure \ac{rtos}, now \myhref{https://threadx.io/}{threadx.io}, with dev documentation at~\myhref{https://projects.eclipse.org/proposals/eclipse-threadx}{eclipse.org}.}, an open-source, vendor neutral \ac{rtos}.

These operating systems, being mostly \ac{rtos} or \ac{rtos}-like, can be evaluated through the worst-case execution time for each task, modularity, response time or jitter~\cite{sym12040592}. At the time of writing, no comparison between some or all of the aforementioned \acp{os} was found.

Vehicular services should be developed independently of the \ac{rtos} and thus an isolation between both is desired. Initiatives such as the \ac{osal} from IBM\footnote{Dev documentation at \myhref{https://www.ibm.com/docs/en/engineering-lifecycle-management-suite/design-rhapsody/9.0.2?topic=references-operating-system-abstraction-layer-osal}{ibm.com}.} or NASA \enquote{isolates embedded application software from a \ac{rtos}} by providing \enquote{a well-defined, generic interface to \ac{rtos} services}\footnote{More information at \myhref{https://opensource.gsfc.nasa.gov/projects/osal/}{opensource.gsfc.nasa.gov}.} and can, therefore, contribute to the necessary portability of services between \acp{os}. Without such initiatives, the definition of the \ac{os} will be mostly related to the associated ecosystem, hindering possiblity interoperability between services developed for each \ac{os}. 

\subsection{Virtualization and orchestration techniques}
\label{sec:blocks:virtualization}
To ensure optimal usage of \acp{ecu}, with concurrent execution of different services with different criticality, partitioning of scheduling of each core per service(s) is required. Virtualization of resources, either memory, buses (\eg \ac{can}) and CPU allow both isolation of services within virtual machines while enabling migration and reusage of legacy software~\cite{Bandur2021}. 

Considering the need for flexibility but also real-time constraints, \acp{ecu} can leverage virtualization principles to support multiple operating systems, both real-time and general purpose \acp{os}, through virtualization. 

Virtualization provides support for running multiple \acp{os} on the same hardware, ensuring the abstraction of hardware and of network topologies~\cite{Ren2022}.One of the key advantages is allowing hardware to be shared and not uniquely assigned to one operating system, while achieving the necessary isolation automakers require between critical and non-critical services.  It is an essential concept in the context of vehicular systems not only because of isolation but to support the aggregation of functions in less \acp{ecu} (per the zonal architectures)  than the number of services and required \acp{os}~\cite{Leonardi2020}.

Virtualization creates a number of challenges, including:
\begin{itemize}
	\item Orchestration: how to distribute resources (both location and quantity) of the virtualization continuum across virtual machines.
	\item Shared I/O: how to distribute input and output devices, such as \ac{usb} or Ethernet buses, across virtual machines. 
	\item Virtualization with real-time performance: since schedulability at hypervisor level is not deterministic~\cite{Bandur2021}.
\end{itemize}

Allowing virtualized services to use and share the same Ethernet port for example requires a scheduler that is capable of communicating with the different virtual machines and the deterministic network configurator. This communication can be achieved through \texttt{hypercalls}, the equivalent of  \texttt{syscalls} but to invoke a hypervisor~\cite{Caruso2021}. This may impact, however, the overall frame transmission delays.

When virtualization is considered, having the scheduler of the traffic can be at either application level, virtual machine level or hypervisor level~\cite{Leonardi2020}. At applicational level, since there’s one scheduler per application, there are going to be race conditions between each scheduler, with non-deterministic behaviors and therefore providing results with high jitter. At virtual machine level, for example leveraging the \textit{Linux Traffic Control} subsystem, while the necessary configurations are transparent to the applications, the hypervisor will not be aware of the special traffic. Working at the hypervisor level solves the previous issue through special \textit{Hypercalls} invoked by the middleware within the Hypervisor level.  \textit{Hypercalls}\footnote{Similarly to \textit{syscalls} concept, \textit{hypercalls} provide an interface between a guest \ac{os} and the hypervisor.} may be used within the middleware to support deterministic networkng within virtualized intra vehicular services~\cite{Caruso2021}.


Vehicular \acp{os} such as Android Automotive provides full support for virtualization through \texttt{\ac{kvm}} and \texttt{virtIO}\footnote{Developer documentation at \myhref{https://wiki.libvirt.org/Virtio.html}{wiki.libvirt.org}.} devices, allowing support at the \ac{vm} side for sensors, graphics, storage, network (including Wi-Fi through \texttt{virtio-net} and \texttt{VirtWifi}\footnote{Developer  documentation at \myhref{https://source.android.com/docs/devices/automotive/virtualization/architecture}{android.com}.}) and with support for communicating with the vehicle through \ac{grpc}\footnote{Details at \myhref{https://source.android.com/docs/devices/automotive/virtualization}{android.com}.}.

Generic Linux-based \acp{os} also support virtualization through containerization. Containerized services have advantages over hypervisor virtualization such as the reduced overhead and reduced storage space usage~\cite{Kugele2018}. 


Eclipse, within the \ac{sdv} foundation proposes Eclipse Ankaios\footnote{Developer documentation at \myhref{https://projects.eclipse.org/projects/automotive.ankaios}{projects.eclipse.org/projects/automotive.ankaios}.} to \enquote{manage multiple nodes and virtual machines with a single unique API in order to start, stop, configure, and update containers and workloads}. Ankaios is tailored towards automotive applications and developed for automotive \acp{ecu} and provides a central place to manage automotive applications, supporting:

\begin{itemize}
	\item Management of multiple nodes and virtual machines with a single unique \ac{api}.
	\item Various container runtimes like Podman and also native applications.
	\item Automotive \ac{spice} process with requirement tracing.
	\item Existing communication frameworks like \ac{someip}, \ac{dds} or \ac{rest} \ac{api} can be used with Ankaios workloads.
\end{itemize}

Kubernetes\footnote{Available at \myhref{https://kubernetes.io/}{kubernetes.io}.}, as a container orchestration tool mainly designed for datacenters, can also be consider to manage vehicular workloads. The work of~\cite{2020-Virtualization-46} proposes vehicles as nodes of a Kubernetes cluster. 

Apart from its inherent advantages as container orchestration and scheduler, Kubernetes is very flexible in networking terms, since networking between containers can be changed through \acp{cni}\footnote{Project available at \myhref{https://github.com/containernetworking/cni}{github.com/containernetworking/cni}.}. \acp{cni} are a specification for configuring network interfaces in containers. This allows considering  deterministic networking concepts (\cref{sec:blocks:deterministic}) in container scenarios. For example, the work of~\cite{Garbugli_2023} proposes support for deterministic networking within Kubernetes containers without modifying the application binaries through a side car container that exposes the standard \ac{posix} socket interface to other containers and redirects traffic  to deterministic configured network interfaces in the host computer. Not modifying the application binaries is essential to ensure compatibility with existing vehicle services but also to simplify the development process of new services.


\subsection{Intra-vehicular communications}
\label{sec:blocks:comms-intra}
Vehicular systems need to communicate with other systems within and outside the vehicle. While most intra-vehicular communication is expected to be through wired physical layers, this does not have to be the norm. \Cref{tab:soa_wiredBusesComparison} provides an overview of the main buses that may be used for intra-vehicular communications. 

\begin{table*}[!ht]
\caption[Overall comparison of main intra vehicular wired buses]{Overall comparison of main intra vehicular wired buses \cite{Bandur2021}.}
\label{tab:soa_wiredBusesComparison}
\centering
\resizebox{\linewidth}{!}{%
	\begin{threeparttable}[t]
		\begin{tabular}
  {lcccc}
		\hline
		\textbf{Characteristic / Bus} &
		  \textbf{\acs*{lin}} &
		  \textbf{\acs*{can}} &
		  \textbf{FlexRay} &
		  \textbf{Ethernet} \\ \hline
		\textbf{Media} &
		  1 wire &
		  \ac*{utp} &
		  2/4 wires &
		  \acs*{utp} \\ \hline
		\textbf{Topology} &
		  Bus &
		  Bus &
		  Star, bus &
		  Star, bus \\ \hline
		\textbf{Access control} &
		  Master-slave &
		  Priority-based messages &
		  \acs*{tdma} &
		  Switched point-to-point link \\ \hline
		\textbf{Max. Baud Rate} &
		  19.2 Kb/s &
		  1 Mb/s, up 16 Mb/s (XL)\tnote{a} &
		  10 Mb/s &
		  10 Mb/s, 100 Mb/s, 1000+ Mb/s \\ \hline
		\textbf{\acs*{mtu} (Bytes)} &
		  8 &
		  8/64 &
		  254 &
		  1472 (UDP), 1460 (TCP), 9000 (Jumbo) \\ \hline
		\textbf{Max nodes} &
		  16 (1 master) &
		  16 (500 Kb/s) &
		  22 (bus), 22/64 (star), 64 (mixed) &
		  256 to theoretically limited \\ \hline
		\textbf{Security Modules} &
		  None &
		  SecOC &
		  SecOC &
		  MACsec, IPsec, TLS/DTLS, SecOC \\ \hline
		\end{tabular}
			\begin{tablenotes}
			\item[a] If disregarding recommenced maximum data-rates~\cite{ASHJAEI2021102137}.
		\end{tablenotes}
	\end{threeparttable}%
}
\end{table*}


With the appearance of the first computational systems in the 1980s~\cite{Datanetw31:online}, the need for a bus capable of joining the different \ac{ee} systems within the vehicle. Bosch developed the \ac*{can}\footnote{Should not be confused with \ac{sae} J1939, which is a higher level protocol that uses \ac{can} as its physical and medium access layers and extends it by supporting multiframe messages~\cite{Creating15:online}.} bus to ensure communication between multiple systems. The \ac{can} main advantage is the reduced cost since it requires only a pair of cables~\cite{2016_Bus_2}. However, it requires a collision detection and mitigation protocols and lacks the scalability and adaptability to new traffic and new messages and the possible starvation in a bus overloaded with high-priority messages. \ac{can} also presents a limitation of \qty{1}{\mega\bit\per\s} of date rate, while variations such as \ac{canfd} raises the practical maximum date to 6 megabits per second with a higher 64 byte data field, while keeping the limitations on number of nodes and cable length~\cite{Bandur2021}. The BMW group, for example, is working towards an \ac{soa} to support for example new solutions of in-vehicle ambient lighting, that is colliding with this limitation on the number of nodes~\cite{IEEE_SA_d1_03}. Other buses, such as \ac{can} can support more LEDs, but Ethernet — in conjunction with \ac{tsn} can better support an accurate \qty{20}{\milli\s} LED update deadline without jitter and necessary synchronizations to support multi LED lighting animations. In addition, the \ac{can} and its variations are based on broadcasting messages. This contributes towards to a safety flaw of the \ac{can} — its fragility to the injection of counterfeit messages\footnote{Example at \myhref{https://kentindell.github.io/2023/04/03/can-injection/}{kentindell.github.io}.}. 

In the past, these limitations could be addressed through static analysis of the overall vehicular systems and \acp{ecu}. However, the increasing number of features —particularly data-intensive functionalities related to semi-autonomous driving— has necessitated a significant expansion in the volume and complexity of the wiring harness. This increase involves not only a greater number of wires but also added complexity, leading to a substantial weight of at least 40-50 kg\footnote{Source: \myhref{https://semiengineering.com/shedding-pounds-in-automotive-electronics/}{semiengineering.com}.}.

\begin{figure}[!ht]
	\centering
	\includegraphics[width=0.8\columnwidth]{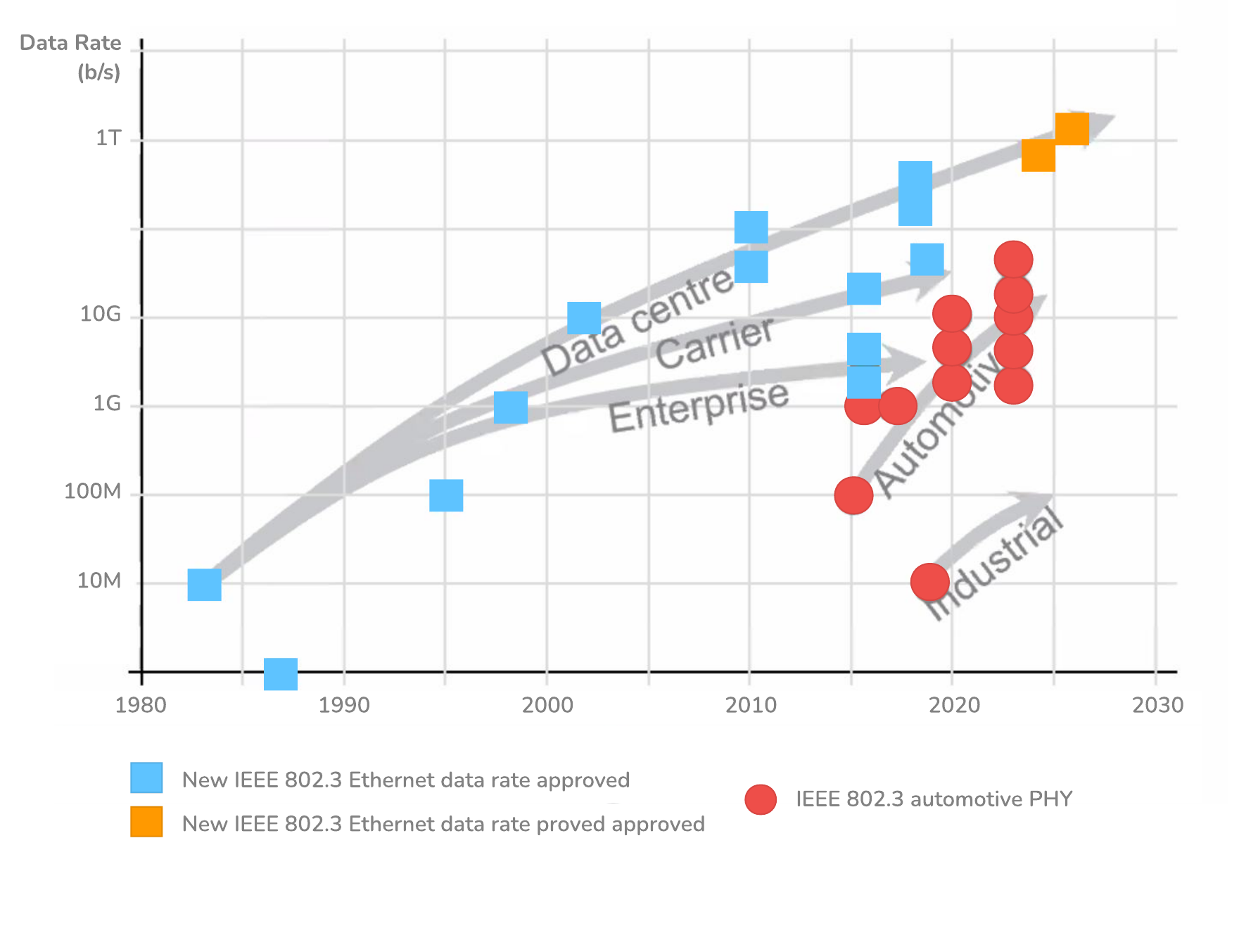}
	\caption[Evolution of Ethernet data rate and standards over time]{Evolution of Ethernet data rate and standards over time~\cite{Nichitelea2019}}
	\label{fig:background_EthernetDataRate}
\end{figure}

While legacy buses like \ac{can} or \ac{lin} can and will continue to be on edge intra vehicular systems, such as smaller more specific \acp{ecu}, sensors and actuators, the long term objective is the edges nodes to be also connected through Ethernet~\cite{IEEE_SA_d1_03}. Within the context of vehicular networks, Ethernet is being more and more considered as the backbone connection between high performance \acp{ecu}, with gateways to legacy buses like \ac{can}. As depicted in~\cref{fig:background_EthernetDataRate}, there are multiple iterations of Ethernet standards, \ie there's not a single cable and solution for all scenarios. Automakers shall take into consideration the characteristics of the cables (such as width, weight or resistance to heat and vibrations)— and not only the theoretical bandwidth — in order to choose the best ones to interconnect vehicular \acp{ecu}. Several physical layer specifications can be considered, including IEEE 100BASE-T1, 1000BASE-T1, and 1000BASE-RH~\cite{Amel2014,Nichitelea2019}. Possible Ethernet backbone upgrades to ensure future scalability are also being explored, including 10BASE-T1S or multi-Gb links.




Common to all its specifications, Ethernet presents several advantages, including longer payloads, higher bandwidth and the possibility to support \ac{qos} or multiple protocol layers~\cite{IEEE_SA_d2_02}. Ethernet is capable of greater bandwidth (compared to \qty{1}{\mega\bit\per\second} /  \qty{16}{\mega\bit\per\second} of the \ac{can} / \ac{canfd} baud-rate) and can transmit data with low latency and jitter~\cite{Nichitelea2019}. The work of~\cite{Takrouni2020} proves the overall better performance of Ethernet comparing with FlexRay in the end-to-end latency for the tested use case (a simulated Electronic Stability Control module).
The BMW Group, addressing the previously mentioned challenge of supporting new in-vehicle ambient lighting solutions, demonstrated that Ethernet combined with \ac{tsn} can effectively meet the stringent \qty{20}{\milli\s} LED update deadline. This approach ensures minimal jitter and provides the necessary synchronization to support multi-LED lighting animations~\cite{IEEE_SA_d1_03}.

Towards the goal of support Ethernet on the vehicles, major automakers and their providers have created the OPEN Alliance (One-Pair Ether-Net)\footnote{At \myhref{https://www.opensig.org/}{opensig.org}.} initiative. Canonical also defends \ac{ee} architectures which combines and consolidates \acp{ecu} and interconnects them using an Ethernet network~\cite{2022-Problem-8}. Such initiatives could also benefit from the resilience proven by industry grade protocols, such as PROFINET\footnote{At \myhref{https://www.profibus.com/technology/profinet/}{profibus.com}.}. 

Ethernet can be integrated with \ac{can} for example through the \ac{can}-XL protocol. This is done by mapping an Ethernet frame into a \ac{can}-XL frame. Since \ac{can}-XL provides 2048 data bytes, Q-tagged or enveloped Ethernet frames fit into a \ac{can}-XL frame~\cite{CANCANFD55:online}. 
Examples of gatewaying strategies from \ac{can} to Ethernet has been presented by~\cite{Zuo2021}, including \ac{fpga} based gateways with suitable latency, routing mechanisms between Ethernet and \ac{can} or FlexRay, with the authors proposing a gatewaying up to the application layer — \ac{someip} in this case through a \ac{someip} service and a \ac{rtos} program that read the \ac{can} messages. 

Alternatives to Ethernet include FlexRay and \ac{most}. Unlike \ac{can}, FlexRay has the unique ability to sync up nodes on a network without an external synchronization clock signal by using two special types of frames~\cite{FlexRay}, being a possible alternative to Ethernet with Deterministic mechanisms (more on~\cref{sec:blocks:deterministic}). The \ac{most} specification defines \enquote{the protocol, hardware and software layers necessary to allow for the efficient and low-cost transport of control, real-time and packet data} using \enquote{different physical layers (fiber optic, UTP, coax)} and supporting 25, 50 and 150 Mbps data transfer speeds~\cite{MOSTTec59:online}, still under the theoretical maximus of Ethernet. 

\subsection{Inter-vehicular communications}
\label{sec:blocks:comms-inter}
For inter vehicular communications, it is impractical to consider wired physical layer. Within these context, the \ac{v2x} paradigm is of increasing importance considering the advent of both \ac{v2v} scenarios and smart cities. The usage of cloud infrastructure is also an integrated part of the \ac{sdv} paradigm, requiring \ac{v2i} wireless communications. For example, CARIAD, the initiative from the Volkswagen group towards \acp{sdv}, launched an uniform application store for all brands within the Volkswagen group~\cite{2023-Vehicle-Abstraction-Layer-23}, that indicates a strong cloud presence within its ecosystem.

Vehicular nodes can use different wireless access technologies to communicate with other road elements, including vehicles, \ac{v2v}; pedestrians, \ac{v2p}; and infrastructure, \ac{v2i}; creating a \ac{vanet}, a sub-type of a \acp{manet}. These technologies can be classified based on their communication range~\cite{dasanayaka2020enhancing}, with a summary of the main characteristics of the most used access technologies described in~\cref{tab:related-work-access_tech}.

\begin{table*}[!ht]
    \centering
    \caption[Example of wireless vehicular communications access technologies and main characteristics]{Example of wireless vehicular communications access technologies and main characteristics.}
    \label{tab:related-work-access_tech}
    \resizebox{0.8\linewidth}{!}{%
    	\begin{threeparttable}[t]
        \begin{tabular}{lccc}
        
        %
       
        \hline
\textbf{Technology}                          & \textbf{Data Rate\tnote{a}}     & \textbf{Range}        & \textbf{Latency}                              \\ \hline

 Cellular~(LTE)~\cite{surveyltevs5g}         & 100/\qty{50}{\mega\bit\per\s}    & Long \; ($>$ \qty{1}{\kilo\meter})  & 10-\qty{100}{\milli\s}                                    \\ \hline

 Cellular~(LTE-A)~\cite{surveyltevs5g}       & 3/\qty{1.5}{\giga\bit\per\s}     & Long \; ($>$ \qty{1}{\kilo\meter})  & \qty{10}{\milli\s}                                              \\ \hline

 Cellular (5G)~\cite{surveyltevs5g,survey5g} & 20/\qty{10}{\giga\bit\per\s}     & Long \; ($>$ \qty{1}{\kilo\meter})  & $\leq$~\qty{4}{\milli\s}      \\ \hline

 Wi-Fi (IEEE~802.11be)\tnote{c}              & $\leq$~\qty{40}{\giga\bit\per\s}~\tnote{b} & Short \; ($<$ \qty{90}{\meter} in \qty{2.4}{\giga\hertz})& \qty{10}{\milli\s}~\cite{9500256}                                                             \\ \hline

 Li-Fi (IEEE~802.11bb)~\cite{10315104}       & \qty{10}{\mega\bit\per\s} up to \qty{9.6}{\giga\bit\per\s}              & Short (few meters)\tnote{d}& Low\tnote{e} \\ \hline

 \acs{3gpp} C-V2X~\cite{10013676}&  $\leq$ \qty{26}{\mega\bit\per\s}~\cite{10056390}\tnote{f}                                    & Short to Long  & Low    \\ \hline
           
 ITS - G5~(IEEE~802.11p)~\cite{10013676}& $\leq$~3 up to \qty{27}{\mega\bit\per\s}~\cite{10056390} & Short \; ($<$ \qty{1}{\kilo\meter}) & Low  \\ \hline

 IEEE~802.11bd~\cite{ieee80211bd}\tnote{g}              &  $2\times$ IEEE~802.11p & $2\times$ IEEE~802.11p & Low \\ \hline

        \end{tabular}
        
        \begin{tablenotes}
            \item[a] Downlink / uplink.
            \item[b] Theoretical.
            \item[c] Details at~\myurl{https://standards.ieee.org/beyond-standards/the-evolution-of-wi-fi-technology-and-standards/}.
            \item[d] As described at \myurl{https://spectrum.ieee.org/li-fi-better-than-wi-fi}.
            \item[e] Up to \qty{1}{\milli\s} as claimed by \myurl{https://emag.directindustry.com/2021/07/20/li-fi-shines-as-the-future-of-short-range-communications-light-led-microchip-radio-waveband/}.
            \item[f] Considering \ac{lte} based C-V2X and as claimed by \myurl{https://eu.mouser.com/applications/new-v2x-architectures}.
            \item[g] Theoretical values.
        \end{tablenotes}
 	\end{threeparttable}%
}

\end{table*}

Cellular based solutions, either through \ac{lte}, \ac{ltea} or 5G, can provide connection to other vehicles and infrastructure, with high coverage and medium latency at a high monetary cost~\cite{10013676}.

Wi-Fi, including its latest iteration (Wi-Fi 7 / IEEE 802.11be), can also theoretically provide connectivity to and from vehicles, as well as IEEE 802.11bb / Li-Fi. Li-Fi adapts the \ac{mac} layer of 802.11 while supporting light (visible or infrared) based physical layers in order to overcome interference in dense deployments (which Wi-Fi is particularly susceptible to) and restrictions or inefficiencies of radio frequencies - potentially increasing the maximum. An evolution of the original infrared physical layer proposed in 1999 for IEEE 802.11, it is targeted at the industrial markets~\cite{9855461}. However, both solutions fail to consider the high mobility of vehicular nodes, which means frequent handovers between access points.

Vehicular targeted standards include \ac{cv2x} and ITS-G5. \ac{cv2x} is cellular based and aims at integrating cellular technologies within \ac{v2v} scenarios. Based on cellular standards it also allows direct device-to-device communication with low latency and high reliability, but in short ranges (\textpm 1 km), with low cost associated with the possible necessary hardware changes. \ac{cv2x} also support device-to-network, enabling cloud services to be part and parcel of the end-to-end solution. On the other hand, ITS-G5 is based on IEEE 802.11p for physical and \ac{mac} layer. It requires existence of road infrastructure — \acp{rsu} — and that vehicles are equipped with \acp{obu}. Problems of this alternative includes handling handovers and scaling to many nodes simultaneously and vulnerability to packet collisions due to concurrent transmissions, and packet collisions due to hidden vehicular nodes. Existing literature shows that \ac{cv2x} offers performance advantages over ITS-G5 in terms of its additional link budget, higher resilience to interference, and better non-line-of-sight capabilities~\cite{Alparslan2021}. 

\ac{rat} evolutions, such as the IEEE 802.11bd and New Radio for \ac{cv2x} can supplement today's vehicular networks in supporting higher bandwidth and \ac{qos} wireless communications. IEEE 802.11bd are more suitable for supporting more \ac{v2x} applications efficiently, with better throughput and low transmission latency, by adapting the physical and \ac{mac} layer techniques introduced in IEEE 802.11n/ac/ax, while ensuring interoperability and backwards compatibility with 802.11p.~\cite{8723326_Naik2019}. 

\subsection{Deterministic intra and inter-vehicular communications}
\label{sec:blocks:deterministic}
Both ITS-G5 and \ac{cv2x} can reliably support safety applications that demand an end-to-end latency of around \qty{100}{\milli\second} as long as the vehicular density is not very high~\cite{Farzaneh2017}. However, as the quality of service requirements of \ac{v2x} use-cases become more stringent, which is the case in many advanced \ac{v2x} applications~\cite{bailleul20towards}, the two current \ac{rat} fall short of providing the desired performance, since some use-cases require the end-to-end latency to be as low as \qty{3}{\milli\second} with a reliability of \qty{99.999}{\percent}~\cite{8723326_Naik2019}.

Services running through a communication network and with requirements of high performance and high precision should consider not only techniques to decrease the overall end-to-end latency of data from/to services — for example by leveragging the edge computing concept~\cite{Malinverno2020,Khakimov_2020} — but also techniques to ensure the latency requirements and real-time hard constraints. The usage of deterministic communications and networks that support such communications are vital when considering vehicular services that are somehow critical. Determinism, per the Merriam-Webster dictionary, \enquote{is the quality or state of being determined}, \ie the quality of being firmly resolved. Within networks this concept can be regarded as the delivery of units of information — either abstracted in frames or packets or other abstraction — in a fully resolved manner in several domains, such as temporal or reliability, in all the steps of the frame/packet from the producer to the consumer(s) of information.

\begin{figure*}[!htp]
	\centering
	\includegraphics[width=1\textwidth]{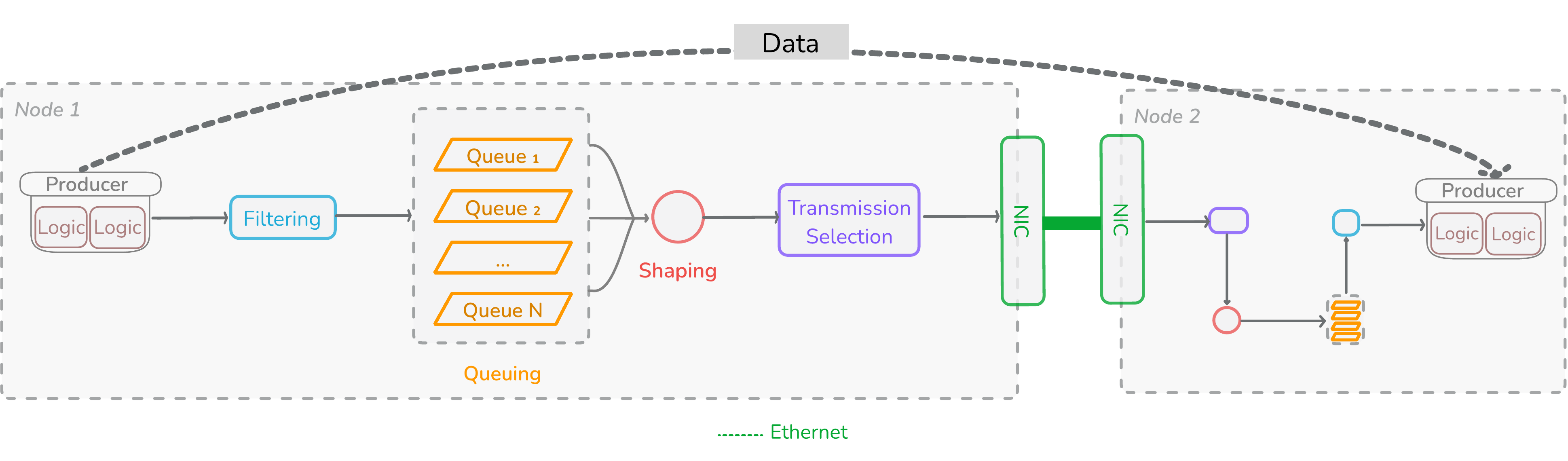}
	\caption[Main steps of a packet switching]{Main steps of a packet switching. Partially adapted from~\cite{lopes2023time}.}
	\label{fig:background-tsn-packet}
\end{figure*}

Regarding the temporal determinism, within the exchange of a unit of data between two network entities one may consider the main steps of a packet switching as presented in~\cref{fig:background-tsn-packet}. Each step introduces delay that, if the network is not properly configured, can be unbounded and potentially even random. Even if high bandwidth is a possibility, transmission, and queuing delays still exist, affecting end-to-end delay~\cite{Alparslan2021,Farzaneh2017}.

Sets of standards such as \ac{tsn}\footnote{At \myhref{https://1.ieee802.org/tsn/}{1.ieee802.org/tsn}.} and Deterministic Networks\footnote{At \myhref{https://datatracker.ietf.org/wg/detnet/about/}{datatracker.ietf.org/wg/detnet}.} can ensure that high reliability and time precision for each step of the packet switching process~\cite{Seol2021}. Current solutions to tackle these issues within the \ac{tsn} and Deterministic Networks are further presented in this section. The next subsections present the state-of-the-art on the main initiatives and sets of standards towards determinism in networks in lower \ac{osi} layers. These initiatives started for wired mediums such as Ethernet but can also be extended to wireless contexts.

\subsubsection{In wired context}
\label{sec:sota-wired}
Methods towards achieving deterministic communications within a wired intra-vehicular context (such as the connectivity between vehicular \acp{ecu}) are being standardised, with two major initiatives to be considered: a) the \ac{ieee}~\ac{tsn} set of standards, operating at \ac{osi} layer 2; and b) the \ac{ietf} \ac{detnet} operating at \ac{osi} layer 3. Both initiatives shall be seen not as competitors but as complementary initiatives.

\label{sec:sota-wired-detnet}~The \ac{ietf} \textit{\acl{detnet}} set of standards aims at defining deterministic — \ie with bounded latency, loss and jitter — paths for the packets~\cite{tutorial40:online}. Those standards operate at layer 3 and can be supported by a \ac{sdn} approach, where one or more controllers are responsible for defining the nodes behavior, to warrant~\cite{rfc8655}:
\begin{itemize}
	\item Explicit routes: to ensure deterministic routing. This may be achieved for example through the usage of \ac{mpls}, a routing technique based not on network addresses but labels.
	\item Resource allocation: to ensure protection against congestion, resources (such as bandwidth) is allocated.
	\item Service paths: to ensure protection against equipment and medium failures packets are replicated through disjoint paths.
\end{itemize}

\Cref{tab:detnet} provides an overview on the \ac{rfc} standards and documents that define the goals, use cases and building blocks behavior. The \ac{detnet} building blocks operate at the \ac{ip} layer and use lower-layer technologies such as \ac{mpls} and \ac{tsn}. These building blocks can and should be controlled based on observable metrics (hops, their state, and end-to-end latency), and on fault detection/prediction and identification~\cite{rfc8655}.
\begin{table*}[!ht]
    \centering
    \caption{Main \acs*{detnet} standards, specifications and documents.}
    \resizebox{0.9\linewidth}{!}{%
        \begin{tabular}{p{0.20\linewidth}p{0.15\linewidth}p{0.65\linewidth}}
            \hline 
            \textbf{Document}& \textbf{Goal}   & \textbf{Importance} \\ \hline
            RFC 8557 & Problem statement        &Contextualizes the need for \ac{detnet} with the need for a deterministic forwarding of packets.\\ \hline
            RFC 8578 & Use cases          &Validates self-driving vehicles as examples of use cases that can be addressable by \ac{detnet} techniques.\\ \hline
            RFC 8655 & Architecture       &Defines the main building blocks of 
\ac{detnet}.\\ \hline
            RFC 8938, 8939, 8964, 9023, 9024, 9025, 9037, 9056 & Data Plane  &Defines how building blocks may be implemented and metadata that packets require. Includes information of integration with \ac{tsn} and \acs*{mpls} for explicit routing.\\ \hline
            RFC 9016
\acs*{yang} model (draft) \cite{ietf-detnet-yang-18} & Flow information model  &Modelling of the configuration information for  configuration of \ac{detnet} building blocks.\\ \hline 
            RFC 9055 & Security                &Considerations on security, including challenges posed by replication of packets and  spoofing.\\ \hline
            RFC 9320 & Bounded Latency         &Provides mathematical modelling of the bounded latency values for a \ac{detnet} architecture.\\\hline
        \end{tabular}
    }
    \label{tab:detnet}
\end{table*}






At a lower \ac{osi} layer, the \textit{\acl{tsn}} group presents an effort to solve transmission, and queuing delays issues. \Cref{tab:tsn} provides an overview on the \ac{ieee}~standards and documents that define the goals, use cases and building blocks behavior. \ac{tsn} standards adds to Ethernet mechanisms to ensure time characteristics, performance, reliability, and security. Mechanisms include time synchronization (\eg gPTP~\cite{bailleul20towards}), replication and elimination for reliability (e.g. \ac{frer})~\cite{8091139}, and bounded low latency (\eg credit-based~\cite{8684664} or time aware shaping algorithms~\cite{8613095}) that can provide the necessary real-time performance and bounding of the latencies~\cite{9738961,Ozawa2022,Ulbricht2022}.

The \ac{tsn} standards are an evolution of the \ac{avb} standards, dedicated to extending the Ethernet standards to provide guaranteed quality of service by ensuring high precision time synchronization between nodes, bounded latency of high throughput streams (at \si{\micro\second} scale) and reservation of bandwidth within audio and video scenarios. The \ac{tsn} toolbox adds reliability concerns, more accurate clock synchronization, bounded low latency even for real-time streams (at \si{\micro\second} scale) and security concerns~\cite{10.1145/3487330}. As a comparison, \enquote{the design goal of \ac{avb} was a maximum latency of \qty{2}{\milli\s} over 7 hops in a \qty{100}{\mega\bit\per\s} network for audio and video traffic}, while \ac{tsn} standards set \enquote{the goal to set the maximum latency to \qty{100}{\micro\s} over 5 hops}~\cite{8412465}. Through the IEEE 802.1Q standard three types of traffic classes for such traffic are defined — Scheduled Traffic (with precise scheduling), Stream Reservation (which needs to be configured in order to ensure fairness) and Best Effort~\cite{Leonardi2020}.

In conjunction with \ac{detnet}, but operating at \ac{osi} layer 2 — and originally on top of wired Ethernet networks — the \ac{ieee}~802.1 \ac{tsn} set of standards provide mechanisms for~\cite{8672474}:
\begin{itemize}
	\item Time synchronization: applications and other network mechanisms require a precise (down to nanosecond accuracy) synchronization between the nodes of the network. Protocols such as \ac{ptp}~\cite{bailleul20towards} can provide such precision within wired and wireless networks. Synchronization must take into account physical layer latencies and their variations, as these can impact propagation delay measurements and overall synchronization accuracy. IEEE 802.3cx proposes a mechanism to report the physical layer delays in a dynamic way~\cite{IEEE_SA_d1_08}.
	
	\item Replication and elimination for reliability: improve the reliability by ensuring that frames are replicated and sent through disjoint paths. This replication needs additional mechanisms to ensure repeated frames are ignored. Solutions such as \ac{frer}~\cite{8091139} can manage this replication.
	
	\item Resource management: Mechanisms to ensure the configuration of \ac{tsn} talker, listener and of equipment between both and the necessary allocation of resources.
	
	\item Bounded latency or bandwidth: ensure the bandwidth or latency and its jitter is deterministic by ensuring transmission happens at a specific time. 
    	
    \end{itemize} 
    
Regarding bounded latency, traffic shaping can provide the necessary real-time performance and bounding of the latencies or bandwidth~\cite{9738961,Ozawa2022,Ulbricht2022}, with shapers such as:
	\begin{itemize}
		\item \ac{cbs}: based on regulating the transmission rate of frames to smooth out traffic, reduce bursts and ensure bandwidth for certain traffic classes, by delaying the transmission of some frames. When the queue contains a frame to be transmitted, a credit counter is checked and the frame is only transmitted if the counter is non-negative (and if it is, is held back until is non-negative). This counter accumulates credits when the transmission queue is idle, and consumes credits when frames are transmitted, with the rate of accumulation and consumption configurable via parameters (\textit{idle slope} and \textit{send slope})\footnote{More details at \myhref{https://inet.omnetpp.org/docs/showcases/tsn/trafficshaping/creditbasedshaper/doc/index.html}{inet.omnetpp.org}.}~\cite{8684664}.
		\item \ac{tas}: works by associating the data to traffic classes and the traffic classes to gates that open and close at specific times and for specific intervals of time, ensuring no unwanted competition between traffic classes that should not compete for transmission, therefore eliminating the jitter associated with that competition~\cite{8613095,PatentChristianMardmoeller2023}.
		\item \ac{ats}: which is based on the leaky bucket principle, with transmission allowed when credit is greater than the size of the frame~\cite{8412465}.
	\end{itemize}

\begin{table*}[!ht]
    \centering
    \caption[Main \acs*{tsn} standards]{Main \acs*{tsn} standards~\cite{8672474} (partially adapted).}
     \label{tab:tsn}
    \resizebox{0.9\linewidth}{!}{%
        \begin{tabular}
         {p{0.15\linewidth}p{0.25\linewidth}p{0.55\linewidth}}
            \hline 
            \textbf{Document}& \textbf{Goal}   & \textbf{Importance} \\ \hline
            802.1Qca& Path control and reservation&Define mechanisms to allocate resources to transmit frames.\\ \hline
            802.1Qbv& Time-aware scheduling&Defining when and what frames are sent by the emissor.\\ \hline
            802.1Qbu, 802.3br& Frame preemption&Defines the possibility of higher priority frames interrupting lower priority, essential in ensuring precise transmission time for higher priority flows.\\ \hline
            802.1Qcc& Central configuration model&Defines the need for coordination nodes (the \acs*{cuc} and \acs*{cnc}). Updates on the \acs*{srp} protocol.\\ \hline
            802.1Qci& Filtering and policing&Defines mechanisms to count and filter frames and detect and mitigate disruptive flows of frames.\\ \hline
            802.1CB& Redundancy through frame replication and elimination&Defines mechanisms to replicate frames and manage the repeated frames to increase reliability of the network and avoid equipment failure impacting \acs*{per} and similar metrics.\\ \hline
            1588, 802.1AS& Precise ($ns$ level) time synchronization between nodes&Defines mechanisms for a nanosecond synchronization between clocks of several network nodes.\\\hline
 P802.1DG (draft)& \acs*{tsn} profile for automotive in-vehicle Ethernet communications.& Recommendations for \acs*{tsn} implementation within vehicular architectures.\\\hline
        \end{tabular}
    }
   
\end{table*}

\begin{figure}[!ht]
	\centering
	\includegraphics[width=1\columnwidth]{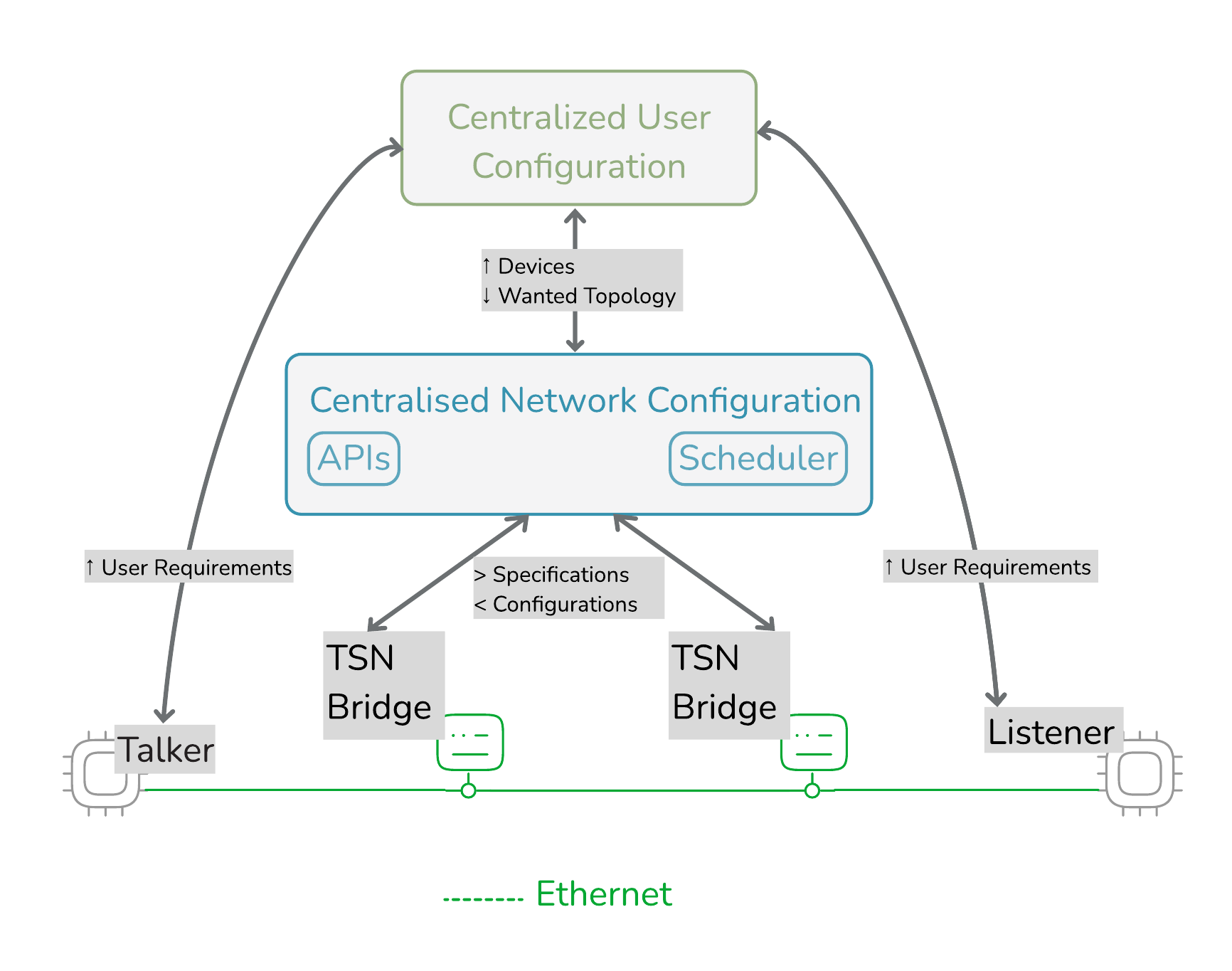}
	\caption[Interaction between \acs*{cuc},  \acs*{cnc} and the remaining \acs*{tsn} topology]{Interaction between \acs*{cuc},  \acs*{cnc} and the remaining \acs*{tsn} topology~\cite{CiscoWhitePaper}. }
	\label{fig:soa-tsn-cuc-cnc}
\end{figure}

How the network should be configured is also addressed through the IEEE 802.1Qcc standard, which describes three possible models~\cite{SATKA2023102852}:
\begin{itemize}
	\item A fully distributed model where all the end stations communicate their requirements — through protocols such as \ac{srp} or \ac{rap}.
	\item A centralized network/distributed user model, where an entity, the \acs*{cnc}, gather the network capabilities.
	\item A fully centralized model, where the \acs*{cnc} is helped by the \acs*{cuc} to ensure proper network equipment configuration.
\end{itemize}
These three models are based on the existence of two entities — the \ac{cnc} and the \ac{cuc}. The \ac{cnc} defines the overall network schedule and is the main controller of the network within the network bridges, dispatching configuration with protocols such as \ac{netconf}. The \ac{cuc} mediates the requests between end devices and their applications~\cite{CiscoWhitePaper}, collects all user requirements and interacts with the \ac{cnc} to decide the topology and scheduling of the network. \Cref{fig:soa-tsn-cuc-cnc} provides an overview of these interactions. 

\ac{tsn} is not however the only standard at layer 2 for deterministic communications. \ac{ttethernet}, a standard  for industrial and avionics applications improved from the Avionics Full Duplex Switched Ethernet, is an alternative\cite{Bandur2021}. While the concepts of deterministic timing of the transmission of frames is inherently the same, both differ in a number of aspects~\cite{8569454,craciunas2017overview}. \ac{tsn} is a set of standards while \ac{ttethernet} is a specific \ac{sae} AS6802 standard. In the former, the schedule is on entire traffic classes, while in latter is on individual frames. In addition, \ac{tsn} supports lower-priority traffic using non-used bandwidth, while \ac{ttethernet} blocks other traffic even when there are no time-scheduled frames to transmit results in a less efficient use of bandwidth.

\subsubsection{In wireless context}
\label{sec:wireless-detnet}

While wired network access mediums have well defined / bounded error rates, based essentially on the electromagnetic noise ratio, wireless mediums (such as inter vehicular communications are prone to the creation of unpredictable non-stochastic jitter and packet losses through additional issues and challenges, including~\cite{8672474}: 
\begin{itemize}
	\item Fading of the transmission: based on distance between nodes, changes in the propagation environment or node mobility.
        \item Shadowing and attenuation: from other physical obstacles and phenomena (\eg reflection, absorption, diffraction)~\cite{goldsmith_2005}.
	\item Interference: from other devices communicating on the same spectrum frequencies. This issue is more pronounced when unlicensed spectrum (\eg 2.4 GHz) is used since its usage is uncontrolled and prone to additional noise from other uncontrollable entities. 
	
	\item Random channel access latency: the randomness associated with existing listen-before-talk procedures for multiple access to the wireless channels contributes to non-stochastic delays.
	\item Integration of synchronization mechanisms: while considering the differences in propagation time in wireless communications.
	\item Bounding an end-to-end latency: or its jitter by integrating scheduling on transmission by each \ac{ap}. 
	\item Integration of reliability technique: while considering and taking maximum advantage of the main different channel characteristics.
	\item Resource management: including the usage of several \acp{ap} for reliability or the dynamic management of the spectrum used at a given moment, which remain open challenges.
\end{itemize}

Achieving determinism within a wireless communication requires that all these challenges are properly addressed. While fading, interference, shadowing and attenuation are mostly unavoidable without changes to the physical environment, interference can be minimised. The usage of other parts of the spectrum, possibly licensed, with wider channels, and frequencies in multiple links working in conjunction (multi link operation) can further decrease the overall noise within the physical layer and therefore decrease the jitter of latencies, and increase the determinism of the communications~\cite{8672474}. In addition, a periodic analysis of the spectrum usage can support a decision on the best frequencies to be used.

Regarding the unpredictable time to access the wireless medium, existing listen-before-talk solutions such as \ac{csma} consider that in the event of a collision, before re-transmission a random amount of time shall be waited. This creates an unpredictable, non-deterministic jitter on the transmission of information. Other channel access solutions must be considered in order to achieve higher determinism on the transmission of the frames through both a wired or wireless medium, such as dividing the channel per frequency, time or both (e.g. \ac{fdma},~\ac{tdma} and~\ac{ofdma}, respectively) or adding scheduling algorithms to listen-before-talk solutions (\ac{csma-ca})~\cite{9512046}. In alternative, solutions that consider collisions to be solved at the receptor side (\eg \ac{noma}) can also be considered.

Another promising solution is a coordination and scheduling of the access to the wireless medium to ensure not only the typical concerns of fairness or throughput but the fulfillment of a specific time of transmission, ensuring more determinism than the previous approaches~\cite{8672474}. Within wireless mediums, the usage of the \ac{ieee}~802.11be (Wi-Fi 7) concept of \ac{rtwt} (originally introduced for power saving) can further decrease conflicts during transmission. By defining when a certain station cannot transmit — and therefore any transmission ends before the start of this time — it is possible to further decrease collisions between critical and less critical frames being transmitted simultaneously. 

Projects such as PREDICT-6G~\cite{predict6g} and DETERMINISTIC-6G~\cite{deterministic6g} defend that, since the wireless environment has a non-stochastic jitter that is unavoidable (at the very least because of fading, interference, shadowing and attenuation), any system should adapt to those variations. 
Such adaptation can be done through online (in runtime) prediction of the variability of latency and jitter through monitoring of the system and prediction on \acp{kpi} (\eg instantaneous availability) to then act on resource allocation and schedules. To support this prediction, models of the channel behaviour can be considered to model the performance of the channel based on the medium conditions, transmission power, an angle or distance between nodes. Such models are not a one size fits all solution, with each different environment requiring a different analysis~\cite{8672474}.

Regarding the integration of synchronization mechanisms, the \ac{ieee} 802.11 standards provides time synchronization capabilities through a mechanism named \ac{tsf} that supports a low precision synchronization with a maximum of 1 second of accuracy~\cite{10034532}. The \ac{tsn} standard for high precision synchronization, the IEEE 802.1AS (\ac{ptp}) defines a profile for 802.11 mediums~\cite{8672474} that considers within the synchronization of the clocks an estimation of the propagation delay and based on the number of neighbours of the nodes.

Regarding the bounding of end-to-end latency and jitter, the integration of shaping techniques, namely \ac{tas}/802.11Qbv or \ac{ats} is possible. Solutions for integration include mapping the scheduling policies to each transmission queue within an \ac{ap} and by adding to each frame fields with the flow identification and characteristics (delay bound, maximum packet size) to map the scheduling policies to the 802.11be \ac{mac}. The overall scheduling would require a full knowledge of the network topology.

Regarding reliability, including \ac{frer} (\ac{tsn} standard \ac{ieee}~802.1CB) through redundant wireless links and channels is possible. The usage of several \acp{ap}, with the necessary seamless predictive handover between \acp{ap}, remains however an open challenge~\cite{10034532}. In addition, management of the transmission power can further improve the reliability by decreasing the impact of interference and obstructions in the transmission.

Finally, regarding the resource management, the need for some sort of centralization in decision making for the configuration of the network nodes — \acp{ap} and switches — is supported both by \ac{tsn} and \ac{detnet} standards. \ac{tsn} \ac{ieee}~802.1Qcc standard define that 2 entities (the previously defined \ac{cuc} and \ac{cnc}) are responsible for monitoring and defining resource allocation and configurations~\cite{10034532}, while the \ac{detnet} Reference Architecture~\cite{rfc8655} considers the existence of a Controller Plane responsible for~\enquote{topology discovery and maintenance, packet route selection and instantiation or path fail-over mechanisms}~\cite{rfc7426}. An architecture should therefore consider the existence at least of a controller that configures and monitors the network nodes and the information flow.

All the previous points may benefit from additional performance  performance with a deployment considering hardware acceleration, for example through the usage of \acp{asic} or \acp{fpga}~\cite{PatentPatel2022}. \acp{fpga}, by allowing the reconfiguration of the hardware to fit the needs of the application(s), would add to the hardware acceleration, the flexibility of deployment and reconfiguring of new wireless \acp{ap} that combine the previous techniques accelerated by hardware. Such acceleration is essential to ensure the processing times are sufficiently small to allow determinism in the overall communication.

Examples of software that may run on top of the assisted solutions is the \texttt{openwifi}\footnote{Source code available at \myhref{https://github.com/open-sdr/openwifi}{github.com/open-sdr/openwifi}.} stack or the work in~\cite{9134382}, which introduces a custom 802.11 and \ac{tsn} inspired solution that allows ultra low control cycles (within \qty{100}{\micro\s}), reduced jitter (within the \si{\nano\second} barrier) and low \ac{per} (less than $< 10^{-7}$ for distances to the \ac{ap} between \qty{6}{\meter} and \qty{12}{\meter} and $5 \times 10^{-6}$ for distances greater than \qty{23}{\meter}) through a \ac{fpga} implemented solution for \acp{ap}. Each \ac{ap} contains several modules implemented in a \ac{fpga}:
\begin{itemize}
	\item Timer: used by the scheduler and synchronized by an external \ac{pps} signal;
	\item Scheduler: operating at \ac{osi} level 2, decides which frames should be sent and when are they due;
	\item Custom \ac{mac}: contains information for controlling frames, with changes from the original 802.11 access scheme to avoid collision between real-time and non-real-time frames and supporting the necessary synchronization between clocks; 
	\item Physical layer: The encoder/decoder and additional hardware modules required for modulation features.
\end{itemize}

While this work offers a promising solution, it lacks full testing in mobility scenarios, integration with \ac{tsn} and integration with other wireless standards such as 5G-NR. However, such integration with cellular communication technologies is possible. \acs{3gpp} Releases 16-18 defends that, ensuring ultra reliable and low latency communications start at the radio level and that \ac{tsn} standards should be added to the overall architecture, with parts of the 5G core acting as \ac{tsn} bridges~\cite{10034532}, one per 5G \ac{upf}~\cite{ETSI_TS_123_501}. This allows a transparent integration of 5G systems with a \ac{tsn} system but require a set of translators between the 5G control plane and \ac{tsn} standards~\cite{ACIA2021}. The configuration of the \ac{tsn} network can then be based on the 5G configuration, with the mapping between the \ac{tsn} network configurations~\cite{PatentJosephVinay2021}(such as priorities or \ac{tas} gate opening times) and 5G \ac{qos} parameters, namely the \ac{5qi}, a scalar that represents characteristics such as~\cite{ETSI_TS_123_501, 9904785}:
\begin{itemize}
	\item Resource Type: distinguishes between \ac{gbr}, \ac{dcgbr} or non-\ac{gbr}, where the first two types of flows can pre-allocate resources;
	\item Priority Level: indicates a priority in scheduling resources. Can be used to prioritize the allocation of resources to certain flows;
	\item Delay Budget: an upper bound of the time a packet can spend inside a 5G system (between the \ac{ue} and a \ac{upf}) without being dropped;
	\item \acl{per}: a ratio between the number of received packets and the number of transmitted packets.
\end{itemize}

Developing algorithms to map \ac{tsn} configurations to 5G configuration values remains an active area of research. The work of~\cite{9779191} proposed mapping the default priority value of a 5G frame to the \ac{pcp}. Other work by~\cite{9212141} consider that \ac{tsn} frames shall be considered \ac{dcgbr}, while~\cite{9904785} propose the mapping of \ac{tsn} traffic into 5G QoS \ac{5qi} characteristics. For example, the \textit{resource type} characteristic is derived from the type of \ac{tsn} flow: deadline constrained, jitter constrained or bandwidth constrained, per the truth table in \cref{tab:sota-tsn-5g}, with similar analysis for the other characteristics within \ac{5qi}.

\begin{table}[!htp]

\centering
\caption[Mapping between \acs*{tsn} flow characteristics and \acs*{5qi} values]{Mapping between \acs*{tsn} flow characteristics (on the left) and \acs*{5qi} values (on the right) ~\cite{9904785}.}
\label{tab:sota-tsn-5g}
\resizebox{\columnwidth}{!}{%
    \begin{tabular}{ccc||ccc} \hline 
    \textbf{Deadline} &  \textbf{Jitter} &  \textbf{Bandwith} & \textbf{\acs*{gbr}} & \textbf{\acs*{dcgbr}} & \textbf{Non-\acs*{gbr}} \\ \hline 
    0  & 0  & X  & 0   & 0      & 1       \\ \hline 
    0  & 1  & 0  & 1   & 0      & 0       \\ \hline 
    0  & 1  & 1  & 0   & 1      & 0       \\ \hline 
    1  & 0  & 0  & 1   & 0      & 0       \\ \hline 
    1  & 0  & 1  & 0   & 1      & 0       \\ \hline 
    1  & 1  & 0  & 1   & 0      & 0       \\ \hline 
    1  & 1  & 1  & 0   & 1      & 0      
     \\ \hline\end{tabular}
 }
\end{table}

While most current work is towards exploring future industrial integration between 5G and \ac{tsn}~\cite{ACIA2021}, real applicability remains limited. This is partially due to the lack of interoperability between 5G and \ac{tsn}-capable equipment, with configuration usually considering propriety mechanisms or at least standard mechanisms with changes~\cite{SATKA2023102852}. In fact, according to~\cite{SATKA2023102852}, no tool has been contributed within the \ac{tsn}-5G interaction, with most work currently working towards high precision synchronization support (which 5G networks support through \ac{ieee} 802.1AS with a time error limited to \qty{900}{\nano\second}~\cite{ACIA2021}), proof-of-concept architectures and mathematical analysis.
A notable exception is the work by~\cite{electronics11111666} which presents \enquote{a prototype 5G system integrated with a \ac{tsn} network for a typical industrial mobile robotics use case}. The prototype is based on commercial solutions for a 5G system and \ac{tsn} switches for deterministic communications between an application in the factory cloud and a mobile robot for a factory floor environment. Results of such integration were promising, with end-to-end latency below \qty{0.8}{\milli\s} and, more importantly, with a 99.9\% reliability and jitter in the order of \qty{500}{\micro\s}.

\subsubsection{In vehicular context}

Solutions to adopt \acp{tsn} within the vehicle’s systems are sparse and moreless recent, with some of the oldest integration within vehicular systems in 2016~\cite{7746299}, and further integration in recent years~\cite{Syed2022,Zou2023}, including with 5G~\cite{Wang2020,Ding2022}. In addition to 5G, both wired and wireless technologies based on IEEE 802.11p or IEEE 802.11bd should be considered within these networks, ensuring that service(s) can communicate with the vehicle’s systems and surroundings with reliability and redundancy. 


The challenges on \acp{tsn} within vehicular contexts are two-fold: for one, the automotive use case is a critical scenario, which requires well defined latencies, ordering and high level reliability; on the other side, the high dynamism of the intra and inter vehicular topology — each vehicle may have a different set of sensors and actuators, creates the need for a more \textbf{automatized} \ac{tsn} configuration that can be adapted to different streams.

The work of~\cite{2021-TSN-18} discusses how to characterize the traffic — frame size, period, deadline constrains, priority, proportion of streams and how to analyze efficiency — overload analysis, scheduling analysis, and spacial complexity/memory usage. This can be a basis for a \ac{tsn} configurator to decide the mapping between the streams within the vehicular network and the  necessary \ac{tsn} shapers configurations.

\begin{figure}[!ht]
\centering
\includegraphics[width=\columnwidth]{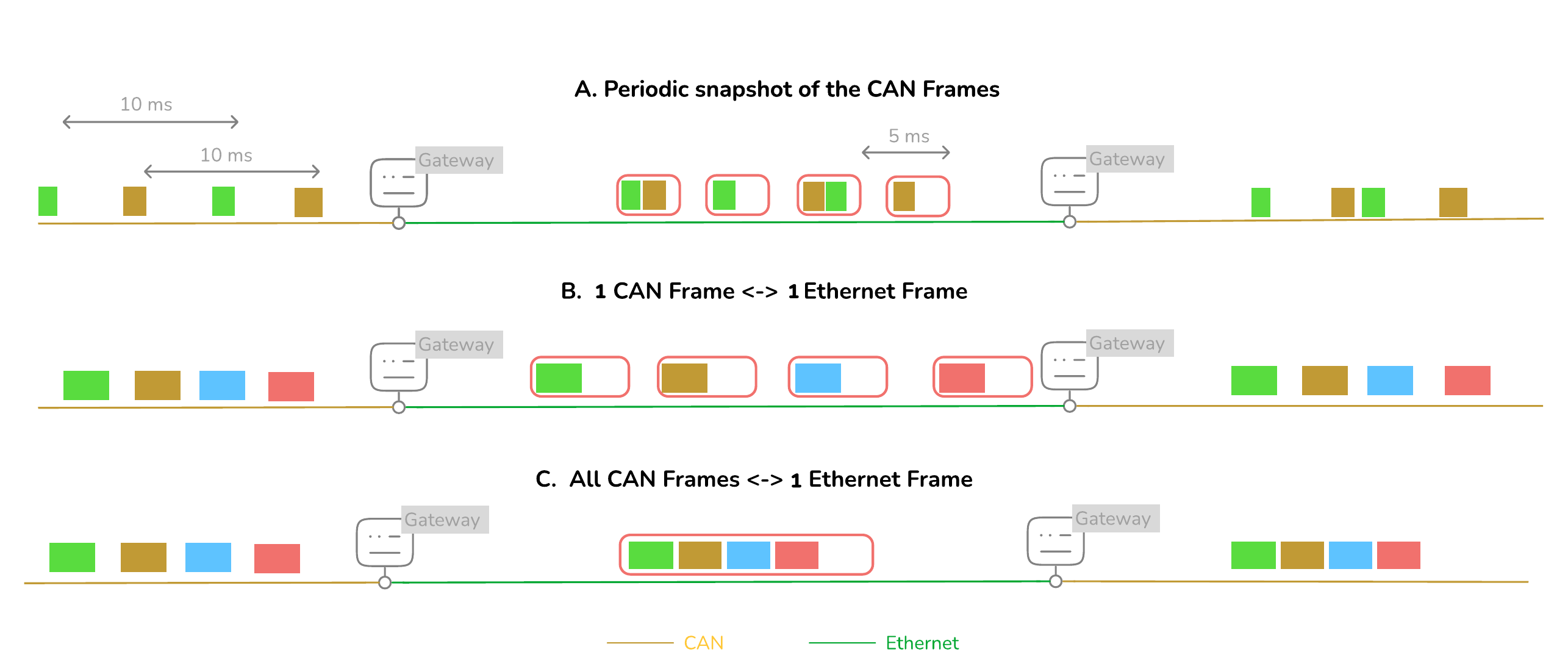}
\caption[Examples of \acs*{can} from and to Ethernet Gatewaying Strategies]{Examples of \acs*{can} from and to Ethernet Gatewaying Strategies~\cite{2023-TSN-19}.}
\label{fig:soa-renault-tsn-can-sync}
\end{figure}

Not only \ac{tsn} has to be properly configured, but interoperability with other protocols must be assured. Ethernet with \ac{tsn} must coexist and be interoperable between with other wired and wireless access technologies, including legacy automotive standards such as \ac{can}. Synchronization strategies between the vehicular Ethernet backbone and \ac{can} are currently being investigated by~\cite{2023-TSN-19,Zuo2021}, with three approaches under consideration as depicted in~\cref{fig:soa-renault-tsn-can-sync}:
\begin{itemize}
\item Periodic snapshot: periodically all original frames are mapped unto 1 or more frames of the destination bus. Within this solution, the main challenge is the computation of the period of snapshot, specially in asynchronous traffic.
\item 1:1 Packing: in this solution, 1 frame from the original bus is mapped onto 1+ frame of the destination bus. This may mean unused space in a frame, particularly in the direction \ac{can} to Ethernet, since the maximum size of a \ac{can} frame is \qty{108}{\bit}\footnote{Considering standard \ac{can} as described in \myhref{https://www.ti.com/lit/an/sloa101b/sloa101b.pdfl}{ti.com}.} vs \qty{1500}{\byte} for the Ethernet \ac{mtu}.
\item All packing: in this solution, all \ac{can} frames up to a point (functionally defined) are grouped into 1 Ethernet frame. This solution optimizes the usage of \ac{mtu} of Ethernet vs the 1:1 packing solution.
\end{itemize}

\begin{figure}[!ht]
	\centering
	\includegraphics[width=1\columnwidth]{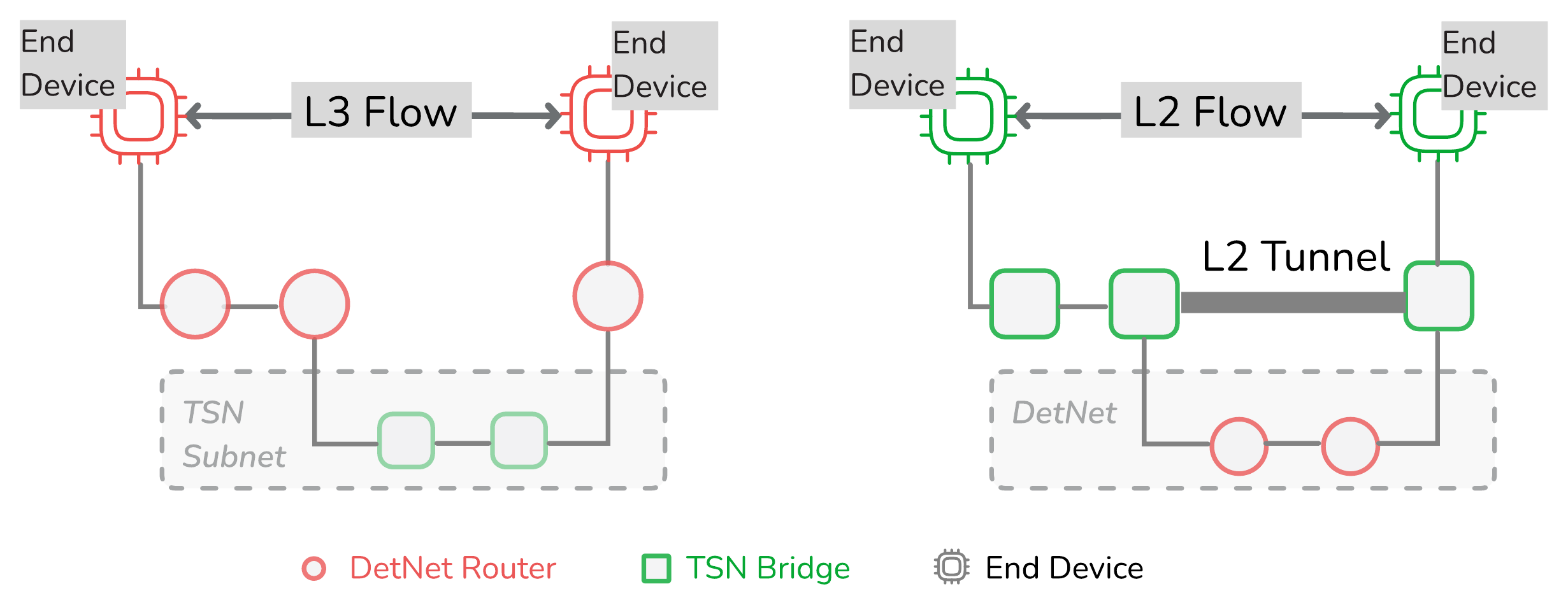}
\caption[Examples of integration of \acs*{tsn} and DetNet may be considered within a 5G system]{Examples of integration of \acs*{tsn} and DetNet may be considered within a 5G system~\cite{10002776}.}
\label{fig:soa_5g_tsn_detnet}
\end{figure}

The integration of deterministic communications in vehicles is not exclusive to \ac{tsn} standards. \ac{detnet} standards can be used in conjunction with \ac{tsn} standards to provide both upper and lower latency boundaries (while \ac{tsn} can only ensure an upper bound), better scalability and interoperability with existing \ac{ip} infrastructure (since it acts on \ac{osi} layer 3)~\cite{10002776}. Both sets of standards can act simultaneously, with~\cite{10002776} proposing that each \ac{tsn} domain would be mapped into a subnet (or a tunnel) of a \ac{detnet} network, as depicted in \cref{fig:soa_5g_tsn_detnet}.

Integration with cloud infrastructure is also possible. For example, within~\cite{2022-Virtualization-47} is proposed a full-fledged dynamic interference-aware cloud scheduling architecture for latency-sensitive workloads.

\section{Open challenges}
\label{sec:sota-sdv-discussion}
\begin{figure}[!ht]
\centering
\includegraphics[width=\columnwidth]{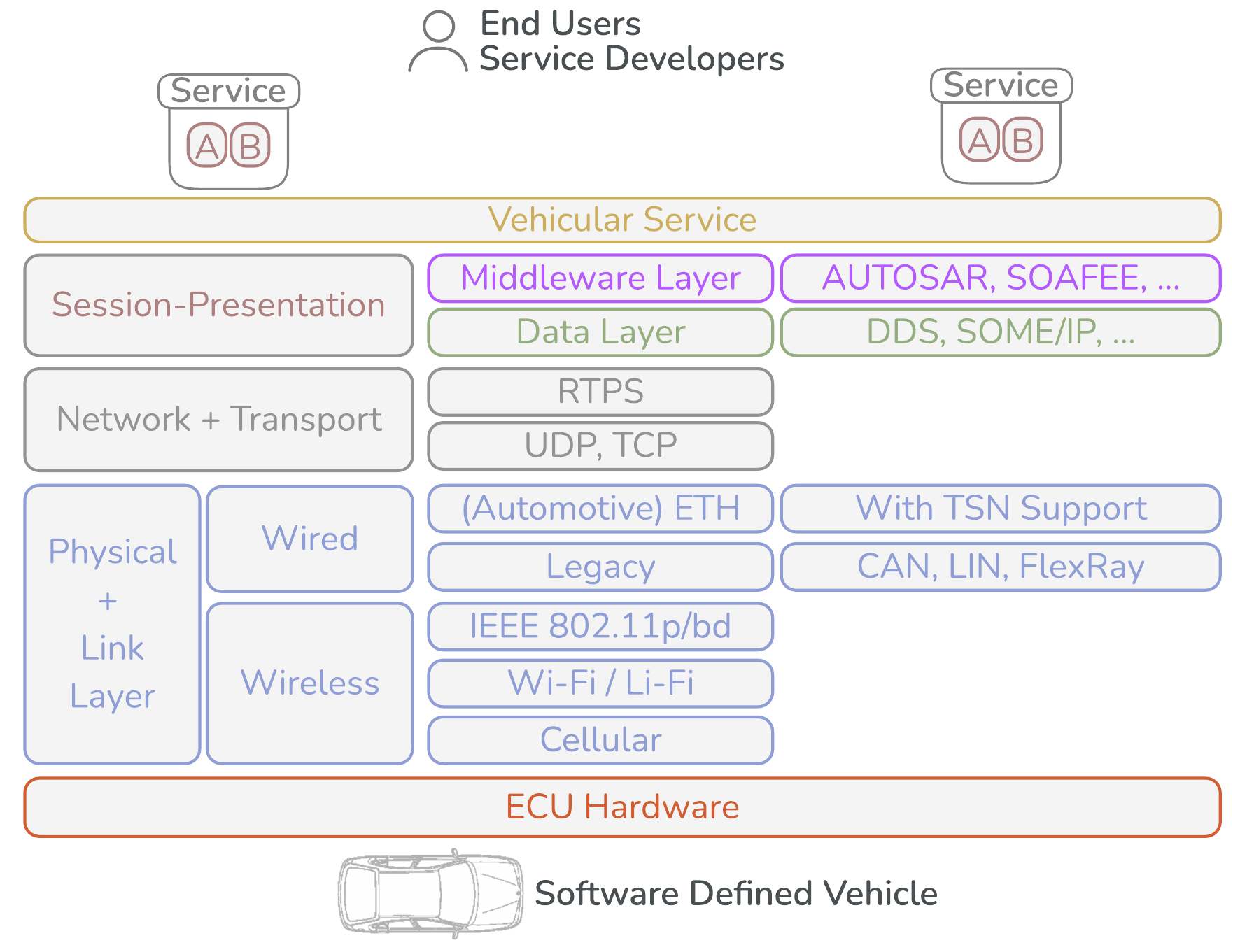} 
\caption[\acs*{osi} layers to have into consideration for an automated configuration of the whole vehicular stack]{\acs*{osi} layers to have into consideration for an automated configuration of the whole vehicular stack~\cite{IEEE_SA_d1_02}.}
\label{fig:soa-problem-osi}
\end{figure}

The previous subsections focused on specific parts of the vehicular stack - either middleware, \acl{os} and virtualization techniques. All these parts need to be configured properly in a holistic way to ensure the services are deployed properly. The work of~\cite{IEEE_SA_d1_02} emphasizes the need for  configurations and reconfigurations of the whole vehicular stack to be dynamic, from the subjacent bus (such as Ethernet), up to the configuration of the \acp{ecu} and services. Such reconfigurations require the integration of multiple levels of configuration — from deterministic networking standards to the appropriated \ac{soa} protocols to higher level as described in \cref{fig:soa-problem-osi}. 

Per the current state-of-the-art, it is essential to create tools to support a more agile service development, that considers the services requirements in terms of bandwidth, safety, criticality and timing of delivery of its information, while ensuring a interoperability and modularity to support (almost) any vehicle. These tools shall take into account not only the orchestration of those services throughout the vehicular zonal \ac{ee} architecture and its \acp{ecu} but also the network interconnecting those \acp{ecu} and connecting the vehicle to other elements on the smart city road. Therefore, determinism communications and its standards, tendentially and historically thought for static well-known environments, need to be able to be dynamically configured to support the highly dynamic environment within vehicles and their equally dynamic surroundings. Other challenges include supporting and improving:

\begin{figure*}[!t]
	\centering
	\includegraphics[width=0.8\textwidth]{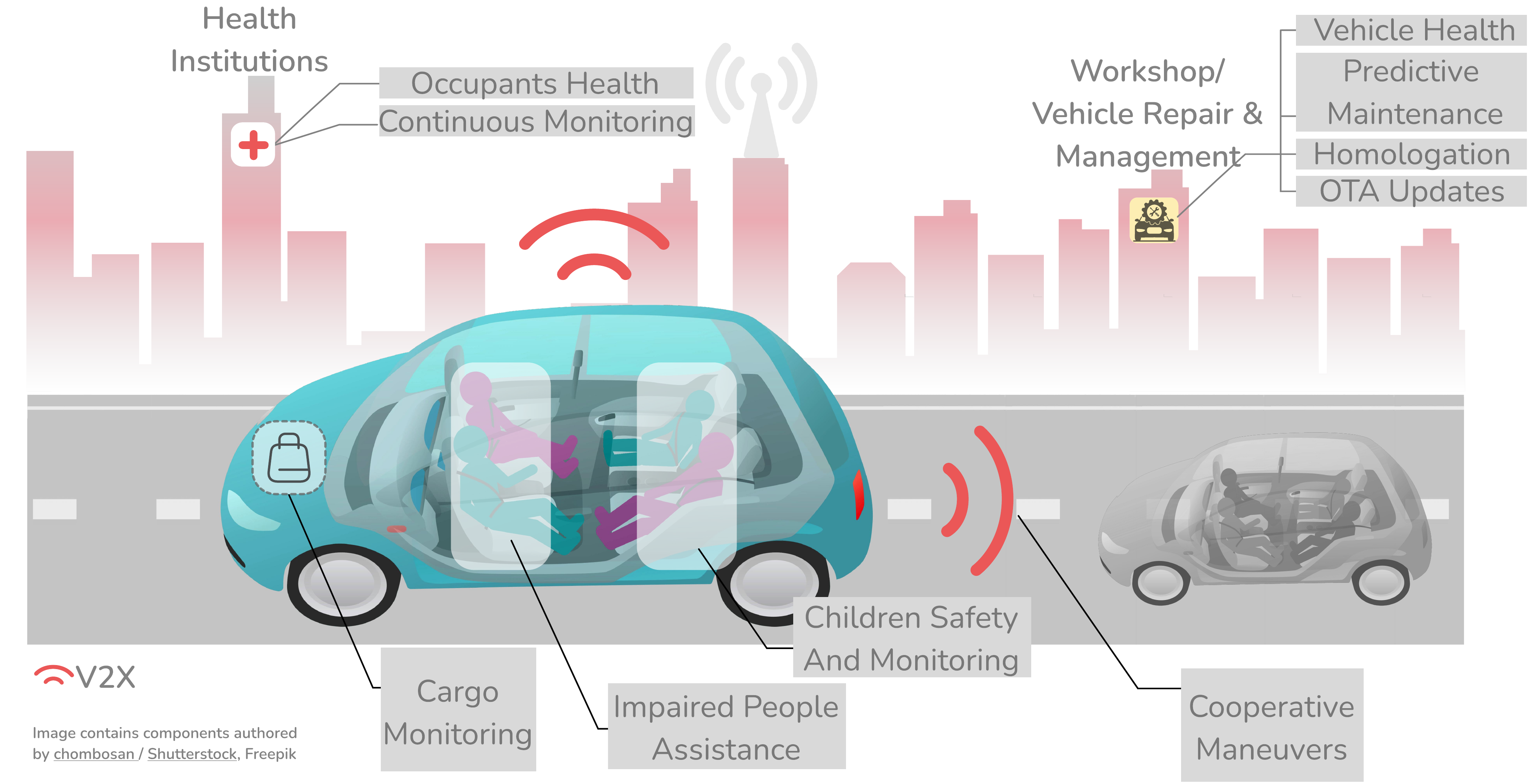}
	\caption[Overview of the potential use cases]{Overview of the potential use cases.}
	\label{fig:arch-use-cases-overview}
\end{figure*}

\begin{enumerate}
\item \textbf{Agility of service development:} simplify and harmonize vehicular development processes — (almost) everything as software and software processes; 
\item \textbf{Data driven services:} vehicular architectures should be data driven in order to process great quantities of data from camera/lidar/sensors streams in use cases like autonomous driving~\cite{Mobileye67:online};
\item \textbf{Safety concerns:} following zero trust policies is essential, as well confidential computing concerns.
\item \textbf{Optimized resource orchestration:} leveraging the unused resources — but not necessarily on a in-vehicle point of view, while considering costs and capabilities (CPU bound, I/O bound) from the hardware into the scheduling and orchestration;
\item \textbf{Interoperability:} multi-vendor, multi-hardware and multi vehicle interoperability;
\item \textbf{Modularity:} towards a more manageable stack where automakers don't do everything;
\item \textbf{Next-gen distributed storage:} considering cloud storage and solutions such as \ac{ipfs} towards use cases such as \ac{ota} updates;
\item \textbf{Clustering:} support the clustering of \acp{ecu} and the deployment of vehicular services across clusters, with reinforcement learning techniques (or equivalent) to choose the best node affinity;
\item \textbf{Automatic generation  of design architectures:} computer generated architectures for zonal  vehicular \acp{ecu};
\item \textbf{Legacy compatibility:} solutions to integrate \acp{ecu} interconnected with deterministic communications in a zonal \ac{soa} architecture;
\item \textbf {Virtualization and automatic scheduling:} reallocate workloads depending of resources with general purpose hardware and virtualized infrastructure;
\item \textbf{Digital twin:} standardize the vehicle state representation~\cite{IEEE_SA_d1_02}.
\end{enumerate}

Considering the current state-of-the-art, within the dynamic deployment of vehicular services focusing on the communication part with \ac{tsn} and deterministic networks, the main challenges include:
\begin{enumerate}
\item \textbf{Easy to use/plug and play:} Auto-configuration of parameters for \ac{tsn} shapers based on the number, characteristics of \acp{ecu} and workloads. For example ~\cite{Caruso2021} propose a \ac{tsn} platform abstraction layer and set of \ac{api} for applications to send/receive frames. The \ac{api} was responsible for managing flow registration, transmit and receive Ethernet frames, creating the Ethernet frames, configuring the Ethernet port and scheduling frames to the appropriate queue.

\item \textbf{Integration of \ac{tsn} within wireless networks:} Some efforts are being done, but it is current work as discussed in \cref{sec:wireless-detnet}~\cite{NextGene78:online}.


\item \textbf{Built-in time awareness:} at the link layer — at the applicational level is more resource intensive, while at link layer could even be accelerated by FPGAs.

\item \textbf{\ac{tsn} shaper automatic selection and configuration:} automatic generation of necessary configurations, with reinforcement learning as a possibility to choose the best configurations.

\end{enumerate}



%

%

The next sections will focus on discussing potential use cases that may benefit from such solution and strategies, while also discussing the proposed solution for a more agile service development, and tools for its evaluation.

\section{Potential use cases}
\label{sec:use-cases}


\Cref{fig:arch-use-cases-overview} lays out the main areas of use cases where the \ac{sdv} paradigm may be useful.

Even if vehicles will tend to become driver-less, autonomous vehicles may also take decades to be fully on the roads, and they will coexist with manual-driven cars. As such, autonomous vehicles shall consider the needs of the other occupants. It is therefore essential to think about vehicles, not only as means of transportation that may tend towards not requiring drivers, but also as adaptive objects, that suit themselves to the needs of occupants. 

Examples of such needs would be the vehicle itself recognizing the needs or potential health issues, and providing personalized profiles and automatic adjustments on seating, cooling systems, driving profiles, infotainment screens, lighting, audio, and other vehicular subsystems that may support some sort of impairment. But this adaptation must be carried out flawlessly, since any failure can mean harm to someone, or even death.

\begin{table*}[!ht]
\caption[Main requirements for some possible use cases.]{Main requirements for some possible use cases (based on \cite{5GAA2019}, with adaptations).}
\label{tab:arch-requirements}
\centering
\resizebox{\textwidth}{!}{%
	\begin{threeparttable}[t]
	\begin{tabular}{p{1.5in}p{.6in}p{1in}p{.7in}p{.8in}p{1in}p{.9in}p{0.8in}p{.7in}}
	\hline
	\textbf{Use Case} &
	  \textbf{Range} &
	  \textbf{Information size} &
	  \textbf{Latency} &
	  \textbf{Reliability} &
	  \textbf{Velocity} &
	  \textbf{Node density} &
	  \textbf{Positioning} &
	  \textbf{Pattern} \\ \hline
	
	\textit{Cross Traffic Left Turn Assist} &
	  \qty{350}{\meter}  &
	  \qty{1000}{\byte}/message\tnote{a} & 
	  \qty{10}{\milli\s} &
	  99.9\% &
	  \qty{33.3}{\meter}/s\tnote{b} &
	  1500 /$km^{2}$ &
	  \qty{1}{\meter} for lane accuracy &
	  Discrete \\ \hline
	
	\textit{Software Update} &
	  \textless \qty{100}{\meter} (V2V) &
	  \textless \qty{10}{\giga\byte}/day, \textless \qty{10}{\giga\byte}/hour (usually/urgent) &
	  \qty{30}{\s} (V2V) &
	  99\% &
	  \qty{25}{\meter}/s\tnote{c}, 70 m/s\tnote{d}  &
	  \textless 1500/$km^{2}$ &
	  30 m for street/road accuracy &
	  Discrete \\ \hline
	
	\textit{Predictive maintenance} &
	  N/A &
	  \textless \qty{1}{\kilo\byte}/message &
	  \textless \qty{30}{\s} &
	  99.99\% &
	  70 m/s\tnote{e} &
	  15000/$km^{2}$\tnote{f} &
	  \qty{1}{\meter} for lane accuracy &
	  Continuous \\ \hline
	
	\textit{Speed Harmonisation\tnote{g}} &
	  \textless \qty{150}{\meter}\tnote{h} &
	  \textless \qty{500}{\byte}/message\tnote{i} &
	  \qty{1400}{\milli\s}\tnote{j} &
	  $\sim$80\% &
	  69.4 / 33.3 / 19.4\tnote{k} &
	  \textless 15000/$km^{2}$ &
	  \qty{1}{\meter} for lane accuracy &
	   Continuous \\ \hline
	
	\textit{Context aware maneuvering} &
	  \qty{100}{\meter} &
	  \qty{100}{\byte}/message\tnote{o} &
	  \qty{50}{\milli\s} &
	  99\% &
	  \qty{33.3}{\meter}/s\tnote{l} &
	  \textless 15000/$km^{2}$\tnote{m} &
	  \qty{1}{\meter} for lane accuracy &
	  Continuous (\qty{10}{\hertz})\tnote{n} \\ \hline


	\end{tabular}%
	\begin{tablenotes}
         \setlength{\columnsep}{0.8cm} \setlength{\multicolsep}{0cm}  
         \begin{multicols}{3} 
		\item[a] ETSI CAM size with extras
		\item[b] considering maximum velocity of 120 km/h 
		\item[c] maximum non highway speed of 90 km/h 
		\item[d] urgent, during a vehicle top speed at certain highways
		\item[e] during a vehicle top speed at certain highways
		\item[f] heavy traffic around the vehicle
		\item[g] worst case scenario, with human reaction
		\item[h] majorant of the provided values
		\item[i] speed, location and additional information
		\item[j] minorant of the provided values
		\item[k] highway / rural / urban
		\item[l] maximum velocity of 120 km/h
		\item[m] considering an urban situation 
            \item[n] worst case scenario in possible range of frequency of transmission of \ac{mcm}, per \cite{ETSI_TS_103_561} 
            \item[o] maximum size of the message \cite{ETSI_TS_103_561}
            \end{multicols} 
	\end{tablenotes}
\end{threeparttable}%
}
\end{table*}


When considering areas of use cases for \ac{sdv}, one can consider both the drivers — for which most use cases, including autonomous driving, tend to focus on — and occupants. Both can benefit from the vehicular technology that adapts to each occupant while increasing its autonomy. \acl{sdv} can use the existing sensors and actuators on vehicles to support new services to support drivers and occupants in situations automakers could not predict during their design phase.

The following subsections present some use cases that are envisioned as possibilities to validate the system(s) to be researched — \textit{intra and inter-vehicle deterministic use cases and critical non-active safety use cases} that require adaptability — both at software level and network level (while ensuring reliability and bounded latency). In all described use cases, the \ac{sdv} paradigm supports some adaptability of the vehicle to the occupants' characteristics and state to support some automated action (instead of a manual process). The occupants, either driving or passengers, can greatly benefit from services for customization of vehicular functions, protection, monitoring, and companionship.

\subsection{Requirements and characterization of vehicular services}

\Cref{tab:arch-requirements} provides an overview of the main requirements envisioned for the use cases. Use cases can be characterized by several types of requirements. The study in~\cite{Pretschner2007} defines several vehicular use cases and their characteristics in terms of deadline (soft/hard and interval), data complexity (simple/complex and large) and pattern (discrete/continuous). 

During the overall definition and characterization of any vehicular service supporting an use case, it is essential to formulate the service and characterize its tasks in terms of those requirements, with particular emphasis on characteristics such as \cite{Bandur2021,Kugele2018,2021:vdi_wissensforum_gmbh:eliv_2021,Chebaane2020}: 
\begin{itemize}
    \item Priority and criticality; 
    \item Data specifications: such as format, size and real time guarantee requirements (periodicity or deadline for receiving/sending data);  
    \item Expected computing location: edge vs cloud;  
    \item Computing resources: either in terms of \ac{cpu}, memory, storage, or graphical resources; 
    \item Required real-time guarantee levels: in hard real-time, weakly hard real-time or soft real-time, depending on the impact of missed deadlines;
    \item Characteristics specific to the vehicular domain: such as the mobility of vehicular nodes and associated communication handovers and failures, and the heterogeneity and overall low specification of the \acp{ecu};
    \item Non-functional requirements: noteworthy examples include high interoperability (including backwards compatibility with older vehicles and other vehicle brands), reliability, availability, scalability, extensibility, adaptability, testability and cost minimization.
\end{itemize}

These characteristics can be seen as the \ac{sla} of the service. Such \acp{sla} should and could be defined by the developer of each vehicular service within a descriptor file for the service to support selecting the best technologies/stacks and configurations. Such configurations would be ensured/enforced by a distributed configurator through a set of \ac{qos} tools to enforce policies on the task scheduling and communication technologies tasks~\cite{Chebaane2020}. On task scheduling, the configurator would be responsible for a dynamic, intelligent, task scheduling and resource estimation, provisioning and allocation, including the offloading to other layers of communication (fog and cloud) while keeping the original non-functional requirements. Such scheduling could include the prediction of the overall resource usage while considering heterogeneity and computational cost of the nodes for optimization purposes~\cite{Chebaane2020.47,Chebaane2020.50}. Estimation and prediction of the mobility of a certain service deployment based on trends of resource usage and prediction of the mobility of the vehicle would be essential in scenarios where the service requires deployment on multiple vehicles and/or external infrastructure to ensure coordination with other vehicles or entities~\cite{Chebaane2020.57}.

Within the communication technologies, the configurator would be responsible for the dynamic management of the communication technologies during the mobility of the vehicle. Solutions would include the federation of multiple physical wireless and wired access layers in order to obtain a multi technology solution. Such management could also include slicing of the communication solution to support different levels of \ac{qos} specified by the service descriptors, as well as the management of the high mobility with the management of reactive and proactive handovers leveraging \ac{sdn}~\cite{SILVA2021100372} based solutions;

\subsection{Services for autonomous driving in all road situations}
\label{sec:arch-use-cases-road}
While autonomous driving is a current and future trend with great focus from automakers, its adaptability to all road and traffic situations is an open problem. Current solutions can handle open road or highway contexts, but higher density (\eg urban) contexts require a more cooperative approach to autonomous driving, with vehicles sharing all the information they possess and coordinating manoeuvres.

\subsubsection{Context Aware Maneuvering}
\label{sec:arch-use-cases-road-cam}
Considering the importance of working towards autonomous driving for everyone, it is vital to ensure that each autonomous driving vehicle is capable of creating the best possible decision not only for itself, but for other vehicles, and road occupants as well. This requires leveraging the concept of cooperation between vehicles and others. While standardization work, for example through \ac{etsi} \acp{mcm} are under way, the cooperation between vehicles, particularly for autonomous driving is in its initial stages. 

Vehicles may consider cooperative manoeuvres between non-autonomous vehicles and/or autonomous vehicles to ensure that complex manoeuvres are done with no harm.  For example, within the context of the cooperation between a personal vehicle public and private transportation, a service can be considered to give priority to bus, ambulance or other public transportation to enter a certain crossroad instead of a privately owned vehicle. This service shall be react to changes on the environment and surroundings of the vehicle and decide the action of each vehicle. Both receive information on the road situation and sending the proposed action to other vehicles would require a minimized bounded latency to ensure a proper decision and that such decision is enforced in the right moment.

When considering the air mobility paradigm, in particular \ac{evtol} vehicles, the cooperation between the vehicle and other vehicles gain an additional importance. New scenarios shall be considered, including cooperation between vehicles and the \ac{evtol} and the road infrastructure (such as landing pads) to ensure that no accident occurs during take off or landing, even in bad visibility (by occlusion or lack of lighting). In these scenarios, coordination between vehicles and infrastructure can avoid a catastrophic air accident between the vehicles or a potentially deadly crash on top of property or people~\cite{Thecasef27:online}.

Autoware, an open source framework considered to be state-of-the-art for autonomous driving\footnote{Apart from closed source solutions from automakers (more details on~\cref{sec:sdv:industrial}), where it is not possible to fully understand the scope of the cooperation between vehicles and its interoperability with other brands.}, considers several services to ensure an autonomous driving — albeit in previously mapped environments — to support manoeuvres such as curb-to-curb, lane following, lane merging, parking or reaction to traffic lights and dynamic entities (\eg \acp{vru}). 

To produce plans that eventually become decisions and actuation to the engines, braking and other vehicular subsystems, Autoware relies on information of the vehicle itself and on a model of the environment. All this planning is therefore very local, with cooperation from other vehicles and road entities proving essential not only to improve Autoware's vision of the world —in a cooperative sensing solution such as the one defined by \ac{etsi} \ac{cps}—, but also to coordinate the manoeuvre that each vehicle will do, ensuring that a safe manoeuvre for one vehicle does not possess danger to another. 

When the planning becomes cooperative and not only local, a new problem arises. Information from and to each vehicle \textit{must} be sent within a well defined time-frame that is compatible with Autoware producing a decision and actuating on the vehicle's subsystems. If the information arrives after this defined time-frame, the decision will not be correct, while if it arrives before this defined time-frame, the decision might be unnecessary — but also potentially dangerous\footnote{For example, a vehicle would end up braking before being necessary, becoming vulnerable to being hit in the back.}.

\begin{figure}
	\centering
	\includegraphics[width=\columnwidth]{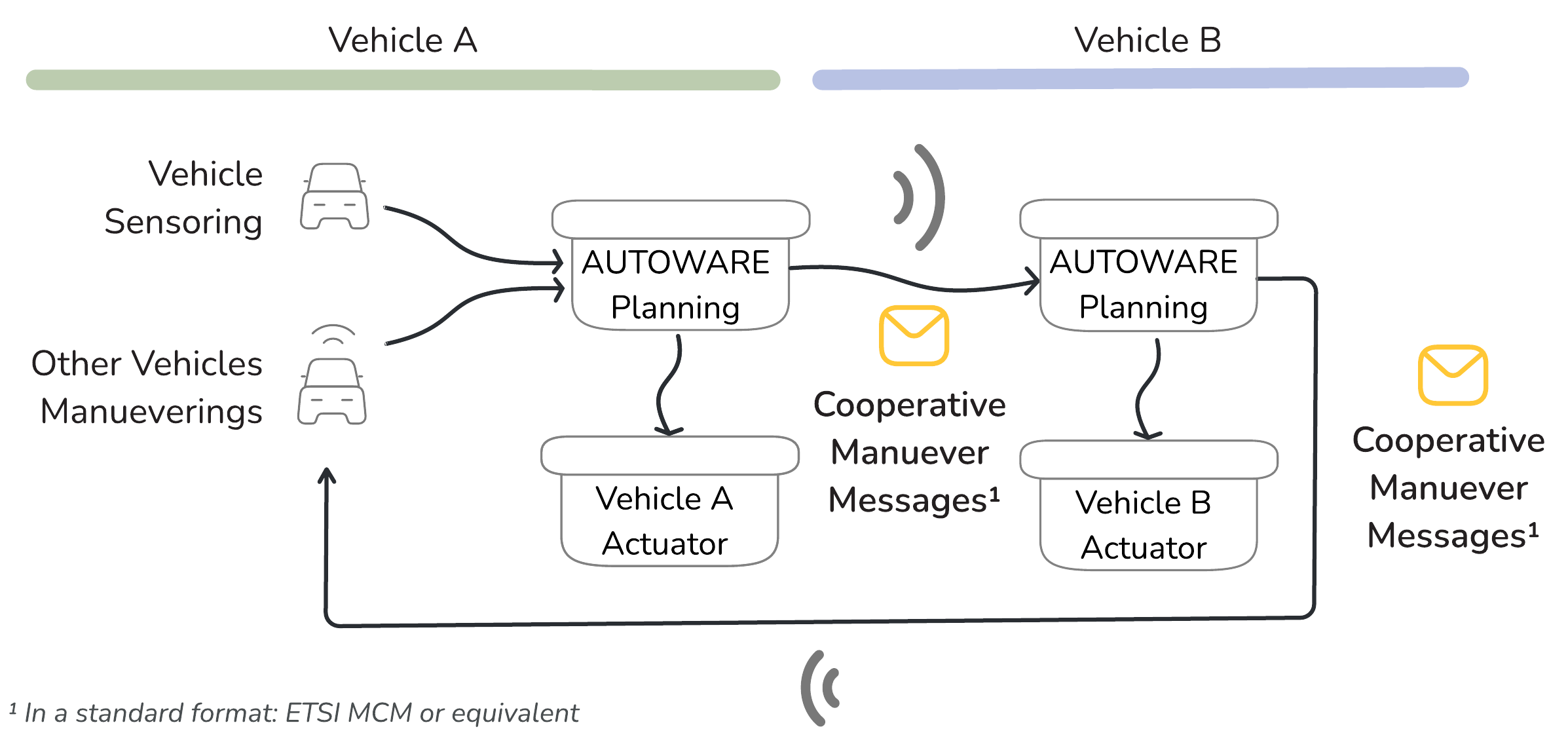}
	\caption[Overview of the Context Aware Maneuvering use case]{Overview of the Context Aware Maneuvering use case.}
	\label{fig:arch:considered-usecase-overview}
\end{figure}

As an example scenario, one may consider coordinating lane merging manoeuvre between an ambulance and a private vehicle. This use case requires the development of services depicted in~\cref{fig:arch:considered-usecase-overview}, namely a) a message creator and producer to support exchanging data between Autoware instances in different vehicles, and b) a parser and injector of such information into Autoware planning. The data format would follow \ac{etsi} \ac{mcm} standards and it would be periodic, with deadline insurance requirements (and therefore both a minimum and bounded latency requirement) with no tolerance to failures of delivery, and with a variable size with an upper bounds corresponding to the maximum size of the \ac{mcm} message. 

The integration of Autoware with cooperative manoeuvering is considered in~\cite{Asabe2023b}\footnote{Additional details at \myhref{https://tlab-wide.github.io/AutowareV2X/pr-48/design/use-cases/}{tlab-wide.github.io/AutowareV2X}.}. However, it fails to consider the deterministic communications requirements or the need for orchestrating the overall deployment in vehicles' \acp{ecu} in a more automated approach.

\subsubsection{Vehicle malfunction situations}
\label{sec:arch-use-cases-road-emergency}

Even when full autonomy of a vehicle is not only a reality but also legal, vehicles can still have unexpected issues that endanger their occupants and surroundings. Examples include a flat tire, a tire explosion, failed waterproofing of the batteries (in the case of electric vehicles), or a failure on the valve timing (in the case of internal combustion engines). In these scenarios, services to \textbf{analyze, monitor and warn} others about the situation should be deployed in a bounded minimal time, and the information to other vehicles shall be sent with both minimized, and bounded latency. 

When considering \aclp{cav}, services that monitor and react when sensors are malfunctioning are critical, and their information must also be bounded in time and transmitted reliably. Without such services, the \acp{cav} cannot react to a malfunctioning or occluded sensor (such as camera, \ac{lidar} or equivalent) and can wrongly assume that their information is correct, with potentially deadly consequences.

\subsection{Services for vehicles tailored for everyone}
While the future of mobility is seemingly more autonomous and more feature-packed, less concern is given to occupants and on their needs, including children or occupants who may have some sort of physical or cognitive restriction. Restrictions considered here may include some minor impairment, vision or hearing issues, concentration issues, and others, that restrict the occupants' capability of reacting to the environment in the same way. 

The following situations are examples where a vehicle has an occupant with some issue/limitation who can greatly benefit from a \ac{sdv} paradigm. In this paradigm, the vehicle can deploy services tailored for the people's specific needs for customization of vehicular functions, protection, monitoring, and companionship. In all situations, like in previous use cases, the services to be developed and deployed shall have their data delivered in a bounded time — so the services are not left in an inconsistent state.

\subsubsection{Automatic vehicle configuration based on occupants characteristics}

Information on the occupants' characteristics and vital parameters can be obtained by intra-vehicular subsystems that allow the vehicle to adapt interior functions (\eg seating position and configurations, lighting, sound, infotainment) to the best comfort of the occupants, therefore creating a set of customizations tailored to each occupant's characteristics.

Not only the services can adapt, but warn if there is an issue (like a non-fixed wheelchair or baby seat). Such warnings can be sent both to the infotainment vehicular subsystems, and also to other vehicles. Vehicles can then use such information, as well as information about the occupants' characteristics, to deploy services to adapt the vehicle driving mode, profile and characteristics. For example, such adaptation could signify an increase cruise control following distances, reduce throttle response, or warn other smart city infrastructures when someone inside the vehicle is ill or impaired.

These services can greatly benefit the healthcare industry in particular. Consider a scenario where an ambulance is transporting a patient. Vehicular services can support scenarios such as this:
\textit{\begin{itemize}
		\item Ambulance: I have a person in risk of life.
		\item Other vehicles: Understood, I need to slow down and provide space around the ambulance.
		\item \ac{sdv} Orchestrator: We'll provide you with specific applications (\eg to control acceleration profiles or climate control settings). We'll deploy an application for monitoring ill people that will run on a cluster of $X$ vehicles. We will coordinate with all the vehicles in the path, and with the city traffic manager, to create a corridor for you to be able to pass between other vehicles in the city.
	\end{itemize}
} 

\subsubsection{Adapting the vehicle's subsystems to impaired occupants}
\label{sec:arch-use-cases-health}

\begin{figure}[!ht]
	\centering
	\begin{subfigure}[b]{0.6\columnwidth}
		\includegraphics[width=\columnwidth]{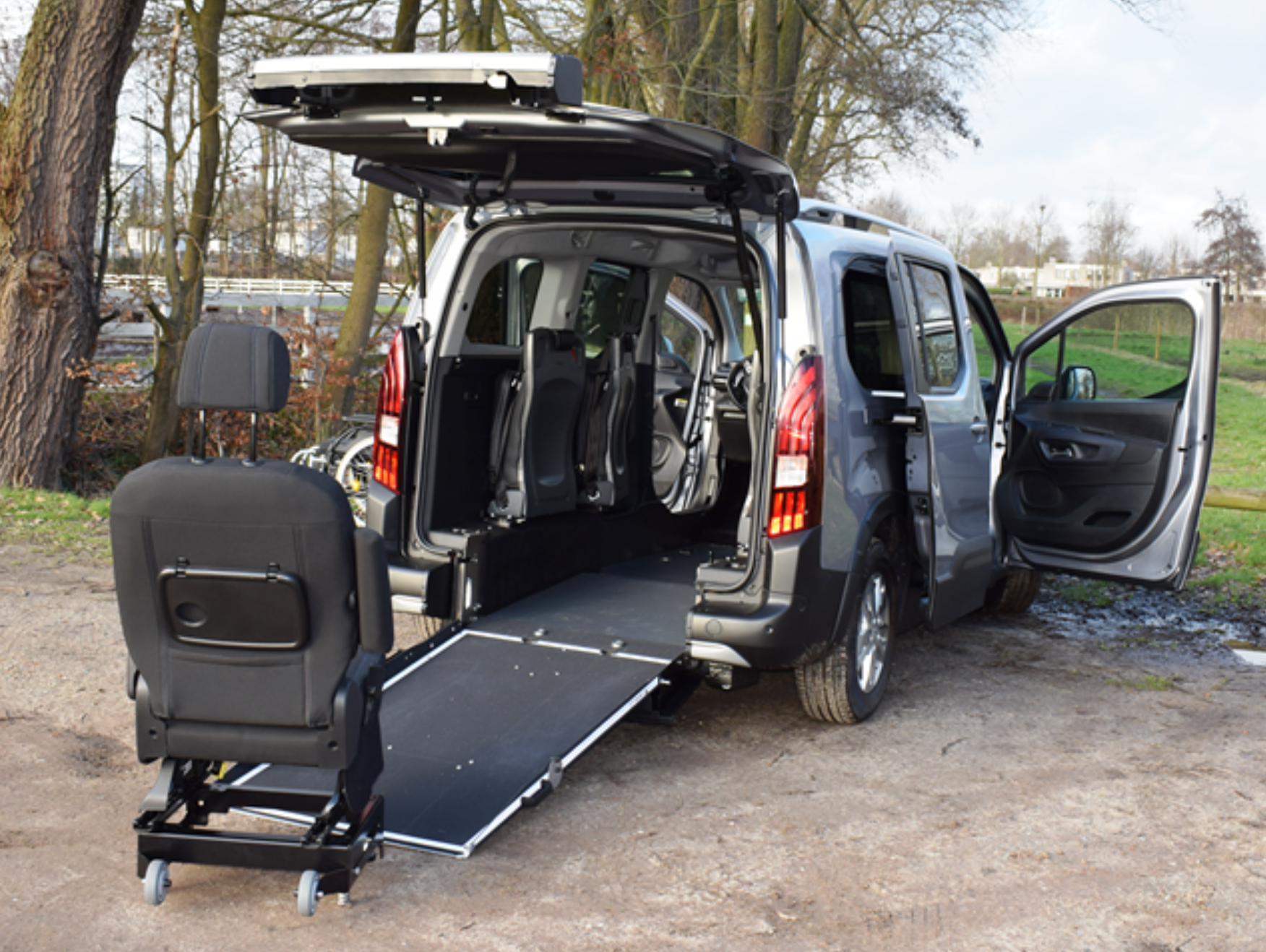}      
		\caption[Example of ramp where wheelchairs need to be pushed manually by the healthcare provider]{Example of ramp where wheelchairs need to be pushed manually by the healthcare provider (courtesy of Tripod Mobility\footnote{More details at \myhref{https://www.tripodmobility.com/en/products/wheelchair-accesible-vehicles/upfront/}{tripodmobility.com}.}).}
		\label{fig:arch-usecase-ramp}
	\end{subfigure}
	\hfill
	\begin{subfigure}[b]{0.6\columnwidth}
		\includegraphics[width=\columnwidth]{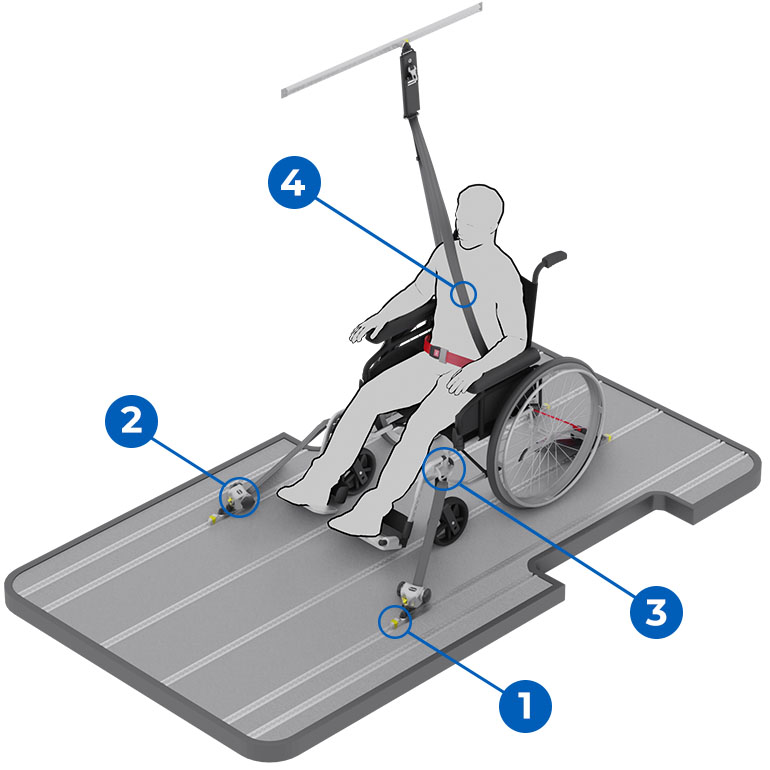}      
		\caption[Example of expected attachment points within a vehicle for a wheelchair]{Example of expected attachment points (4 in this case) within a vehicle for a wheelchair (courtesy of BraunAbility\footnote{Available at \myhref{https://www.braunability.eu/en/products/tie-downs-and-seatbelts/wtors/}{braunability.eu}.}).}
		\label{fig:arch-usecase-seatbelts}
	\end{subfigure}
	\caption[Detail view on the main steps for transporting impaired people on a vehicle]{Detail view on the main steps for transporting impaired people on a vehicle.}
\end{figure}

Consider a situation where, in order to transport impaired people in a wheelchair, a healthcare provider  must\footnote{From an informal interview with a specialist nurse in medical-surgical nursing.}:
\begin{enumerate}
		\item Get the wheel chair up a moving ramp or lift, per depicted in~\cref{fig:arch-usecase-ramp}\footnote{Another example of such ramps and lifts in \myhref{https://www.gilaniengineering.com.au/product/hydraulic-wheelchair-platform-accessible-lift-hoist-for-cars-and-vans/}{gilaniengineering.com.au}.}.
		\item Attach the wheelchair to the vehicle through a set of several belts fixed to specific points within the vehicle, as depicted in~\cref{fig:arch-usecase-seatbelts}.
\end{enumerate}

The healthcare provider must do such \textbf{manually}, multiple times (for multiple people) and in a short time-frame and without little to no feedback on whether the person and the chair are actually fixed properly. Mistakes include fixing outside of the fixing point(s), twisting the straps and/or simply not fixing the person and/or chair. In addition to that, specific formation is required to do such tasks. 

A vehicular service (or set of services) that is capable of communicating with other in-vehicle systems (\eg infotainment),  but also with the healthcare provider registry systems, would greatly decrease the potential human errors of the current solution, and since it would minimize the dependency on humans, it could even support an increase on the patient's privacy. One could even consider a more intelligent connected wheelchair directly coordinating with the vehicle all the necessary movements. Such systems would not require a minimized latency, but guarantees of delivery of the information within well-known time bounds and with low jitter in order to ensure no unwanted movements of the wheelchair that may harm the person.

The adaptability of the vehicle does not need to end on improving or automate mechanical subsystems. In fact, greater challenges can be considered. Depending on the profile of the person (whether it's any sort of impairment, special needs or old age), the vehicle could further adapt itself for example by choosing a softer, appropriated music (supporting music theraphy\footnote{More details at \myhref{https://www.britannica.com/topic/music-therapy}{britannica.com/topic/music-therapy}.}), adapt the climate subsystems to ensure the adequate temperature, tint the glass to avoid too much sun exposure. 

Other scenarios may include the protection of other type of vulnerable humans, specifically babies:
\begin{itemize}
    \item Managing situations were babies are left unattended in cars. It is a widely known issue\footnote{Unfortunate examples at \myhref{https://edition.cnn.com/2024/07/17/us/hot-car-deaths-charges/index.html}{cnn.com}.}. Babies, when left in vehicles and especially in hot weather, where vehicle temperatures can rise rapidly, are at risk of heatstroke and potential death. Vehicular subsystems and services could detect the presence of the baby and alert caregivers or authorities if the child is left inside under dangerous conditions.
    \item Managing the set of airbags that should or should not be deployed based on the position or age of the baby. 
    \item Monitoring a baby who is ill or very restless and provide such information directly to a hospital. 
\end{itemize}

\subsubsection{Personal assistants}

Vehicles can include humans in new ways within the driving experience, even when considering the \ac{sdv} paradigm is parallel to the advent of autonomous vehicles. The unparallel amount of information and sensing the vehicle can obtain from its environment, driver and occupants, can be used to support personal assistants that give occupants the sensation of being in control — even if the vehicle is self-driving, potentially increasing not only the enjoyment of the vehicle but more importantly the perceived trustworthiness on the vehicle.

Such assistants can be developed as a set of subscribing services within the vehicle's \acp{ecu} dedicated to infotainment subsystems and, if using speech or touch interaction modalities, it can greatly benefit from a well defined timely delivery of information which decreases the overall jitter visible to the end users.




\subsection{Services for assisted fleet management and maintenance}
\acp{sdv} can also support scenarios related to management of vehicle(s), including managing its health, maintenance (predicatively or not) and homologation, both for individual vehicles or for a fleet. 

\subsubsection{Fleet management}

\begin{figure}[!ht]
	\centering
	\includegraphics[width=\columnwidth]{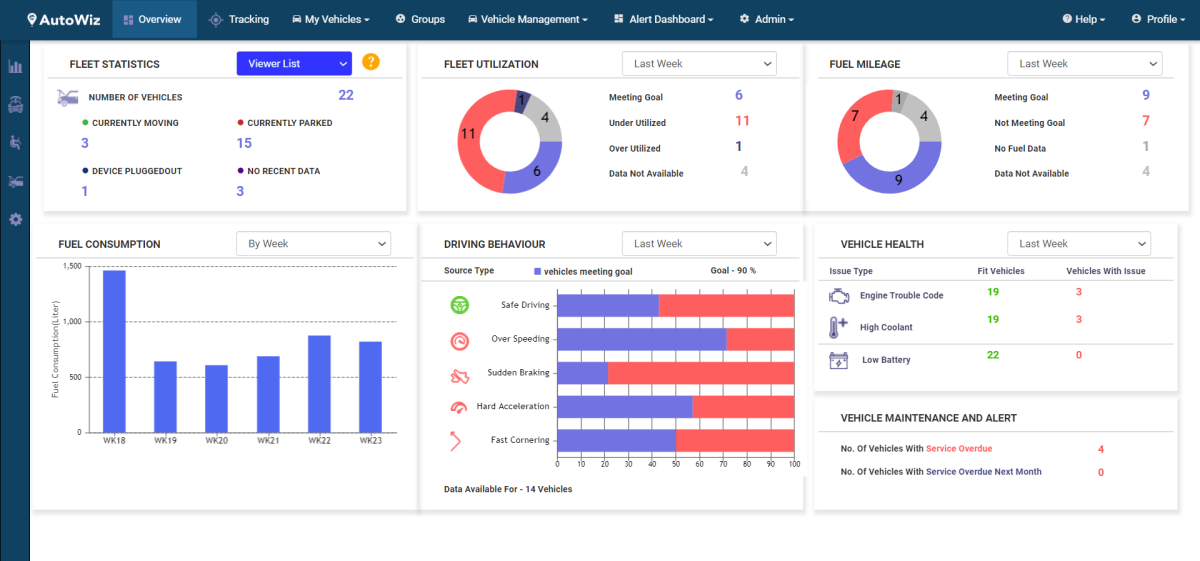}
	\caption[Example of vehicle fleet management system]{Example of vehicle fleet management system (courtesy of AutoWiz).}
	\label{fig:arc:usecase-fleet}
\end{figure}

Fleet management scenarios and remote services, such as the one presented at~\cref{fig:arc:usecase-fleet}\footnote{More details at \myhref{https://www.autowiz.in/fleet-features.html}{autowiz.in}.}, are examples of use cases that can benefit from a set of services deployed across multiple vehicles. They consider a set of services containing the logic required to monitor, in real-time, the status (\emph{e.g. location or number of hours the vehicle is driven)} of a vehicle within a fleet. This set of services, if orchestrated throughout a cluster of vehicles, can support systems for enterprises managing a vehicular fleet, homologating each vehicle, or even for state entities to perform the routine inspection. The information from/to those services, while it does not necessarily require an extremely low latency, it must be delivered and delivered within known time bounds and with low jitter.

\subsubsection{Remote vehicle health and predictive maintenance}

Constant monitoring of the main vital parameters of a vehicle state — such as tire condition, brake system condition, engine cooling system status (in the case of internal combustion engines) or the power delivery subsystems (in the case of electric vehicles) — requires not only a reduced latency but also, in case the vehicle detects an issue, a guarantee of delivery in a bounded time. This is required so that, at the very least, other vehicles are aware of the problem before it has impact on the road. In the case of emergency scenarios — as previously described in \cref{sec:arch-use-cases-road-emergency}, the latency shall also be minimized.

Predictive maintenance scenarios, while not necessarily requiring a bounded latency within its communication, still require a dynamic deployment of services that considers the vehicular \acp{ecu}.

\subsubsection{Over-the-air updates and real-time homologation}
Changes in the vehicles’ software — either via \ac{ota} updates, through the addition of new services or after a vehicle is sold to ensure no changes — should, in an automated way, be reflected in a re-homologation process. This process, while not requiring a minimized latency, shall be done in a bounded timely manner, with a guarantee of reliability; otherwise, vehicles may be driving on public roads with an unwanted, unsafe state~\cite{Webinar2023HumanVSDigitalDriver}.

\section{Proposed vision}
\label{sec:architecture}
The \ac{sdv} concept is capable of supporting several services spanning several different types of use cases. All the previously defined use cases require $N$ time critical services and $K$ services. In all these use cases, a certain vehicular service can either be fully deployed on one or more nodes, or divided through multiple nodes / cluster of vehicles, with both scenarios considering the cloud as an option. Also common to all use cases is their need for time criticality. Within the \ac{sdv} paradigm, the vehicle may send the information about its occupant to other subsystems, or vehicles through communications, which must be deterministic in order to ensure the information is obtained not too soon nor too late, therefore risking lives. While the use cases do not necessarily focus on controlling the critical systems intra-vehicle such as power-train and braking subsystems, they are essential for other passive safety functionalities. 

\begin{figure}[!ht]
	\centering
	\includegraphics[width=\columnwidth]{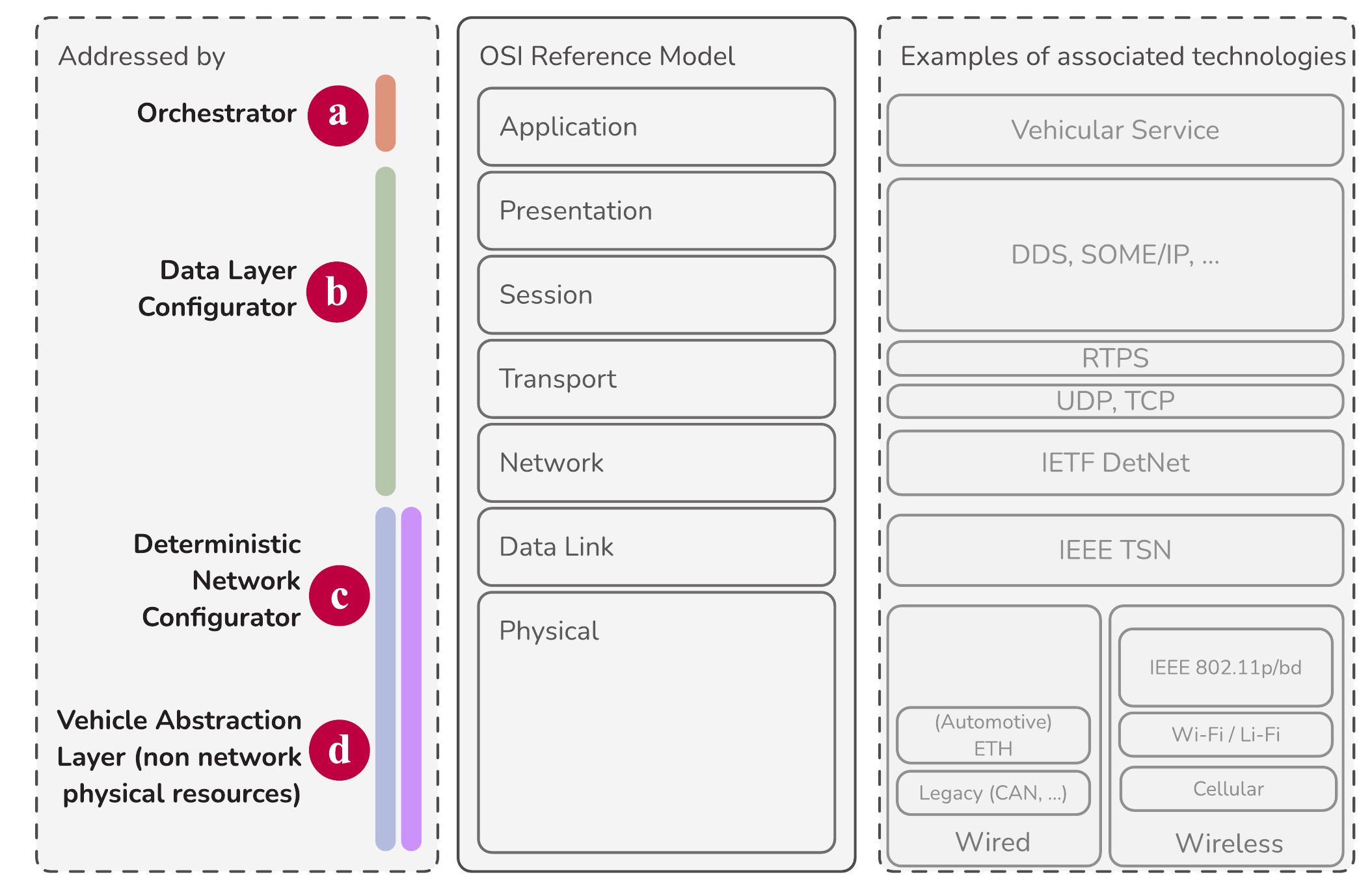}
	\caption[Overall layered view of the architecture main modules framed within the OSI model]{Overall layered view of the architecture main modules framed within the \ac{osi} model.}
	\label{fig:arch-layers}
\end{figure}

Given the criticality on the one side of the use cases and associated services, and the increased complexity of development and deployment of services, it is essential to consider a dynamic management of services. Towards such management, several layers must be considered as depicted in \cref{fig:arch-layers}. These layers should act at least in the following domains~\cite{IEEE_SA_d1_02}:
\begin{itemize}
	\item Orchestration of the services: where they should be run and policies to decide the node(s) and the number of replicas. Such domain is the responsibility of the orchestrator, \pcircle{a}.
    \item Configuration of the abstraction layers required to support exchanging data in predefined formats and protocols. Such domain is the responsibility of the data layer configuration (\pcircle{b}).
    \item Management of the deterministic network: including automatic configuration of routing and deterministic traffic shapers and reliability processes. This domain is the responsibility of the deterministic network configurator (\pcircle{c}).
    \item Virtualization of the hardware in order to support better resource usage and decoupling of services and operating systems from each \ac{ecu}. This domain is the responsibility of the vehicle abstraction layer (\pcircle{d}).
\end{itemize}

\begin{figure}[!ht]
	\centering
	\includegraphics[width=\columnwidth]{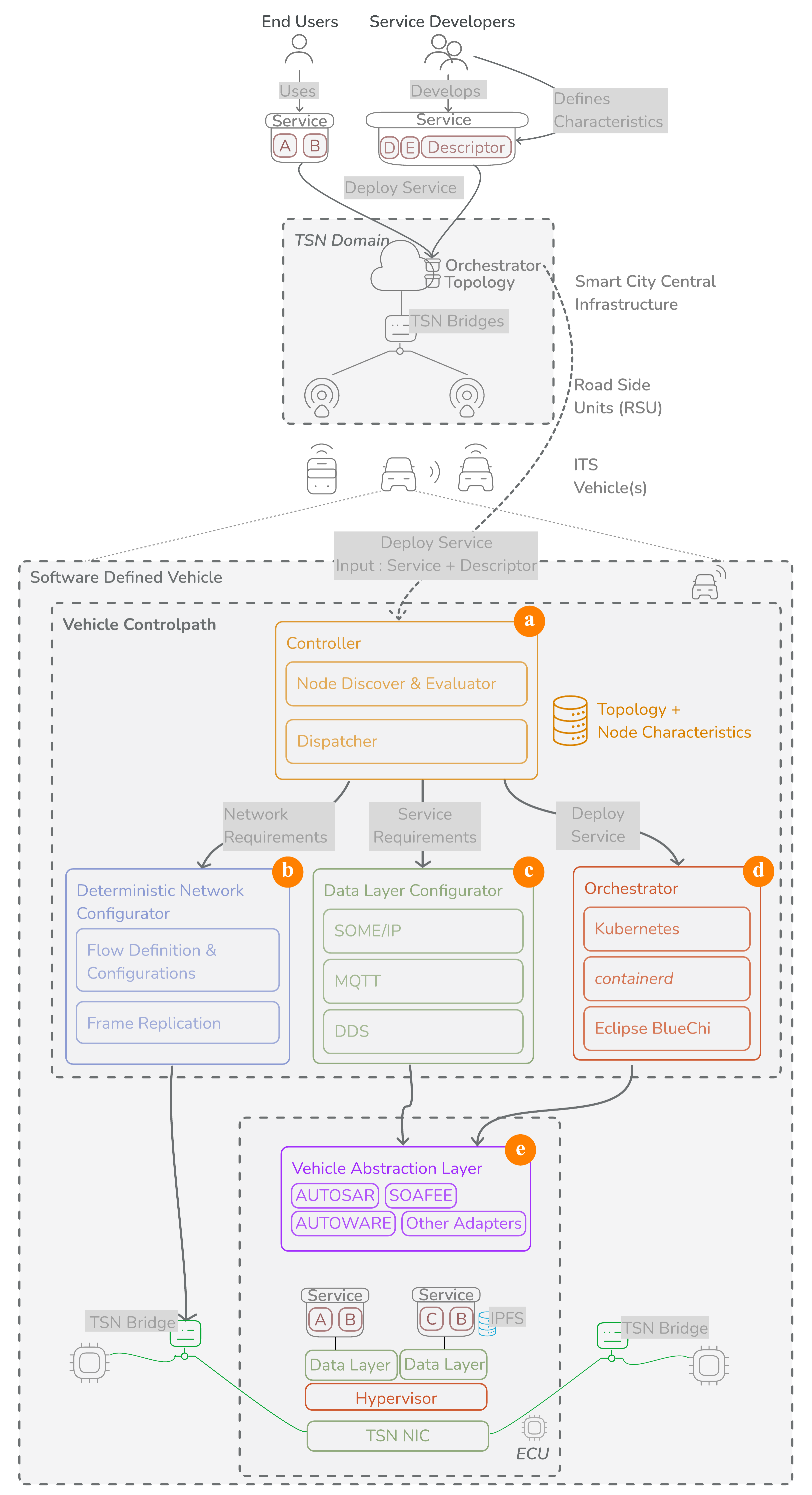}
	\caption[Overall high level architecture of the proposed solution, depicting the service development and deployment flow from the service developers down to the vehicular hardware]{Overall high level architecture of the proposed solution, depicting the service development and deployment flow from the service developers down to the vehicular hardware.}
	\label{fig:arch}
\end{figure}

\Cref{fig:arch} presents the overview of an initial proposal of architecture towards such a dynamic management of vehicular services. This architecture supports support service scheduling, redundancy, and healing based on the vehicles' computing capabilities.  

The \textbf{Controller} (\orcircle{a}) is the main entry-point of the system to be researched and ensures the first aforementioned layer. Through user-defined and system-defined metrics within a Service Descriptor, it decides the amount of resources and the placement of those resources throughout the computing continuum and the different vehicles throughout the smart city, while considering possible handovers and optimal usage of the computing and network resources. The controller is responsible for controlling 4 verticals.

The first vertical, the \textbf{Deterministic Network} (\orcircle{b}), ensures that messages – either control messages to actuate or status messages containing information from vehicles and the smart city – are delivered in a well-specified time bound and with fault-proof reliability. Research topics here include, but are not limited to, integrating clock synchronization mechanisms into wireless vehicular networks and mechanisms towards dynamic automatic definition, and optimal configuration of flows of packets, their queue priorities and characteristics. The module within the architecture, the \textit{Deterministic Network Configurator}, is responsible for creating the necessary traffic flows, their parameters and configuring the necessary hardware, namely the \ac{nic} within each \ac{ecu} and the \ac{tsn} capable Ethernet switches interconnecting them. 
It is expected at least 3 traffic classes to be configured: a) system control messages with maximum criticality; b) high size (non) periodic streams of data and c) periodic or non periodic service messages.

The \textbf{Data Layer} vertical (\orcircle{c}) is responsible for ensuring the data is exchanged between the vehicular services in the most appropriated set of \ac{soa} protocols — such as \ac{someip}, \ac{mqtt} or \ac{dds}. Research topics here include, but are not limited to, the design of a abstraction layer. Such abstraction layer would increase interoperability between all these protocols while ensuring deterministic communication through integration with the underlying deterministic network mechanisms.

The \textbf{Orchestrator} vertical (\orcircle{d}) is responsible for ensuring the deployment of the services, either within a Kubernetes worker node, within a virtual machine managed for example by \ac{kvm} or through a \textit{containerd} or equivalent container runtime. Research topics here include ensuring deterministic communication through integration with the underlying deterministic network mechanisms (as previously mentioned in the Data Layer vertical). 

\begin{figure}[!ht]
	\centering
	\includegraphics[width=\columnwidth]{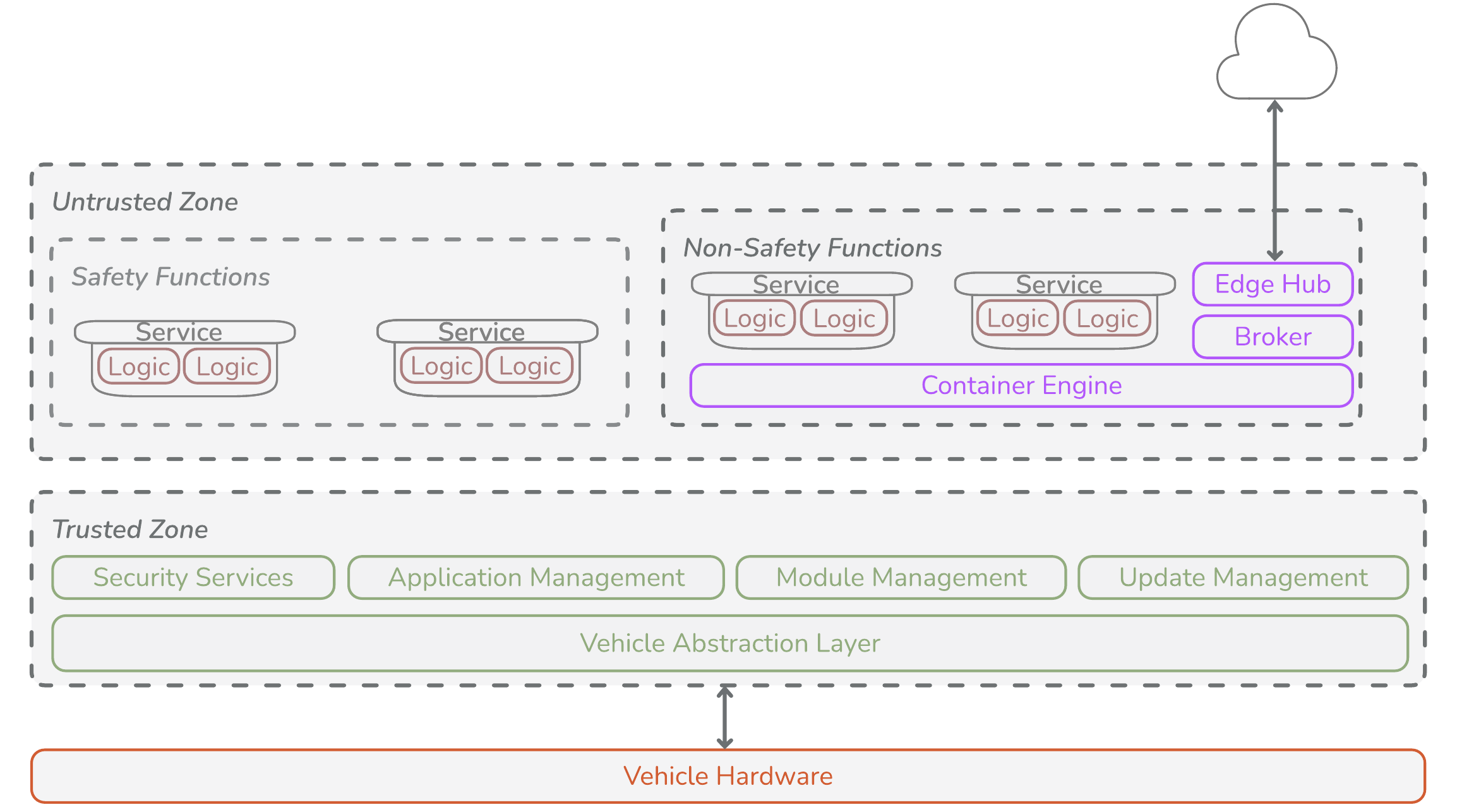}
	\caption[The \acs*{val} based architecture for vehicles, where the divisions between trusted and untrusted zones and between safety and non-safety services are evident]{The \acs*{val} based architecture for vehicles,as described by~\cite{2021:vdi_wissensforum_gmbh:eliv_2021}, where the divisions between trusted and untrusted zones and between safety and non-safety services are evident}
	\label{fig-arch-winnt-based-arch}
\end{figure}

These 3 verticals dispatch orders to the \textbf{\ac{val}} (\orcircle{e}) of each \ac{ecu}, which abstracts different types and models of vehicles' \acp{ecu} – both simulators and real vehicles – their actuators, sensors, and \acp{ecu} through the abstraction of existing frameworks, such as \ac{autosar}, Autoware or others. Conceptually, the \ac{val} is no different to the standardized abstractions operating systems created during the mid 1980s and 1990s to abstract the hardware from the remaining kernel, with~\cite{2021:vdi_wissensforum_gmbh:eliv_2021} proposing an extension of such a paradigm into the vehicular paradigm, as depicted in~\cref{fig-arch-winnt-based-arch}.

The research on the \ac{val} may also consider the creation of a digital twin of the vehicle and the research of the impact of sensors and services on the vehicle networks.

\subsection{Overall end-to-end interaction}
\begin{figure}[!ht]
	\centering
	\includegraphics[width=\columnwidth]{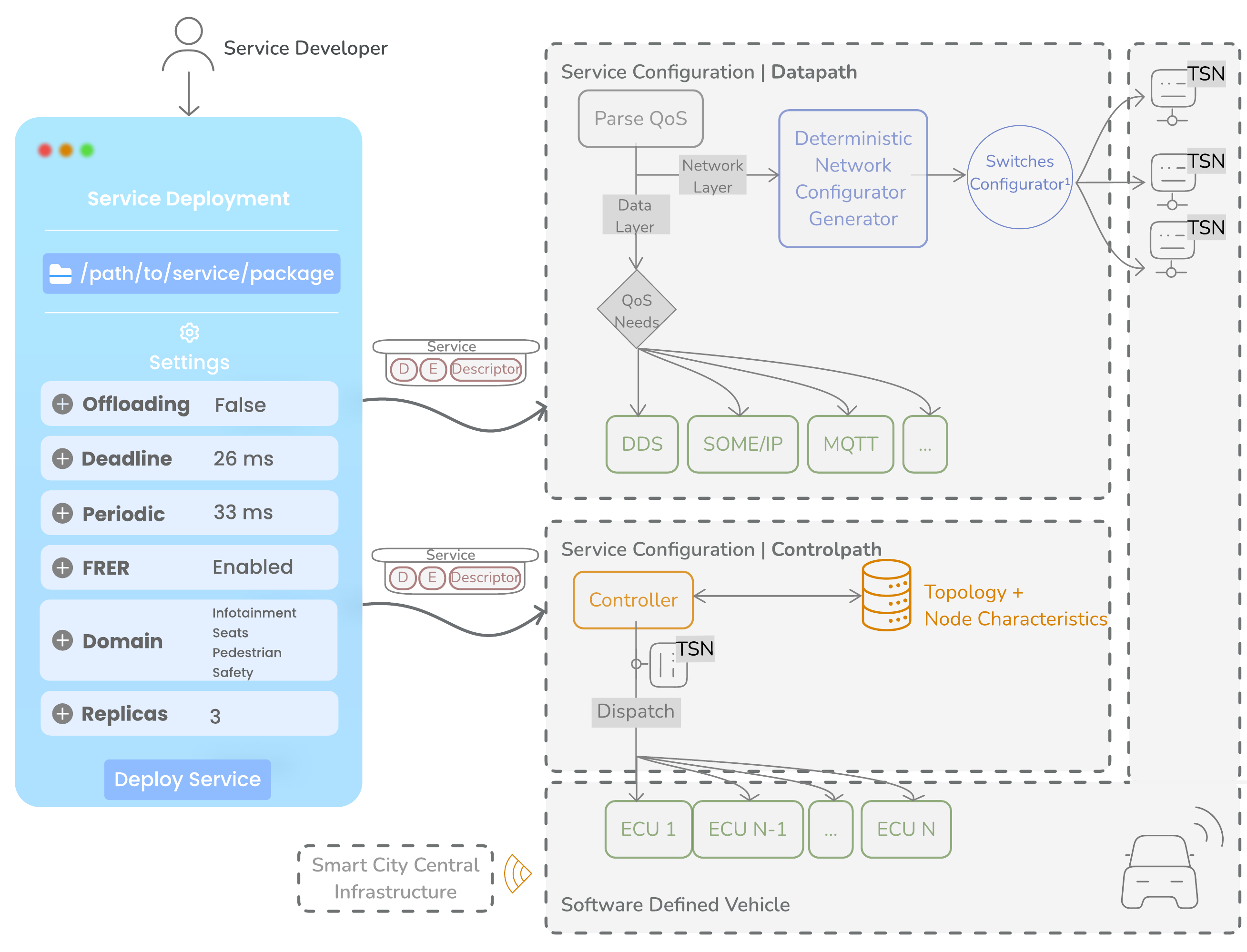}      
	\caption[Overall end-to-end interaction model of the proposed solution]{Overall end-to-end interaction model of the proposed solution.}
	\label{fig:arch-ee-interaction}
\end{figure}

\Cref{fig:arch-ee-interaction} provides an overview of the end-to-end interaction of the proposed system, covering everything from the features available to the end-user to the configuration of the required hardware needed to run the services and meet their requirements.

Our vision includes a system that manages service deployment, execution and evaluation of intra- and inter- vehicular services. Services can be associated with domains, that define not only requirements but the universe of information the service can subscribe. Service developers define a service descriptor (an example of the current vision for its structure is detailed in listing~\ref{listing:arch-use-case-service descriptor}) to describe the main service characteristics. This set of requirements is used to configure and dispatch the services to the appropriated orchestrator in the appropriated \acp{ecu} and configure the \textit{datapath}, \ie the network and data layer resources required for the service to communicate with others.


Services must be deployed by the controller which must decide the best node(s) for the service characteristics, either automatically, manually, or based on criteria (\eg time, energy, cost). The controller must further ensure latency bounds for the deployment of the service and for the exchange of messages between them if the service descriptor requires criticality and/or \ac{qos} levels on the data exchange and processing. The controller is the module responsible for a) evaluating the nodes on the match between the service characteristics and the node characteristics, and b) dispatching the configurations to the $N$ nodes best fulfilling the service characteristics. Receiving as input both the service descriptor and the node characteristics, the algorithm outputs a score to each node and if the node may be used or not (to support requirements that \textbf{must} be fulfilled).         


The vehicular service developer shall interact with the system through a declarative paradigm, \ie set the wanted state\footnote{Kubernetes is a widely known example of such declarative paradigm.}. A service descriptor file — or equivalent structure — describing the necessary characteristics of the vehicular service can support this paradigm. This information is defined by the service developer and can be used by the system to be developed to support the service deployment.

Associated with each field is not only the value, with the type prescribed, but also its criticality. Since not all conditions can be met by a node, it is essential for the controller module to know which metrics can be unmet and by what margin.

For example, when considering the use cases in~\cref{sec:arch-use-cases-health}, one can imagine the service developer creating a service descriptor such as the one in listing~\ref{listing:arch-use-case-service descriptor}. This service descriptor is then submitted to the controller within the target vehicle(s). Within each vehicle a controller is responsible fo evaluating the \acp{ecu} and configuring them and the necessary network and computing components to ensure that the service requirements are properly considered. In the case of the considered service, this entails at least: 
\begin{itemize}
    \item Deploying the necessary containers; 
    \item Allocating and configuring the necessary resources within an \ac{ecu} connected to the necessary actuators (in this case the motors for the fixing points and the ramp);  
    \item Configuring the switches and other network equipment connecting such actuators and the \acp{ecu} where the service connect with the necessary deterministic networking mechanisms (in this case \ac{tsn} \ac{tas} scheduling could be a possibility).
\end{itemize}
while keeping fulfilling the other already running services and their requirements.

\begin{code}
        \raggedleft
        \begin{lstlisting}[language=Toml, caption=Example of possible service descriptor., label=listing:arch-use-case-service descriptor]
        title = "WheelchairDriver"          

        [ServiceMetadata]
        Author = "TheWheelChairCompany"        
        Version = "1.0"                  
        Domain = "safety"                
        
        [Flows]                      
        [Flows.Flow1]            
        [Flows.Flow1.NodeSpecs]         
        [Flows.Flow1.NodeSpecs.NodeA]
        Image = "thewheelchairservice"          
        ImageType = "docker"            
        Replicas = 2                     
        CPU = 2
        Memory = 1024                 
        Storage = 51200000000             
        GPU = false                       
        Energy = 1
        Offloading = false     
        
        [Flows.Flow1.DataSpecs]        
        DataFormat = "json"            
        DataSize = 8096
        
        [Flows.Flow1.TrafficSpecs]       
        Guarantee = 4                    
        Reliability = true               
        Delivery = true
        Wired = true
        
        [Flows.Flow1.TrafficSpecs.TrafficTimeSpecs]               
        MaxLatency = 50                 
        Periodicity = 100
        TransmitOffset = 0               
        Jitter = 10                      
        \end{lstlisting}
\end{code}

\subsection{Proposal of an evaluation architecture and metrics}
\label{sec:arch-testbed}

\begin{figure}
	\centering
	\includegraphics[width=\columnwidth]{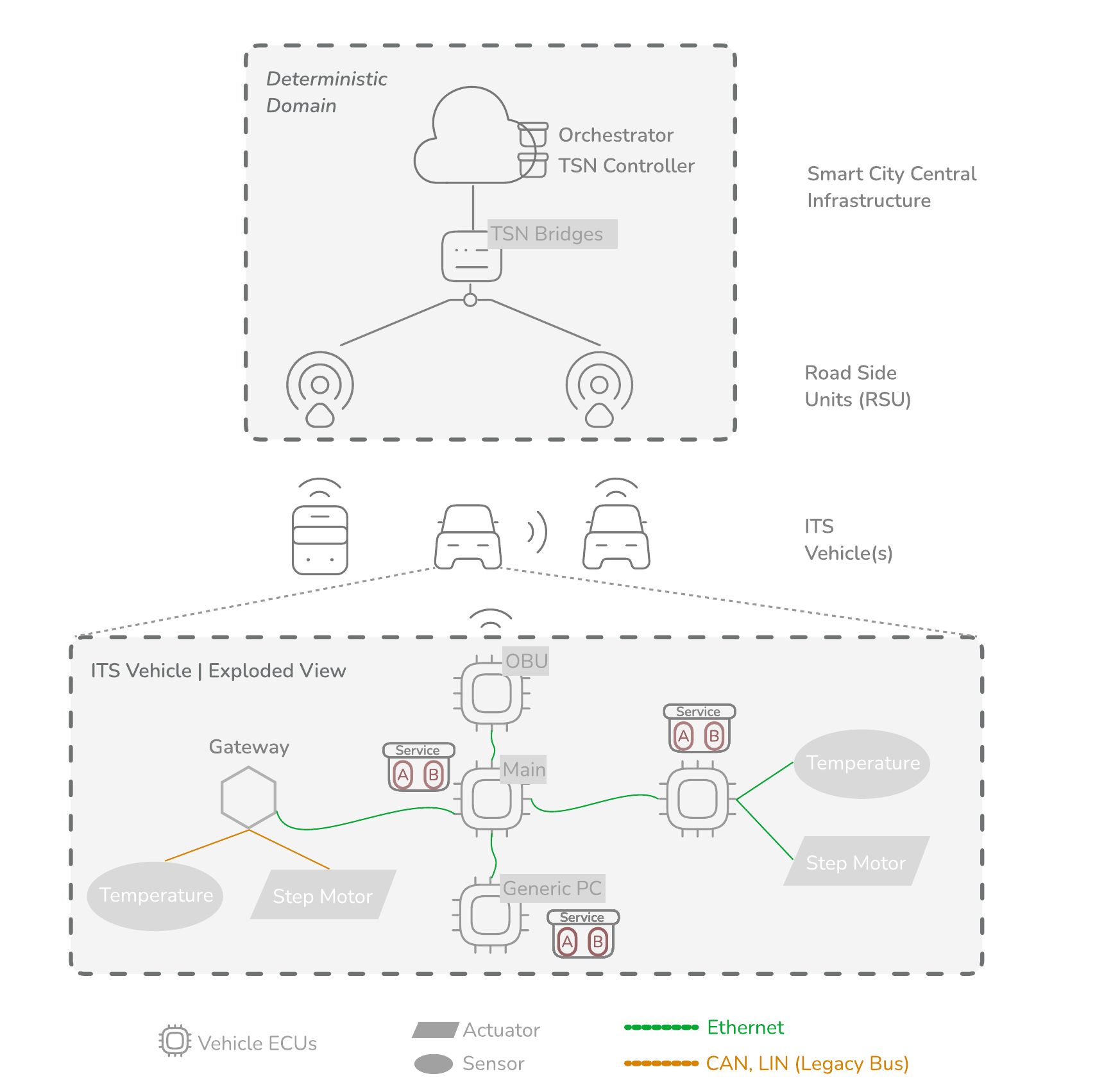} 
	\caption[Proposed testbed solution for evaluation of the system to be developed]{Proposed testbed solution for evaluation of the system to be developed.}
	\label{fig:arch-testbed}
\end{figure}

The evaluation of the proposed system requires at least a testbed with a topology equivalent to the one depicted in \cref{fig:arch-testbed}. In this topology, a vehicle form a deterministic network domain and can connect to other vehicles and smart city infrastructure. A smart city infrastructure, including its cloud and \acp{rsu} form another deterministic network domain~\cite{lopes2023time}. Each domain is responsible for ensuring a deterministic end-to-end latency for messages of vehicular services by configuring appropriately the necessary traffic shapers and synchronization between \ac{tsn} capable switches. These switches are connected through Ethernet and ensure connectivity between the \acp{ecu} (in the vehicle domain) or between \acp{rsu}, and between \acp{rsu} and the smart city cloud (in the smart city cloud). 

In terms of required equipment, such a testbed shall include at least one or more \acp{ecu}, a data center~\cite{Laaroussi2018}, a vehicular edge computing area~\cite{Laaroussi2018}, one vehicular sensor, one vehicular actuator, a Ethernet backbone with \ac{tsn} capable switch, and a \ac{can} legacy bus. The testbed may be deployed either by software or hardware. Software simulators can be used to mimic: a) the network, b) the vehicular topology, and c) the inter vehicular domain, while hardware can be used to function/emulate \acp{ecu}, Ethernet switches with \ac{tsn} support, \ac{can} based sensors and actuators.

Analysis of overall performance of the system should be at least in terms of throughput and latency for edge (intra-vehicular), cloud, and hybrid approaches but also consider additional metrics, such as, inter-frame/packet latency~\cite{TimeSens56:online}, data rate, transmission queue length~\cite{TimeAwar24:online}, percentage of missed deadline or offset of period and worst frame/packet latency. In terms of metrics it can include average node (\ac{ecu}) pressure, uptime and downtime of each service (and associated error) and time until recovery. 


\subsubsection{Hardware equipment}
The hardware equipment that may be considered within the testbed in \cref{fig:arch-testbed} shall consider hardware equipment in two different contexts: the network infrastructure, comprised by the switches, routers and access points that shall be compatible with deterministic communication standards, such as \ac{tsn} standards; and the \ac{its} intra vehicle hardware, including generic computers or specific computers (\ie \acp{ecu}).

\begin{figure}
	\centering
	\includegraphics[width=.8\columnwidth]{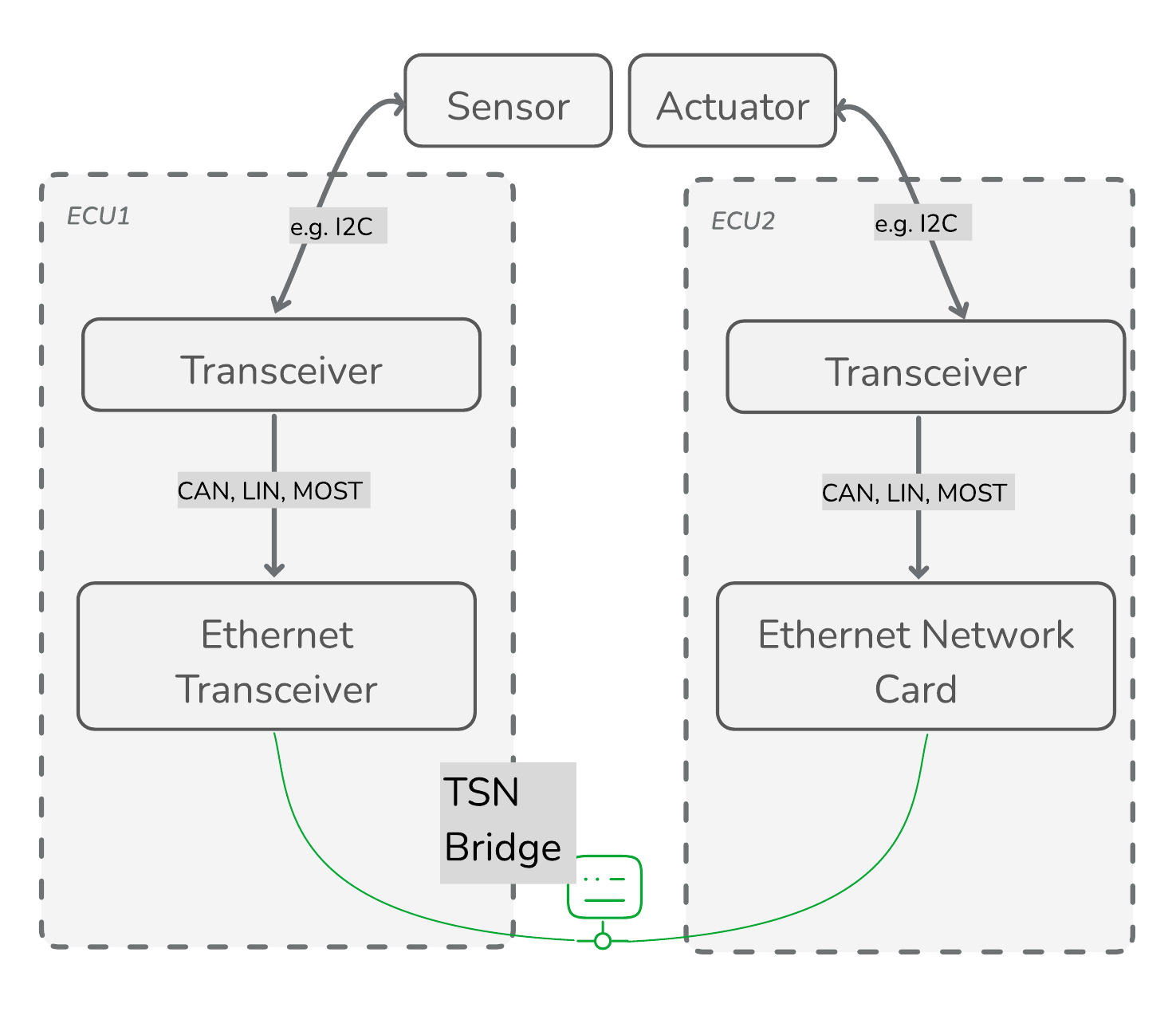}
	\caption[Basic topology to simulate the vehicle]{Basic topology to simulate the vehicle.}
	\label{fig:arch-hw}
\end{figure}

A basic topology to simulate an entire vehicle is depicted in \cref{fig:arch-hw} and requires at least two computational nodes, one per vehicle, connected through a network backbone (initially considered as a Ethernet Switch with support for deterministic communications), one sensor, one actuator, and necessary transceivers to bridge the sensors and the actuators to the Ethernet/Wireless domain.

Regarding the network infrastructure, solutions from closed source from typical network manufacturers to open source solutions are available. In principle all equipment supporting the basic architecture within~\cite{PatentKishore2020,PatentBush2021} are enough to support the basic testing of the proposed architecture.
Cisco possesses \ac{tsn} capable switches, such as the IE 4000 Series Industrial Switch. Within its operating system, the IOS, it is possible to install support to \ac{tsn} standards\footnote{Details at \myhref{https://www.cisco.com/c/en/us/td/docs/switches/lan/cisco_ie4000/tsn/b_tsn_ios_support/b_tsn_ios_support_chapter_010.html}{cisco.com}.}. 

Closed source alternatives include for example the NXP SJA1105TEL\footnote{Details at \myhref{https://www.nxp.com/products/interfaces/ethernet-/automotive-ethernet-switches/five-ports-avb-and-tsn-automotive-ethernet-switch:SJA1105TEL}{nxp.com}.} switch, specifically target at vehicular \acp{ecu}, or the Broadcom  BCM53570 \ac{soc}.

Relyum offers closed-sourced \ac{fpga} based alternative solutions\footnote{More details at \myhref{https://www.relyum.com/web/rely-tsn-lab/}{relyum.com}.} for \ac{tsn} switches, with theoretical support for configuration through \ac{netconf}, with alternatives from SoC-e\footnote{Such as the alternative at \myhref{https://soc-e.com/avb-automotive-ethernet-switch-ip-core/}{soc-e.com}.} also available. 

In addition, several \ac{soc} alternatives also support \ac{tsn}. Xilinx \ac{soc}, such as the KR260 series\footnote{Details at \myhref{https://www.xilinx.com/content/dam/xilinx/publications/product-briefs/kr260-product-brief.pdf}{xilinx.com}.} contain support for \ac{tsn}, integrate a \ac{fpga} that may support the hardware acceleration of parts of the proposed architecture in \cref{fig:arch} and support the interconnection with existing legacy vehicular systems through its support for \ac{can} bus, while also keeping an open architecture having in mind the reconfigurability that \acp{fpga} support.

Regarding vehicular hardware, two alternatives may be considered: an approach based on simulating parts of the vehicle; and an entire development vehicle.

PixKit\footnote{Details at \myhref{https://pixmoving-moveit.github.io/pixkit-documentation-en/pix-chassis-user-manual/}{pixmoving-moveit.github.io}.} is a possibility currently under study. Alternatives such as the The Car Lab's HYDRAONE\footnote{Available at \myhref{http://weisongshi.org/hydraone/}{weisongshi.org/hydraone}.} could also be considered, even if less advanced than the PixKit alternative, in particular without out-of-the-box semi autonomous driving\footnote{Semi autonomous since the PixKit solution is fully dependent on previously mapped environments.}.
Within the vehicle, computation systems such as NXP, through its BlueBox Development Kit\footnote{Details at \myhref{https://www.nxp.com/design/design-center/development-boards-and-designs/bluebox-3-0-automotive-high-performance-compute-ahpc-development-platform:BlueBox}{nxp.com}.}  or Renesas through its R-Car H3e-2G Starter Kit\footnote{Details at \myhref{https://www.renesas.com/in/en/products/automotive-products/automotive-system-chips-socs/rtp8j779m1askb0sk0sa003-r-car-h3e-2g-starter-kit}{renesas.com}.} can be considered as \ac{ecu} for testing purposes. Well known alternatives such as the Raspberry Pi and equivalents can also be considered.

It is important to point out the usage of existing infrastructure, such as \ac{atcll} smart city infrastructure — namely \acp{rsu} — to ensure connectivity of the testbed vehicular systems with the remaining smart city and other vehicles.

Finally, one should consider hardware equipment related to the use cases themselves, namely sensors and actuators. For example, the use case in~\cref{sec:arch-use-cases-health}) includes step motors — for the ramp or fixing the seat belts  — or sensors to detect unbuckled seat belts, weight sensors to detect used wheelchairs and screens for the direct interaction with the users.

\subsubsection{Software and simulators}
When hardware equipment is not available or if it does not correspond to the requirements of the testbed as depicted, software simulation alternatives may be considered.

While solutions such as \ac{autosar} simulate and model the automotive Ethernet, they lack \ac{tsn} support for such modelation~\cite{ASHJAEI2021102137}.
To simulate the network, both in its wired and wireless components, one can use use case agnostic solutions such as OMNET\footnote{At \myhref{http://omnetpp.org/}{omnetpp.org}.}, that with INET\footnote{Available at \myhref{https://inet.omnetpp.org/}{inet.omnetpp.org}.} and with CoRE4INET\footnote{Albeit with notable limitations, such as no simulation of frame preemption or clock synchronization protocols, with TSimNet model being an alternative for both frame preemption and replication and elimination~\cite{ASHJAEI2021102137}. } or NeSTiNg\footnote{Most logic related to NeSTiNg is now integrated within INET.} can be used to simulate networks with \ac{ttethernet} and \ac{tsn} protocols
\cite{2023-TSN-16}. 

OMNET can also be extended with SUMO\footnote{Available at \myhref{https://eclipse.dev/sumo/}{eclipse.dev/sumo}.} through frameworks such as Veins\footnote{At \myhref{https://veins.car2x.org/}{veins.car2x.org}.}, an open source framework for running vehicular network simulations~\cite{2023_Simulation_15}. It can also be extended for service-oriented communication, through the SOA4CoRE\footnote{Service-Oriented Architecture for Communication over Realtime Ethernet, at \myhref{https://github.com/CoRE-RG/SOA4CoRE}{github.com/CoRE-RG/SOA4CoRE}.} initiative. 
Mininet, a popular alternative to create virtual networks with complex typologies and greatly used in evaluation for \ac{sdn} contexts, can also be extended to consider wireless communications through the Mininet-WiFi\footnote{At \myhref{https://mininet-wifi.github.io/}{mininet-wifi.github.io}.} initiative and can therefore also be an alternative situation.

Matlab, within its Vehicle Network Toolbox, also allows the modeling of vehicle networks. Middleware layers, such as \ac{dds} can be integrated within MATLAB and Simulink tools as a block-set~\cite{Takrouni2017}

RealTime-At-Work Pegase\footnote{At \myhref{https://www.realtimeatwork.com/rtaw-pegase/}{realtimeatwork.com/rtaw-pegase}.} provides a closed proprietary tool for the equivalent purpose of testing \acp{ecu} communications. Renault, for example, considers within their \ac{tsn} tests, an \ac{ecu} based on \ac{autosar} classic software and interconnected with a physical and \ac{mac} Ethernet layer with \ac{tsn} capable hardware~\cite{2023-TSN-19}. The RTaW tool provides the \ac{yang} files for the hardware, with Bosch creating plugins to emulate detailed modeling of the \acp{ecu} integrate into the RTaW Pegase simulation~\cite{2023-TSN-19}.

Tools like Autoware\footnote{At \myhref{https://autowarefoundation.github.io/autoware-documentation/main/}{autowarefoundation.github.io}.} and Carla\footnote{At \myhref{https://carla.org}{carla.org}.} can then be used to simulate vehicular use cases, particularly in autonomous driving scenarios. Autoware in particular is considered a good solution both for being based on the widely used \ac{ros} and for offering hardware\footnote{Available at \myhref{https://www.autoware.org/autoware-open-ad-kit}{autoware.org/autoware-open-ad-kit}.} kits that can be integrated within their software. However, while it has monitoring tools for \ac{cpu} and \ac{gpu} usage and some redundancy in sensor fault management, it lacks any in-vehicle networking, both wired and wireless\footnote{Current focus of Autoware V2X module, details at \myhref{https://tlab-wide.github.io/AutowareV2X/main/}{tlab-wide.github.io/AutowareV2X}.}.


CSS Electronics CANedge2\footnote{Details at \myhref{https://www.csselectronics.com/products/can-bus-data-logger-wifi-canedge2}{csselectronics.com}.} and equivalent equipment can be used to obtain real frames from \ac{can} bus, either simulated or real, and send it to another computer. This can be used to obtain data to support the simulations with the aforementioned tools.
If not possible, buses like \ac{can} can also be simulated for example using Linux support for \ac{vxcan}.\footnote{Source code available at \myhref{https://github.com/torvalds/linux/blob/master/drivers/net/can/vxcan.c}{github.com/torvalds/linux/blob/master/ drivers/net/can/vxcan.c}.}.



\section{Conclusions}
\label{sec:conclusions}
In this article, the concept of Software Defined Vehicles was presented and framed as essential to support the development of the next generation of vehicular services. Considering the criticality of the future \ac{sdv} services, as well as the increased complexity of development and deployment of services, it is more and more critical to consider a dynamic management of services.

The main building blocks of \ac{sdv} to be considered in such a dynamic management were listed and detailed, as well as the main associated challenges. Vehicles will tend to a physical zonal architecture, supporting a set of services running in a set of higher performance \acp{ecu} (in a \ac{soa} paradigm). Existing operating systems will have to evolve towards interoperability and supporting virtualization and orchestration techniques, as well as integrating inter- and intra-vehicular communications with determinism techniques to support critical services that can have direct impact of the occupants and surroundings of the vehicles. 

While the industry, including automakers and their suppliers, are working towards \acp{sdv}, their initiatives are still tending to be proprietary or closed but the trend is towards greater interoperability with holistic solutions such as "the collaboration of collaborations" between AUTOSAR, \ac{covesa}, Eclipse \ac{sdv} and \ac{soafee} announced in CES 2024.

A vision towards the dynamic management of services within \ac{sdv}. Based on a deterministic network configurator, a data layer configurator, a orchestrator and a vehicle abstraction layer, all coordinated by a controller, it supports an easier dynamic management of vehicular services, easing the development burden of such artifacts.

Future work towards such vision includes the further specification, service orchestration, implementation and the proper evaluation of its scheduling algorithms, as well as an end-to-end acceptance tests.


\bibliographystyle{elsarticle-num} 
\bibliography{main}

\begin{thebibliography}{100}
\expandafter\ifx\csname url\endcsname\relax
  \def\url#1{\texttt{#1}}\fi
\expandafter\ifx\csname urlprefix\endcsname\relax\def\urlprefix{URL }\fi
\expandafter\ifx\csname href\endcsname\relax
  \def\href#1#2{#2} \def\path#1{#1}\fi

\bibitem{H1stvisi11:online}
{Renault Group}, \href{{https://media.renaultgroup.com/h1st-vision-the-concept-car-designed-by-software-republique-a-human-centred-vision-of-mobility-for-tomorrow/}}{H1stvision, the concept car designed by software r{\'{e}}publique: a human-centred vision of mobility for tomorrow - renault group global media website}, (Accessed on 07/01/2023) (Jun. 2023).
\newline\urlprefix\url{{https://media.renaultgroup.com/h1st-vision-the-concept-car-designed-by-software-republique-a-human-centred-vision-of-mobility-for-tomorrow/}}

\bibitem{ASHJAEI2021102137}
M.~Ashjaei, L.~L. Bello, M.~Daneshtalab, G.~Patti, S.~Saponara, S.~Mubeen, \href{https://www.sciencedirect.com/science/article/pii/S1383762121001028}{Time-sensitive networking in automotive embedded systems: State of the art and research opportunities}, Journal of Systems Architecture 117 (2021) 102137.
\newblock \href {https://doi.org/10.1016/j.sysarc.2021.102137} {\path{doi:10.1016/j.sysarc.2021.102137}}.
\newline\urlprefix\url{https://www.sciencedirect.com/science/article/pii/S1383762121001028}

\bibitem{2022-Problem-8}
B.~B. Gordan~Markus, \href{https://pages.ubuntu.com/rs/066-EOV-335/images/SDV_WP_Final_27062022.pdf}{{A CTO's Guide to software-defined vehicles}}, Canonical (June 2022).
\newline\urlprefix\url{https://pages.ubuntu.com/rs/066-EOV-335/images/SDV_WP_Final_27062022.pdf}

\bibitem{6183198}
C.~Buckl, A.~Camek, G.~Kainz, C.~Simon, L.~Mercep, H.~St{\"{a}}hle, A.~Knoll, The software car: Building {ICT} architectures for future electric vehicles, in: 2012 {IEEE} International Electric Vehicle Conference, 2012, pp. 1--8.
\newblock \href {https://doi.org/10.1109/IEVC.2012.6183198} {\path{doi:10.1109/IEVC.2012.6183198}}.

\bibitem{2021-Problem-5}
R.~N. Charette, \href{{https://spectrum.ieee.org/software-eating-car}}{{How Software Is Eating the Car - IEEE Spectrum}}, (Accessed on 04/28/2023) (Jun. 2021).
\newline\urlprefix\url{{https://spectrum.ieee.org/software-eating-car}}

\bibitem{AUTOSARE36:online}
{Vector}, \href{{https://www.youtube.com/watch?v=W5F8nQuwuWY}}{{AUTOSAR} essentials | \#engineeringthejigsaw | episode f6 - youtube}, (Accessed on 07/01/2023) (Aug. 2021).
\newline\urlprefix\url{{https://www.youtube.com/watch?v=W5F8nQuwuWY}}

\bibitem{Thecasef27:online}
McKinsey, \href{{https://www.mckinsey.com/industries/automotive-and-assembly/our-insights/the-case-for-an-end-to-end-automotive-software-platform}}{The case for an automotive software platform}, (Accessed on 05/05/2023) (Jan. 2020).
\newline\urlprefix\url{{https://www.mckinsey.com/industries/automotive-and-assembly/our-insights/the-case-for-an-end-to-end-automotive-software-platform}}

\bibitem{moritz}
M.~Neukirchner, \href{{https://www.elektrobit.com/tech-corner/demystifying-the-software-defined-vehicle/}}{Demystifying the software-defined vehicle}, (Accessed on 11/05/2024) (11 2023).
\newline\urlprefix\url{{https://www.elektrobit.com/tech-corner/demystifying-the-software-defined-vehicle/}}

\bibitem{historyOfCars}
E.~Halshaw, \href{{https://www.evanshalshaw.com/blog/the-history-of-cars/}}{The history of car technology}, (Accessed on 11/05/2024) (05 2022).
\newline\urlprefix\url{{https://www.evanshalshaw.com/blog/the-history-of-cars/}}

\bibitem{Zhang2018}
S.~Zhang, O.~Makke, O.~Gusikhin, A.~Shah, A.~Vasilakos, A security model for dependable vehicle middleware and mobile applications connection, in: Proceedings of the 4\textsuperscript{th} International Conference on Vehicle Technology and Intelligent Transport Systems, {SCITEPRESS} - Science and Technology Publications, 2018.
\newblock \href {https://doi.org/10.5220/0006704903790386} {\path{doi:10.5220/0006704903790386}}.

\bibitem{1998-Vehicle-Abstraction-Layer-40}
M.~Kais, N.~Hafez, M.~Parent, An intelligent vehicle architecture for automated transportation in cities, in: 2001 European Control Conference (ECC), 2001, pp. 277--282.
\newblock \href {https://doi.org/10.23919/ECC.2001.7075919} {\path{doi:10.23919/ECC.2001.7075919}}.

\bibitem{1998-Vehicle-Abstraction-Layer-41}
P.~Fromm, P.~Drews, Modular, service-oriented design and architecture of smart vehicles for short distance person and freight transport, in: IECON '98. Proceedings of the 24\textsuperscript{th} Annual Conference of the {IEEE} Industrial Electronics Society (Cat. No.98CH36200), Vol.~4, 1998, pp. 2192--2197.
\newblock \href {https://doi.org/10.1109/IECON.1998.724061} {\path{doi:10.1109/IECON.1998.724061}}.

\bibitem{1992-Vehicle-Abstraction-Layer-39}
J.~Albus, M.~Juberts, S.~Szabo, A reference model architecture for intelligent vehicle and highway systems, in: Proceedings of the Intelligent Vehicles `92 Symposium, 1992, pp. 378--384.
\newblock \href {https://doi.org/10.1109/IVS.1992.252289} {\path{doi:10.1109/IVS.1992.252289}}.

\bibitem{2023-TSN-19}
J.~M. Josetxo~Villanueva, Bouchra~Achemlal, D.~Martini, {Strategies for End to End Timing Guarantees in a Centralized Software Defined Vehicle Architecture Combining {CAN} With {TSN} Backbone}, in: Automotive Ethernet Congress, 2023, (Accessed on 07/04/2023).

\bibitem{mckinsey}
M.~B. e.~a. Alexander~Baule, Florian~Garms, \href{{https://www.mckinsey.com/industries/automotive-and-assembly/our-insights/car-connectivity-what-consumers-want-and-are-willing-to-pay#/}}{Car connectivity: What consumers want and are willing to pay}, (Accessed on 11/05/2024) (01 2024).
\newline\urlprefix\url{{https://www.mckinsey.com/industries/automotive-and-assembly/our-insights/car-connectivity-what-consumers-want-and-are-willing-to-pay#/}}

\bibitem{2020-Problem-11}
S.~Corwin, N.~Jameson, P.~Willigman, \href{https://www2.deloitte.com/content/dam/insights/us/articles/3367_Future-of-mobility-whats-next/DUP_Future-of-mobility-whats-next.pdf}{{The future of mobility: What's next?}}, Deloitte (2016) 28.
\newline\urlprefix\url{https://www2.deloitte.com/content/dam/insights/us/articles/3367_Future-of-mobility-whats-next/DUP_Future-of-mobility-whats-next.pdf}

\bibitem{2020-Problem-6}
J.~W. Edward~Taylor, Norihiko~Shirouzu, \href{{https://www.reuters.com/article/us-autos-tesla-newera-insight-idUSKCN24N0GB}}{How {Tesla} defined a new era for the global auto industry}, (Accessed on 04/28/2023) (Jul. 2020).
\newline\urlprefix\url{{https://www.reuters.com/article/us-autos-tesla-newera-insight-idUSKCN24N0GB}}

\bibitem{2010-Vehicle-Abstraction-Layer-30}
S.~Voget, {AUTOSAR} and the automotive tool chain, in: 2010 Design, Automation \& Test in Europe Conference \& Exhibition (DATE 2010), 2010, pp. 259--262.
\newblock \href {https://doi.org/10.1109/DATE.2010.5457202} {\path{doi:10.1109/DATE.2010.5457202}}.

\bibitem{Bandur2021}
V.~Bandur, G.~Selim, V.~Pantelic, M.~Lawford, Making the case for centralized automotive {E/E} architectures, {IEEE} Transactions on Vehicular Technology 70~(2) (2021) 1230--1245.
\newblock \href {https://doi.org/10.1109/tvt.2021.3054934} {\path{doi:10.1109/tvt.2021.3054934}}.

\bibitem{2016-Vehicle-Abstraction-Layer-26}
S.~F{\"{u}}rst, M.~Bechter, {AUTOSAR} for connected and autonomous vehicles: The {AUTOSAR} adaptive platform, in: 2016 46\textsuperscript{th} Annual IEEE/IFIP International Conference on Dependable Systems and Networks Workshop (DSN-W), 2016, pp. 215--217.
\newblock \href {https://doi.org/10.1109/DSN-W.2016.24} {\path{doi:10.1109/DSN-W.2016.24}}.

\bibitem{2021-TSN-17}
J.~M. Josetxo~Villanueva, N.~Navet, {QoS-Predictable SOA on TSN: Insights from a Case-Study}, in: Automotive Ethernet Congress, 2021, (Accessed on 07/04/2023).

\bibitem{2023-TSN-20}
N.~N. Hoai Hoang~Bengtsson, Time-predictable communication in service-oriented architecture - what are the challenges?, in: Automotive Ethernet Congress, 2023, (Accessed on 07/04/2023).

\bibitem{2020-TSN-22}
P.~K. Oliver~Creighton, Nicolas~Navet, J.~Migge, {Towards Computer-Aided, Iterative TSN-and Ethernet-based {E/E} Architecture Design}, in: 2020 {IEEE} Standards Association (IEEE-SA) Ethernet \& IP @ Automotive Technology Day, 2020, (Accessed on 07/04/2023).

\bibitem{Vector2020}
Vector, {Middleware Protocols in the Automobile: Service-Oriented, Data-Centric or RESTful? These}, Elektronik automotive (3 2020).

\bibitem{2023-Vehicle-Abstraction-Layer-23}
CARIAD, \href{{https://cariad.technology/de/en/news/stories/launch-application-store-for-volkswagen-group.html}}{Cariad launches application store for the volkswagen group}, (Accessed on 07/04/2023) (Mar. 2023).
\newline\urlprefix\url{{https://cariad.technology/de/en/news/stories/launch-application-store-for-volkswagen-group.html}}

\bibitem{2019-TSN-23}
J.~M. Nicolas~Navet, Josetxo~Villanueva, {Early-stage topological and technological choices for TSN-based communication}, in: 2019 {IEEE} Standards Association (IEEE-SA) Ethernet \& IP @ Automotive Technology Day, 2019, (Accessed on 07/04/2023).

\bibitem{2023-Vehicle-Abstraction-Layer-43}
V.~C.~D. Portal, \href{{https://developer.volvocars.com/apis/}}{{Overview APIs}}, (Accessed on 06/28/2023) (2023).
\newline\urlprefix\url{{https://developer.volvocars.com/apis/}}

\bibitem{MercedesBenz}
P.~Hanse, \href{{https://www.vector.com/int/en/news/news/mercedes-benz-mbos-base-layer/}}{{The Hansen Report | Software Development – MB.OS Base Layer}}, (Accessed on 10/07/2024) (2023).
\newline\urlprefix\url{{https://www.vector.com/int/en/news/news/mercedes-benz-mbos-base-layer/}}

\bibitem{Unveiled86:online}
{Junko Yoshida}, \href{{https://www.eetimes.com/unveiled-bmws-scalable-av-architecture/}}{Unveiled: {BMW}{\textquoteright}s scalable av architecture}, (Accessed on 05/04/2023) (Apr. 2020).
\newline\urlprefix\url{{https://www.eetimes.com/unveiled-bmws-scalable-av-architecture/}}

\bibitem{IEEE_SA_d1_02}
P.~Laclau, \href{https://standards.ieee.org/events/automotive/presentations-2022/}{{Ethernet-as-a-Service for Software Defined Vehicles: Design objectives and orientations for an Ethernet-based network stack}}, {IEEE} SA Ethernet \& IP @ Automotive Technology Day (2022) 1--22.
\newline\urlprefix\url{https://standards.ieee.org/events/automotive/presentations-2022/}

\bibitem{GMnowhas36:online}
A.~J. Hawkins, \href{{https://www.theverge.com/2023/10/12/23914060/gm-uservices-api-software-apps-car-sdv}}{{GM now has its own API for software developers to make cool apps for its cars}}, (Accessed on 04/16/2024) (Oct. 2023).
\newline\urlprefix\url{{https://www.theverge.com/2023/10/12/23914060/gm-uservices-api-software-apps-car-sdv}}

\bibitem{2023-Vehicle-Abstraction-Layer-33}
{SOME/IP}, \href{{https://some-ip.com/}}{{Scalable service-Oriented MiddlewarE over IP}}, (Accessed on 04/28/2023) (2023).
\newline\urlprefix\url{{https://some-ip.com/}}

\bibitem{202401SD57:online}
{Eclipse Software Defined Vehicle Working Group}, \href{{https://docs.google.com/presentation/d/1uU4_BevDGuxFmEVvxt5-u99UZELF7BA2bEp2TneFZGw}}{Software defined vehicle working group member’s update}, (Accessed on 04/16/2024) (Jan. 2023).
\newline\urlprefix\url{{https://docs.google.com/presentation/d/1uU4_BevDGuxFmEVvxt5-u99UZELF7BA2bEp2TneFZGw}}

\bibitem{Liu2022}
Z.~Liu, W.~Zhang, F.~Zhao, Impact, challenges and prospect of software-defined vehicles, Automotive Innovation 5~(2) (2022) 180--194.
\newblock \href {https://doi.org/10.1007/s42154-022-00179-z} {\path{doi:10.1007/s42154-022-00179-z}}.

\bibitem{2020-Problem-7}
H.~Proff, T.~Pottebaum, P.~Wolf, \href{https://www2.deloitte.com/content/dam/insights/us/articles/22951\_software-is-transforming-the-automotive-world/DI\_Software-is-transforming-the-automotive-world.pdf}{{Software is transforming the automotive world}}, Deloitte Insights (2020) 20.
\newline\urlprefix\url{https://www2.deloitte.com/content/dam/insights/us/articles/22951\_software-is-transforming-the-automotive-world/DI\_Software-is-transforming-the-automotive-world.pdf}

\bibitem{HowHaveV59:online}
{Vector}, \href{https://www.youtube.com/watch?v=OjnQ52f1SEE}{How have vehicle {E/E} systems evolved? | \#engineeringthejigsaw | episode f10 - youtube}, (Accessed on 07/01/2023) (Nov. 2021).
\newline\urlprefix\url{https://www.youtube.com/watch?v=OjnQ52f1SEE}

\bibitem{HowAreVe60:online}
{Vector}, \href{https://www.youtube.com/watch?v=XpvTOuOgGWY}{How are vehicle {E/E} systems evolving? | \#engineeringthejigsaw | f11 - youtube}, (Accessed on 07/01/2023) (Dec. 2021).
\newline\urlprefix\url{https://www.youtube.com/watch?v=XpvTOuOgGWY}

\bibitem{TheEEarc76:online}
{Bosch}, \href{{https://www.bosch-mobility.com/en/mobility-topics/ee-architecture/}}{The {E/E} architecture of the future}, (Accessed on 05/05/2023) (2023).
\newline\urlprefix\url{{https://www.bosch-mobility.com/en/mobility-topics/ee-architecture/}}

\bibitem{Kugele2018}
S.~Kugele, D.~Hettler, J.~Peter, Data-centric communication and containerization for future automotive software architectures, in: 2018 {IEEE} International Conference on Software Architecture ({ICSA}), IEEE, 2018.
\newblock \href {https://doi.org/10.1109/icsa.2018.00016} {\path{doi:10.1109/icsa.2018.00016}}.

\bibitem{2021-Vehicle-Abstraction-Layer-38}
B.~Murphy, \href{https://www.aptiv.com/docs/default-source/white-papers/2021_aptiv_whitepaper_cicd.pdf?sfvrsn=5fbbe33b_24}{{What the Next Phase of Automotive Software Development Looks Like}}, (Accessed on 04/28/2023) (2021).
\newline\urlprefix\url{https://www.aptiv.com/docs/default-source/white-papers/2021_aptiv_whitepaper_cicd.pdf?sfvrsn=5fbbe33b_24}

\bibitem{OASISSOA53:online}
{OASIS}, \href{{https://www.oasis-open.org/committees/soa-rm/faq.php}}{{OASIS SOA Reference Model TC}}, (Accessed on 20/12/2024) (2023).
\newline\urlprefix\url{{https://www.oasis-open.org/committees/soa-rm/faq.php}}

\bibitem{richards2015microservices}
M.~Richards, Microservices vs. service-oriented architecture, O'Reilly Media Sebastopol, 2015.

\bibitem{NXPVehicleA52:online}
{NXP Semiconductors}, \href{{https://www.nxp.com/design/training/vehicle-architectures-and-networking-training-academy:TS-VEHICLE-ARCHITECTURES-AND-NETWORKING-ACADEMY}}{Vehicle architectures and networking training academy}, (Accessed on 07/01/2023) (2023).
\newline\urlprefix\url{{https://www.nxp.com/design/training/vehicle-architectures-and-networking-training-academy:TS-VEHICLE-ARCHITECTURES-AND-NETWORKING-ACADEMY}}

\bibitem{2022-TSN-21}
H.~H. Bengtsson, M.~Hiller, J.~Migge, N.~NAVET, Signal-oriented {ECU}s in a centralized service-oriented architecture: Scalability of the layered software architecture (01 June 2022).

\bibitem{liang2023performance}
W.-Y. Liang, Y.~Yuan, H.-J. Lin, \href{https://arxiv.org/abs/2303.09419}{{A Performance Study on the Throughput and Latency of Zenoh, MQTT, Kafka, and {DDS}}} (2023).
\newblock \href {http://arxiv.org/abs/2303.09419} {\path{arXiv:2303.09419}}.
\newline\urlprefix\url{https://arxiv.org/abs/2303.09419}

\bibitem{HowDoesD67:online}
{{DDS} Foundation}, \href{{https://www.dds-foundation.org/features-benefits/}}{How does {DDS} compare to other {IoT} technologies?}, (Accessed on 07/12/2023) (2023).
\newline\urlprefix\url{{https://www.dds-foundation.org/features-benefits/}}

\bibitem{SASG72Ro66:online}
J.~S. Andrei~Terechko, Yuting~Fu, \href{https://www.sasg.nl/SASG_72_Roly-poly_on_middleware.pdf}{Middleware for safe software-defined cars}, (Accessed on 04/18/2024) (Oct. 2021).
\newline\urlprefix\url{https://www.sasg.nl/SASG_72_Roly-poly_on_middleware.pdf}

\bibitem{Deployme63:online}
Zenoh, \href{{https://zenoh.io/docs/overview/what-is-zenoh/}}{Developer documentation}, (Accessed on 04/20/2024) (Jan. 2023).
\newline\urlprefix\url{{https://zenoh.io/docs/overview/what-is-zenoh/}}

\bibitem{gutiérrez2018realtime}
C.~S.~V. Guti{\'{e}}rrez, L.~U.~S. Juan, I.~Z. Ugarte, V.~M. Vilches, Real-time linux communications: an evaluation of the linux communication stack for real-time robotic applications (2018).
\newblock \href {http://arxiv.org/abs/1808.10821} {\path{arXiv:1808.10821}}, \href {https://doi.org/10.48550/arxiv.1808.10821} {\path{doi:10.48550/arxiv.1808.10821}}.

\bibitem{Nxpcorpo93:online}
C.~U. Marius~Rotaru, \href{{https://www.nxp.com.cn/docs/en/training-reference-material/DESIGN-AND-IMPLEMENTATION-USING-AUTOMOTIVE-LINUX-BSP.pdf}}{Service oriented architecture: Design and implementation using automotive linux bsp}, (Accessed on 05/24/2023) (Jun. 2019).
\newline\urlprefix\url{{https://www.nxp.com.cn/docs/en/training-reference-material/DESIGN-AND-IMPLEMENTATION-USING-AUTOMOTIVE-LINUX-BSP.pdf}}

\bibitem{2022-Virtualization-44}
P.~Kleiner, {AUTOSAR} adaptive deployment with {OCI} containers for embedded using {VxWorks}, (Accessed on 07/04/2023) (Sep. 2019).

\bibitem{sym12040592}
I.~Ungurean, \href{https://www.mdpi.com/2073-8994/12/4/592}{Timing comparison of the real-time operating systems for small microcontrollers}, Symmetry 12~(4) (2020).
\newblock \href {https://doi.org/10.3390/sym12040592} {\path{doi:10.3390/sym12040592}}.
\newline\urlprefix\url{https://www.mdpi.com/2073-8994/12/4/592}

\bibitem{Ren2022}
J.~Ren, M.~Zhao, H.~Wang, T.~Gao, P.~Yao, Z.~Shao, A time sensitive network-based data distribution service implementation method, in: 2022 International Symposium on Advances in Informatics, Electronics and Education ({ISAIEE}), IEEE, 2022.
\newblock \href {https://doi.org/10.1109/isaiee57420.2022.00040} {\path{doi:10.1109/isaiee57420.2022.00040}}.

\bibitem{Leonardi2020}
L.~Leonardi, L.~L. Bello, G.~Patti, Towards time-sensitive networking in heterogeneous platforms with virtualization, in: 2020 25\textsuperscript{th} {IEEE} International Conference on Emerging Technologies and Factory Automation ({ETFA}), IEEE, 2020.
\newblock \href {https://doi.org/10.1109/etfa46521.2020.9212116} {\path{doi:10.1109/etfa46521.2020.9212116}}.

\bibitem{Caruso2021}
B.~Caruso, L.~Leonardi, L.~L. Bello, G.~Patti, Design of a framework for enabling {TSN} support in heterogeneous platforms with virtualization and preliminary experimental results, in: 2021 26\textsuperscript{th} {IEEE} International Conference on Emerging Technologies and Factory Automation ({ETFA} ), IEEE, 2021.
\newblock \href {https://doi.org/10.1109/etfa45728.2021.9613442} {\path{doi:10.1109/etfa45728.2021.9613442}}.

\bibitem{2020-Virtualization-46}
H.~Sami, A.~Mourad, W.~El-Hajj, {Vehicular-OBUs-As-On-Demand-Fogs: Resource and Context Aware Deployment of Containerized Micro-Services}, IEEE/ACM Transactions on Networking 28~(2) (2020) 778--790.
\newblock \href {https://doi.org/10.1109/TNET.2020.2973800} {\path{doi:10.1109/TNET.2020.2973800}}.

\bibitem{Garbugli_2023}
A.~Garbugli, L.~Rosa, A.~Bujari, L.~Foschini, {KuberneTSN: a Deterministic Overlay Network for Time-Sensitive Containerized Environments}, in: ICC 2023 - {IEEE} International Conference on Communications, IEEE, 2023.
\newblock \href {https://doi.org/10.1109/icc45041.2023.10279214} {\path{doi:10.1109/icc45041.2023.10279214}}.

\bibitem{Datanetw31:online}
D.~Kuhlgatz, \href{{https://www.bosch.com/stories/the-controller-area-network/}}{{Data network for the car: The Controller Area Network {CAN}}}, (Accessed on 05/08/2023) (Nov. 2016).
\newline\urlprefix\url{{https://www.bosch.com/stories/the-controller-area-network/}}

\bibitem{Creating15:online}
{NI}, \href{{https://www.ni.com/pt-pt/innovations/case-studies/19/creating-a-can-bus-communication-platform-based-on-the-sae-j1939-protocol.html}}{Creating a {CAN} bus communication platform based on the {SAE} {J1939} protocol and {NI PXI}}, (Accessed on 07/01/2023) (2019).
\newline\urlprefix\url{{https://www.ni.com/pt-pt/innovations/case-studies/19/creating-a-can-bus-communication-platform-based-on-the-sae-j1939-protocol.html}}

\bibitem{2016_Bus_2}
S.~Corrigan, \href{http://masters.donntu.ru/2005/fvti/trofunenko/library/sloa101.pdf}{{Introduction to the controller area network ({CAN})}}, Application Report, Texas Instruments (2002) 1--15.
\newline\urlprefix\url{http://masters.donntu.ru/2005/fvti/trofunenko/library/sloa101.pdf}

\bibitem{IEEE_SA_d1_03}
P.~Wilms, F.~Hurley, \href{https://standards.ieee.org/events/automotive/presentations-2022/}{{Centralised software and zonal architectures for future innovative ambient lighting enabled by E2B 10Base-T1S}}, {IEEE} SA Ethernet \& IP @ Automotive Technology Day (2022).
\newline\urlprefix\url{https://standards.ieee.org/events/automotive/presentations-2022/}

\bibitem{Amel2014}
B.~N. Amel, B.~Rim, J.~Houda, H.~Salem, J.~Khaled, {FlexRay versus Ethernet for vehicular networks}, in: 2014 {IEEE} International Electric Vehicle Conference ({IEVC}), IEEE, 2014.
\newblock \href {https://doi.org/10.1109/ievc.2014.7056123} {\path{doi:10.1109/ievc.2014.7056123}}.

\bibitem{Nichitelea2019}
T.-C. Nichitelea, M.-G. Unguritu, Automotive ethernet applications using scalable service-oriented middleware over {IP}: Service discovery, in: 2019 24\textsuperscript{th} International Conference on Methods and Models in Automation and Robotics ({MMAR}), IEEE, 2019.
\newblock \href {https://doi.org/10.1109/mmar.2019.8864701} {\path{doi:10.1109/mmar.2019.8864701}}.

\bibitem{IEEE_SA_d2_02}
B.~Petersen, \href{https://standards.ieee.org/events/automotive/presentations-2022/}{{Signal to Service : Remote Control of Legacy Networks}}, {IEEE} SA Ethernet \& IP @ Automotive Technology Day (November 2022).
\newline\urlprefix\url{https://standards.ieee.org/events/automotive/presentations-2022/}

\bibitem{Takrouni2020}
M.~Takrouni, A.~Hasnaoui, I.~Mejri, S.~Hasnaoui, A new methodology for implementing the data distribution service on top of gigabit ethernet for automotive applications, Journal of Circuits, Systems and Computers 29~(13) (2020) 2050210.
\newblock \href {https://doi.org/10.1142/s0218126620502102} {\path{doi:10.1142/s0218126620502102}}.

\bibitem{CANCANFD55:online}
{Vector}, \href{{https://www.vector.com/int/en/know-how/can/}}{{CAN}/{CAN} fd/{CAN} xl}, (Accessed on 05/09/2023) (2023).
\newline\urlprefix\url{{https://www.vector.com/int/en/know-how/can/}}

\bibitem{Zuo2021}
Z.~Zuo, S.~Yang, B.~Ma, B.~Zou, Y.~Cao, Q.~Li, S.~Zhou, J.~Li, Design of a {{CAN}FD} to {SOME}/{IP} gateway considering security for in-vehicle networks, Sensors 21~(23) (2021) 7917.
\newblock \href {https://doi.org/10.3390/s21237917} {\path{doi:10.3390/s21237917}}.

\bibitem{FlexRay}
{NI}, \href{https://www.ni.com/pt-pt/shop/seamlessly-connect-to-third-party-devices-and-supervisory-system/flexray-automotive-communication-bus-overview.html}{{FlexRay Automotive Communication Bus Overview}} (2023).
\newline\urlprefix\url{https://www.ni.com/pt-pt/shop/seamlessly-connect-to-third-party-devices-and-supervisory-system/flexray-automotive-communication-bus-overview.html}

\bibitem{MOSTTec59:online}
Microchip, \href{{https://www.microchip.com/en-us/solutions/automotive-and-transportation/automotive-products/connectivity/most-technology}}{Most technology}, (Accessed on 07/03/2023) (2023).
\newline\urlprefix\url{{https://www.microchip.com/en-us/solutions/automotive-and-transportation/automotive-products/connectivity/most-technology}}

\bibitem{dasanayaka2020enhancing}
N.~Dasanayaka, K.~F. Hasan, C.~Wang, Y.~Feng, {Enhancing Vulnerable Road User Safety: A Survey of Existing Practices and Consideration for Using Mobile Devices for {V2X} Connections}, arXiv preprint arXiv:2010.15502 (2020).

\bibitem{surveyltevs5g}
R.~Mishra, \href{https://doi.org/10.1007/s41324-022-00474-1}{An overview of backbone technology behind the latest advanced gadgets in use: {4G} \& 5g}, Spatial Information Research 31~(1) (2023) 15--26.
\newblock \href {https://doi.org/10.1007/s41324-022-00474-1} {\path{doi:10.1007/s41324-022-00474-1}}.
\newline\urlprefix\url{https://doi.org/10.1007/s41324-022-00474-1}

\bibitem{survey5g}
H.~Fourati, R.~Maaloul, L.~Chaari, \href{https://doi.org/10.1007/s13042-020-01178-4}{A survey of {5G} network systems: challenges and machine learning approaches}, International Journal of Machine Learning and Cybernetics 12~(2) (2021) 385--431.
\newblock \href {https://doi.org/10.1007/s13042-020-01178-4} {\path{doi:10.1007/s13042-020-01178-4}}.
\newline\urlprefix\url{https://doi.org/10.1007/s13042-020-01178-4}

\bibitem{9500256}
G.~Naik, D.~Ogbe, J.-M.~J. Park, Can {Wi-Fi} 7 support real-time applications? on the impact of multi link aggregation on latency, in: ICC 2021 - {IEEE} International Conference on Communications, 2021, pp. 1--6.
\newblock \href {https://doi.org/10.1109/ICC42927.2021.9500256} {\path{doi:10.1109/ICC42927.2021.9500256}}.

\bibitem{10315104}
IEEE, {IEEE} standard for information technology--telecommunications and information exchange between systems local and metropolitan area networks--specific requirements part 11: Wireless {LAN} medium access control ({MAC} ) and physical layer ({PHY} ) specifications amendment 6: Light communications, {IEEE} Std 802.11bb-2023 (Amendment to {IEEE} Std 802.11-2020 as amended by {IEEE} Std 802.11ax-2021, {IEEE} Std 802.11ay-2021, {IEEE} Std 802.11ba-2021, {IEEE} Std 802.11az-2022, {IEEE} Std 802.11-2020/Cor 1-2022, and {IEEE} Std 802.11bd-2023) (2023) 1--37\href {https://doi.org/10.1109/IEEESTD.2024.10315104} {\path{doi:10.1109/IEEESTD.2024.10315104}}.

\bibitem{10013676}
P.~Teixeira, S.~Sargento, P.~Rito, M.~Lu{\'{i}}s, F.~Castro, A sensing, communication and computing approach for vulnerable road users safety, {IEEE} Access 11 (2023) 4914--4930.
\newblock \href {https://doi.org/10.1109/ACCESS.2023.3235863} {\path{doi:10.1109/ACCESS.2023.3235863}}.

\bibitem{10056390}
E.~Moradi-Pari, D.~Tian, M.~Bahramgiri, S.~Rajab, S.~Bai, {{DSRC} Versus LTE-V2X}: Empirical performance analysis of direct vehicular communication technologies, {IEEE} Transactions on Intelligent Transportation Systems 24~(5) (2023) 4889--4903.
\newblock \href {https://doi.org/10.1109/TITS.2023.3247339} {\path{doi:10.1109/TITS.2023.3247339}}.

\bibitem{ieee80211bd}
B.~Y. Yacheur, T.~Ahmed, M.~Mosbah, Analysis and comparison of {IEEE} 802.11p and {IEEE} 802.11bd, in: F.~Krief, H.~Aniss, L.~Mendiboure, S.~Chaumette, M.~Berbineau (Eds.), Communication Technologies for Vehicles, Springer International Publishing, Cham, 2020, pp. 55--65.

\bibitem{9855461}
E.~Khorov, I.~Levitsky, Current status and challenges of {Li-Fi}: {IEEE} 802.11bb, {IEEE} Communications Standards Magazine 6~(2) (2022) 35--41.
\newblock \href {https://doi.org/10.1109/MCOMSTD.0001.2100104} {\path{doi:10.1109/MCOMSTD.0001.2100104}}.

\bibitem{Alparslan2021}
O.~Alparslan, S.~Arakawa, M.~Murata, \href{https://ieeexplore.ieee.org/document/9481803/}{{Next Generation Intra-Vehicle Backbone Network Architectures}}, in: 2021 {IEEE} 22\textsuperscript{nd} International Conference on High Performance Switching and Routing (HPSR), IEEE, 2021, pp. 1--7.
\newblock \href {https://doi.org/10.1109/HPSR52026.2021.9481803} {\path{doi:10.1109/HPSR52026.2021.9481803}}.
\newline\urlprefix\url{https://ieeexplore.ieee.org/document/9481803/}

\bibitem{8723326_Naik2019}
G.~Naik, B.~Choudhury, J.-M. Park, \href{https://ieeexplore.ieee.org/document/8723326/}{{IEEE} 802.11bd \& {5G} {NR V2X}: Evolution of radio access technologies for {V2X} communications}, {IEEE} Access 7 (2019) 70169--70184.
\newblock \href {https://doi.org/10.1109/ACCESS.2019.2919489} {\path{doi:10.1109/ACCESS.2019.2919489}}.
\newline\urlprefix\url{https://ieeexplore.ieee.org/document/8723326/}

\bibitem{Farzaneh2017}
M.~H. Farzaneh, A.~Knoll, \href{http://ieeexplore.ieee.org/document/8275648/}{{Time-sensitive networking ({TSN}): An experimental setup}}, in: 2017 {IEEE} Vehicular Networking Conference (VNC), IEEE, 2017, pp. 23--26.
\newblock \href {https://doi.org/10.1109/VNC.2017.8275648} {\path{doi:10.1109/VNC.2017.8275648}}.
\newline\urlprefix\url{http://ieeexplore.ieee.org/document/8275648/}

\bibitem{bailleul20towards}
Q.~Bailleul, K.~Jaffres-Runser, J.-L. Scharbarg, P.~Cuenot, Towards robust network synchronization with {IEEE} 802.1 as, ETR 20 (2020) 25.

\bibitem{Malinverno2020}
M.~Malinverno, G.~Avino, C.~Casetti, C.~F. Chiasserini, F.~Malandrino, S.~Scarpina, {Edge-Based Collision Avoidance for Vehicles and Vulnerable Users: An Architecture Based on MEC}, {IEEE} Vehicular Technology Magazine 15~(1) (2020) 27--35.
\newblock \href {https://doi.org/10.1109/MVT.2019.2953770} {\path{doi:10.1109/MVT.2019.2953770}}.

\bibitem{Khakimov_2020}
A.~Khakimov, A.~Loborchuk, I.~Ibodullokhodzha, D.~Poluektov, I.~A. Elgendy, A.~Muthanna, Edge computing resource allocation orchestration system for autonomous vehicles, in: The 4\textsuperscript{th} International Conference on Future Networks and Distributed Systems ({ICFNDS}), ACM, 2020.
\newblock \href {https://doi.org/10.1145/3440749.3442594} {\path{doi:10.1145/3440749.3442594}}.

\bibitem{lopes2023time}
R.~Lopes, D.~Raposo, S.~Sargento, \href{https://arxiv.org/abs/2312.03635}{Towards time sensitive networking on smart cities: Techniques, challenges, and solutions} (2023).
\newblock \href {http://arxiv.org/abs/2312.03635} {\path{arXiv:2312.03635}}.
\newline\urlprefix\url{https://arxiv.org/abs/2312.03635}

\bibitem{Seol2021}
Y.~Seol, D.~Hyeon, J.~Min, M.~Kim, J.~Paek, \href{https://ieeexplore.ieee.org/document/9576720/}{{Timely Survey of Time-Sensitive Networking: Past and Future Directions}}, {IEEE} Access 9 (2021) 142506--142527.
\newblock \href {https://doi.org/10.1109/ACCESS.2021.3120769} {\path{doi:10.1109/ACCESS.2021.3120769}}.
\newline\urlprefix\url{https://ieeexplore.ieee.org/document/9576720/}

\bibitem{tutorial40:online}
J.~F. Carlos J.~Bernardos, \href{{https://www.ieee802.org/1/files/public/docs2023/tutorial-bernardos-farkas-RAW-0723-v01.pdf}}{{IETF Reliable Available Wireless (RAW)}}, (Accessed on 12/21/2023) (Jul. 2023).
\newline\urlprefix\url{{https://www.ieee802.org/1/files/public/docs2023/tutorial-bernardos-farkas-RAW-0723-v01.pdf}}

\bibitem{rfc8655}
N.~Finn, P.~Thubert, B.~Varga, J.~Farkas, \href{https://www.rfc-editor.org/info/rfc8655}{{Deterministic Networking Architecture}} (Oct. 2019).
\newblock \href {https://doi.org/10.17487/RFC8655} {\path{doi:10.17487/RFC8655}}.
\newline\urlprefix\url{https://www.rfc-editor.org/info/rfc8655}

\bibitem{ietf-detnet-yang-18}
X.~Geng, Y.~Ryoo, D.~Fedyk, R.~Rahman, Z.~Li, \href{https://datatracker.ietf.org/doc/draft-ietf-detnet-yang/18/}{{Deterministic Networking (DetNet) YANG Model}}, Internet-Draft draft-ietf-detnet-yang-18, Internet Engineering Task Force, work in Progress (Jul. 2023).
\newline\urlprefix\url{https://datatracker.ietf.org/doc/draft-ietf-detnet-yang/18/}

\bibitem{8091139}
{IEEE}, {IEEE} standard for local and metropolitan area networks--frame replication and elimination for reliability, {IEEE} Std 802.1CB-2017 (2017) 1--102\href {https://doi.org/10.1109/IEEESTD.2017.8091139} {\path{doi:10.1109/IEEESTD.2017.8091139}}.

\bibitem{8684664}
{IEEE}, {IEEE} standard for local and metropolitan area networks-- virtual bridged local area networks amendment 12: Forwarding and queuing enhancements for time-sensitive streams, {IEEE} Std 802.1Qav-2009 (Amendment to {IEEE} Std 802.1Q-2005) (2010) 1--72\href {https://doi.org/10.1109/IEEESTD.2010.8684664} {\path{doi:10.1109/IEEESTD.2010.8684664}}.

\bibitem{8613095}
{IEEE}, {IEEE} standard for local and metropolitan area networks -- bridges and bridged networks - amendment 25: Enhancements for scheduled traffic, {IEEE} Std 802.1Qbv-2015 (2016) 1--57\href {https://doi.org/10.1109/IEEESTD.2016.8613095} {\path{doi:10.1109/IEEESTD.2016.8613095}}.

\bibitem{9738961}
H.~Wang, Z.~Zhao, J.~Wei, Research on the application of time sensitive networking in unmanned in-vehicle network, in: ISCTT 2021; 6\textsuperscript{th} International Conference on Information Science, Computer Technology and Transportation, 2021, pp. 1--8.

\bibitem{Ozawa2022}
Y.~Ozawa, Y.~Ito, \href{https://ieeexplore.ieee.org/document/10014097/}{{Evaluation of TCP Performance Under In-Vehicle Networks With {IEEE} 802.1 TSN}}, in: 2022 {IEEE} 11\textsuperscript{th} Global Conference on Consumer Electronics (GCCE), IEEE, 2022, pp. 668--669.
\newblock \href {https://doi.org/10.1109/GCCE56475.2022.10014097} {\path{doi:10.1109/GCCE56475.2022.10014097}}.
\newline\urlprefix\url{https://ieeexplore.ieee.org/document/10014097/}

\bibitem{Ulbricht2022}
M.~Ulbricht, S.~Senk, H.~K. Nazari, H.-H. Liu, M.~Reisslein, G.~T. Nguyen, F.~H.~P. Fitzek, {TSN-FlexTest: Flexible {TSN} Measurement Testbed (Extended Version)} (Nov. 2022).
\newblock \href {http://arxiv.org/abs/2211.10413} {\path{arXiv:2211.10413}}, \href {https://doi.org/10.48550/arxiv.2211.10413} {\path{doi:10.48550/arxiv.2211.10413}}.

\bibitem{10.1145/3487330}
L.~Deng, G.~Xie, H.~Liu, Y.~Han, R.~Li, K.~Li, {A Survey of Real-Time Ethernet Modeling and Design Methodologies: From {AVB} to {TSN}}, ACM Comput. Surv. 55~(2) (Jan. 2022).
\newblock \href {https://doi.org/10.1145/3487330} {\path{doi:10.1145/3487330}}.

\bibitem{8412465}
S.~Samii, H.~Zinner, Level 5 by layer 2: Time-sensitive networking for autonomous vehicles, {IEEE} Communications Standards Magazine 2~(2) (2018) 62--68.
\newblock \href {https://doi.org/10.1109/MCOMSTD.2018.1700079} {\path{doi:10.1109/MCOMSTD.2018.1700079}}.

\bibitem{8672474}
D.~Cavalcanti, J.~Perez-Ramirez, M.~M. Rashid, J.~Fang, M.~Galeev, K.~B. Stanton, Extending accurate time distribution and timeliness capabilities over the air to enable future wireless industrial automation systems, Proceedings of the IEEE 107~(6) (2019) 1132--1152.
\newblock \href {https://doi.org/10.1109/JPROC.2019.2903414} {\path{doi:10.1109/JPROC.2019.2903414}}.

\bibitem{IEEE_SA_d1_08}
A.~Regev, M.~Gubow, \href{https://standards.ieee.org/events/automotive/presentations-2022/}{{{PHY} latency and its effects on {TSN} performance}}, {IEEE} SA Ethernet \& IP @ Automotive Technology Day (2022) 1--18.
\newline\urlprefix\url{https://standards.ieee.org/events/automotive/presentations-2022/}

\bibitem{PatentChristianMardmoeller2023}
T.~H. Christian~Mardmoeller, Time-sensitive networking (Nov. 2023).

\bibitem{SATKA2023102852}
Z.~Satka, M.~Ashjaei, H.~Fotouhi, M.~Daneshtalab, M.~Sj{\"{o}}din, S.~Mubeen, \href{https://www.sciencedirect.com/science/article/pii/S1383762123000310}{A comprehensive systematic review of integration of time sensitive networking and {5G} communication}, Journal of Systems Architecture 138 (2023) 102852.
\newblock \href {https://doi.org/10.1016/j.sysarc.2023.102852} {\path{doi:10.1016/j.sysarc.2023.102852}}.
\newline\urlprefix\url{https://www.sciencedirect.com/science/article/pii/S1383762123000310}

\bibitem{CiscoWhitePaper}
Cisco, \href{https://www.cisco.com/c/dam/en/us/solutions/collateral/industry-solutions/white-paper-c11-738950.pdf}{Time-sensitive networking: A technical introduction} (2017).
\newline\urlprefix\url{https://www.cisco.com/c/dam/en/us/solutions/collateral/industry-solutions/white-paper-c11-738950.pdf}

\bibitem{8569454}
L.~Zhao, F.~He, E.~Li, J.~Lu, Comparison of time sensitive networking {(TSN)} and ttethernet, in: 2018 IEEE/AIAA 37\textsuperscript{th} Digital Avionics Systems Conference (DASC), 2018, pp. 1--7.
\newblock \href {https://doi.org/10.1109/DASC.2018.8569454} {\path{doi:10.1109/DASC.2018.8569454}}.

\bibitem{craciunas2017overview}
S.~S. Craciunas, R.~S. Oliver, T.~Ag, An overview of scheduling mechanisms for time-sensitive networks, Proceedings of the Real-time summer school L'École d'Été Temps R{\'{e}}el (ETR) (2017) 1551--3203.

\bibitem{goldsmith_2005}
A.~Goldsmith, Path Loss and Shadowing, Cambridge University Press, 2005, Ch. 2 - Path Loss and Shadowing, pp. 27--63.
\newblock \href {https://doi.org/10.1017/CBO9780511841224.003} {\path{doi:10.1017/CBO9780511841224.003}}.

\bibitem{9512046}
A.~F. M.~S. Shah, A.~N. Qasim, M.~A. Karabulut, H.~Ilhan, M.~B. Islam, Survey and performance evaluation of multiple access schemes for next-generation wireless communication systems, {IEEE} Access 9 (2021) 113428--113442.
\newblock \href {https://doi.org/10.1109/ACCESS.2021.3104509} {\path{doi:10.1109/ACCESS.2021.3104509}}.

\bibitem{predict6g}
J.~V. e.~a. Luis M.~Contreras, Antonio de la~Oliva, \href{https://predict-6g.eu/deliverables/}{{Analysis of use cases and system requirements}}, Report D11, PREDICT6G, published (Jun. 2023).
\newline\urlprefix\url{https://predict-6g.eu/deliverables/}

\bibitem{deterministic6g}
H.~N. N. e.~a. Dhruvin~Patel, Edgardo Montes de~Oca, \href{https://deterministic6g.eu/images/deliverables/DETERMINISTIC6G-D1.1-v1.0.pdf}{{Use Cases and Architecture Principles}}, Report D11, DETERMINISTIC6G, published (Jun. 2023).
\newline\urlprefix\url{https://deterministic6g.eu/images/deliverables/DETERMINISTIC6G-D1.1-v1.0.pdf}

\bibitem{10034532}
D.~Cavalcanti, C.~Cordeiro, M.~Smith, A.~Regev, {WiFi TSN: Enabling Deterministic Wireless Connectivity over 802.11}, {IEEE} Communications Standards Magazine 6~(4) (2022) 22--29.
\newblock \href {https://doi.org/10.1109/MCOMSTD.0002.2200039} {\path{doi:10.1109/MCOMSTD.0002.2200039}}.

\bibitem{rfc7426}
E.~Haleplidis, K.~Pentikousis, S.~Denazis, J.~H. Salim, D.~Meyer, O.~Koufopavlou, \href{https://www.rfc-editor.org/info/rfc7426}{{Software-Defined Networking (SDN): Layers and Architecture Terminology}} (Jan. 2015).
\newblock \href {https://doi.org/10.17487/RFC7426} {\path{doi:10.17487/RFC7426}}.
\newline\urlprefix\url{https://www.rfc-editor.org/info/rfc7426}

\bibitem{PatentPatel2022}
D.~P. Hubertus~Munz, Torsten~Dudda, Technique for time-sensitive networking over a radio access network (Mar. 2022).

\bibitem{9134382}
O.~Seijo, J.~A. L{\'{o}}pez-Fern{\'{a}}ndez, I.~Val, {w-SHARP: Implementation of a High-Performance Wireless Time-Sensitive Network for Low Latency and Ultra-low Cycle Time Industrial Applications}, {IEEE} Transactions on Industrial Informatics 17~(5) (2021) 3651--3662.
\newblock \href {https://doi.org/10.1109/TII.2020.3007323} {\path{doi:10.1109/TII.2020.3007323}}.

\bibitem{ETSI_TS_123_501}
{ETSI (European Telecommunications Standards Institute)}, {ETSI TS 123 501 V16.5.1 / 3GPP TS 23.501 version 16.5.1 Release 16}, Tech. rep., {ETSI (European Telecommunications Standards Institute)} (2020).

\bibitem{ACIA2021}
{{5G} Alliance for Connected Industries and Automation}, \href{https://5g-acia.org/whitepapers/integration-of-5g-with-time-sensitive-networking-for-industrial-communications/}{Integration of {5G} with time-sensitive networking for industrial communications} (Feb. 2021).
\newline\urlprefix\url{https://5g-acia.org/whitepapers/integration-of-5g-with-time-sensitive-networking-for-industrial-communications/}

\bibitem{PatentJosephVinay2021}
R.~P. Joseph~Vinay, Quality of service mapping for time-sensitive network traffic in a wireless communication system (Dec. 2021).

\bibitem{9904785}
Z.~Satka, M.~Ashjaei, H.~Fotouhi, M.~Daneshtalab, M.~Sj{\"{o}}din, S.~Mubeen, {QoS-MAN: A Novel QoS Mapping Algorithm for TSN-{5G} Flows}, in: 2022 {IEEE} 28\textsuperscript{th} International Conference on Embedded and Real-Time Computing Systems and Applications (RTCSA), 2022, pp. 220--227.
\newblock \href {https://doi.org/10.1109/RTCSA55878.2022.00030} {\path{doi:10.1109/RTCSA55878.2022.00030}}.

\bibitem{9779191}
Z.~Satka, D.~Pantzar, A.~Magnusson, M.~Ashjaei, H.~Fotouhi, M.~Sj{\"{o}}din, M.~Daneshtalab, S.~Mubeen, Developing a translation technique for converged {TSN}-{5G} communication, in: 2022 {IEEE} 18\textsuperscript{th} International Conference on Factory Communication Systems (WFCS), 2022, pp. 1--8.
\newblock \href {https://doi.org/10.1109/WFCS53837.2022.9779191} {\path{doi:10.1109/WFCS53837.2022.9779191}}.

\bibitem{9212141}
A.~Larra{\~{n}}aga, M.~C. Lucas-Esta{\~{n}}, I.~Martinez, I.~Val, J.~Gozalvez, Analysis of {5G}-{TSN} integration to support industry 4.0, in: 2020 25\textsuperscript{th} {IEEE} International Conference on Emerging Technologies and Factory Automation (ETFA), Vol.~1, 2020, pp. 1111--1114.
\newblock \href {https://doi.org/10.1109/ETFA46521.2020.9212141} {\path{doi:10.1109/ETFA46521.2020.9212141}}.

\bibitem{electronics11111666}
P.~Kehl, J.~Ansari, M.~H. Jafari, P.~Becker, J.~Sachs, N.~K{\"{o}}nig, A.~G{\"{o}}ppert, R.~H. Schmitt, \href{https://www.mdpi.com/2079-9292/11/11/1666}{Prototype of {5G} integrated with {TSN} for edge-controlled mobile robotics}, Electronics 11~(11) (2022).
\newblock \href {https://doi.org/10.3390/electronics11111666} {\path{doi:10.3390/electronics11111666}}.
\newline\urlprefix\url{https://www.mdpi.com/2079-9292/11/11/1666}

\bibitem{7746299}
M.~H. Farzaneh, A.~Knoll, An ontology-based plug-and-play approach for in-vehicle time-sensitive networking ({TSN}), in: 2016 {IEEE} 7\textsuperscript{th} Annual Information Technology, Electronics and Mobile Communication Conference (IEMCON), 2016, pp. 1--8.
\newblock \href {https://doi.org/10.1109/IEMCON.2016.7746299} {\path{doi:10.1109/IEMCON.2016.7746299}}.

\bibitem{Syed2022}
A.~A. Syed, S.~Ayaz, T.~Leinmuller, M.~Chandra, \href{https://ieeexplore.ieee.org/document/9998340/}{{Fault-Tolerant Static Scheduling and Routing for In-vehicle Networks}}, in: 2022 32\textsuperscript{nd} International Telecommunication Networks and Applications Conference (ITNAC), IEEE, 2022, pp. 273--279.
\newblock \href {https://doi.org/10.1109/ITNAC55475.2022.9998340} {\path{doi:10.1109/ITNAC55475.2022.9998340}}.
\newline\urlprefix\url{https://ieeexplore.ieee.org/document/9998340/}

\bibitem{Zou2023}
J.~Zou, X.~Dai, J.~A. McDermid, \href{https://ieeexplore.ieee.org/document/9826899/}{{reTSN: Resilient and Efficient Time-Sensitive Network for Automotive In-Vehicle Communication}}, {IEEE} Transactions on Computer-Aided Design of Integrated Circuits and Systems 42~(3) (2023) 754--767.
\newblock \href {https://doi.org/10.1109/TCAD.2022.3190235} {\path{doi:10.1109/TCAD.2022.3190235}}.
\newline\urlprefix\url{https://ieeexplore.ieee.org/document/9826899/}

\bibitem{Wang2020}
D.~Wang, T.~Sun, \href{https://ieeexplore.ieee.org/document/9284334/}{{Leveraging {5G} {TSN} in {V2X} Communication for Cloud Vehicle}}, in: 2020 {IEEE} International Conference on Edge Computing (EDGE), IEEE, 2020, pp. 106--110.
\newblock \href {https://doi.org/10.1109/EDGE50951.2020.00023} {\path{doi:10.1109/EDGE50951.2020.00023}}.
\newline\urlprefix\url{https://ieeexplore.ieee.org/document/9284334/}

\bibitem{Ding2022}
P.~Ding, D.~Liu, Y.~Shen, H.~Duan, Q.~Zheng, \href{https://ieeexplore.ieee.org/document/9828687/}{{Edge-to-Cloud Intelligent Vehicle-Infrastructure Based on {5G} Time-Sensitive Network Integration}}, in: 2022 {IEEE} International Symposium on Broadband Multimedia Systems and Broadcasting (BMSB), IEEE, 2022, pp. 1--5.
\newblock \href {https://doi.org/10.1109/BMSB55706.2022.9828687} {\path{doi:10.1109/BMSB55706.2022.9828687}}.
\newline\urlprefix\url{https://ieeexplore.ieee.org/document/9828687/}

\bibitem{2021-TSN-18}
C.~Mauclair, M.~Guti{\'{e}}rrez, J.~Migge, N.~Navet, Do we really need {TSN} in next-generation helicopters? insights from a case-study, in: 2021 IEEE/AIAA 40\textsuperscript{th} Digital Avionics Systems Conference (DASC), 2021, pp. 1--7.
\newblock \href {https://doi.org/10.1109/DASC52595.2021.9594349} {\path{doi:10.1109/DASC52595.2021.9594349}}.

\bibitem{10002776}
N.~Ambrosy, T.~Kampa, U.~Jumar, D.~Gro{\ss}mann, {5G} and detnet: Towards holistic determinism in industrial networks, in: 2022 {IEEE} International Conference on Industrial Technology (ICIT), 2022, pp. 1--6.
\newblock \href {https://doi.org/10.1109/ICIT48603.2022.10002776} {\path{doi:10.1109/ICIT48603.2022.10002776}}.

\bibitem{2022-Virtualization-47}
V.~Meyer, M.~L. {da Silva}, D.~F. Kirchoff, C.~A. {De Rose}, \href{https://www.sciencedirect.com/science/article/pii/S0164121222001698}{Iada: A dynamic interference-aware cloud scheduling architecture for latency-sensitive workloads}, Journal of Systems and Software 194 (2022) 111491.
\newblock \href {https://doi.org/10.1016/j.jss.2022.111491} {\path{doi:10.1016/j.jss.2022.111491}}.
\newline\urlprefix\url{https://www.sciencedirect.com/science/article/pii/S0164121222001698}

\bibitem{Mobileye67:online}
{Intel Corporation}, \href{{https://www.intel.com/content/www/us/en/newsroom/news/mobileye-ces-2022-self-driving-secret-data.html}}{Mobileye{\textquoteright}s self-driving secret? {200 PB} of data}, (Accessed on 07/20/2023) (Jan. 2022).
\newline\urlprefix\url{{https://www.intel.com/content/www/us/en/newsroom/news/mobileye-ces-2022-self-driving-secret-data.html}}

\bibitem{NextGene78:online}
{Mathworks}, \href{{https://www.mathworks.com/content/dam/mathworks/mathworks-dot-com/images/events/matlabexpo/us/2019/next-generation-wi-fi-networks-for-time-critical-applications.pdf}}{{Next Generation Wi Fi Networks for Time Critical Applications}}, (Accessed on 05/08/2023) (Nov. 2019).
\newline\urlprefix\url{{https://www.mathworks.com/content/dam/mathworks/mathworks-dot-com/images/events/matlabexpo/us/2019/next-generation-wi-fi-networks-for-time-critical-applications.pdf}}

\bibitem{5GAA2019}
5GAA, \href{https://5gaa.org/content/uploads/2019/07/5GAA_191906_WP_CV2X_UCs_v1-3-1.pdf}{{C-{V2X} Use Cases : Methodology, Examples and Service Level Requirements}}, Tech. rep., 5GAA (2019).
\newline\urlprefix\url{https://5gaa.org/content/uploads/2019/07/5GAA_191906_WP_CV2X_UCs_v1-3-1.pdf}

\bibitem{ETSI_TS_103_561}
{ETSI (European Telecommunications Standards Institute)}, {TS 103 561 V0.0.2 (2022-09) - Manoeuvres{\textquoteright} Coordination Service; Release 2}, Tech. rep., {ETSI (European Telecommunications Standards Institute)} (2022).

\bibitem{Pretschner2007}
A.~Pretschner, M.~Broy, I.~H. Kruger, T.~Stauner, Software engineering for automotive systems: A roadmap, in: Future of Software Engineering ({FOSE} {'}07), IEEE, 2007.
\newblock \href {https://doi.org/10.1109/fose.2007.22} {\path{doi:10.1109/fose.2007.22}}.

\bibitem{2021:vdi_wissensforum_gmbh:eliv_2021}
M.~O.-C. Sean~Selitrennikoff, Architecting for Secure, Safe and Agile Software Defined Vehicles - Learnings from the Personal Computer industry, 1st Edition, Vol. 2384 of VDI-Berichte, VDI Verlag, D{\"{u}}sseldorf, 2021, Ch. End-2-End Architecture, pp. 395--408.
\newblock \href {https://doi.org/10.51202/9783181023846-395} {\path{doi:10.51202/9783181023846-395}}.

\bibitem{Chebaane2020}
A.~Chebaane, A.~Khelil, N.~Suri, Time-critical fog computing for vehicular networks (Apr. 2020).
\newblock \href {https://doi.org/10.1002/9781119551713.ch17} {\path{doi:10.1002/9781119551713.ch17}}.

\bibitem{Chebaane2020.47}
R.~Deng, R.~Lu, C.~Lai, T.~H. Luan, H.~Liang, Optimal workload allocation in fog-cloud computing toward balanced delay and power consumption, {IEEE} Internet of Things Journal 3~(6) (2016) 1171--1181.
\newblock \href {https://doi.org/10.1109/JIOT.2016.2565516} {\path{doi:10.1109/JIOT.2016.2565516}}.

\bibitem{Chebaane2020.50}
S.~Park, Y.~Yoo, Real-time scheduling using reinforcement learning technique for the connected vehicles, in: 2018 {IEEE} 87\textsuperscript{th} Vehicular Technology Conference (VTC Spring), IEEE, 2018, pp. 1--5.

\bibitem{Chebaane2020.57}
I.~Farris, T.~Taleb, M.~Bagaa, H.~Flick, Optimizing service replication for mobile delay-sensitive applications in {5G} edge network, in: 2017 {IEEE} International Conference on Communications (ICC), IEEE, 2017, pp. 1--6.

\bibitem{SILVA2021100372}
M.~Silva, P.~Teixeira, C.~Gomes, D.~Dias, M.~Lu{\'{i}}s, S.~Sargento, \href{https://www.sciencedirect.com/science/article/pii/S2214209621000413}{Exploring software defined networks for seamless handovers in vehicular networks}, Vehicular Communications 31 (2021) 100372.
\newblock \href {https://doi.org/10.1016/j.vehcom.2021.100372} {\path{doi:10.1016/j.vehcom.2021.100372}}.
\newline\urlprefix\url{https://www.sciencedirect.com/science/article/pii/S2214209621000413}

\bibitem{Asabe2023b}
Y.~Asabe, E.~Javanmardi, J.~Nakazato, M.~Tsukada, H.~Esaki, {AutowareV2X}: Reliable {V2X} communication and collective perception for autonomous driving, in: The 2023 {IEEE} 97\textsuperscript{th} Vehicular Technology Conference (VTC2023-Spring), Florence, Italy, 2023.

\bibitem{Webinar2023HumanVSDigitalDriver}
A.~L. Andreas~Herzig, Daniel~Gamber, \href{{https://engagestandards.ieee.org/Human-vs-AI-Driver-Webinar-Registration.html}}{Human vs. "digital driver" - compliance and digital homologation challenges in the automotive industry}, (Accessed on 07/04/2023) (Jun. 2023).
\newline\urlprefix\url{{https://engagestandards.ieee.org/Human-vs-AI-Driver-Webinar-Registration.html}}

\bibitem{Laaroussi2018}
Z.~Laaroussi, R.~Morabito, T.~Taleb, Service provisioning in vehicular networks through edge and cloud: An empirical analysis, in: 2018 {IEEE} Conference on Standards for Communications and Networking ({CSCN}), IEEE, 2018.
\newblock \href {https://doi.org/10.1109/cscn.2018.8581855} {\path{doi:10.1109/cscn.2018.8581855}}.

\bibitem{TimeSens56:online}
StarlingX, \href{{https://docs.starlingx.io/operations/tsn.html}}{Time sensitive networking}, (Accessed on 04/16/2024) (Apr. 2024).
\newline\urlprefix\url{{https://docs.starlingx.io/operations/tsn.html}}

\bibitem{TimeAwar24:online}
INET, \href{{https://inet.omnetpp.org/docs/showcases/tsn/trafficshaping/timeawareshaper/doc/index.html}}{Time-aware shaping}, (Accessed on 04/16/2024) (Feb. 2023).
\newline\urlprefix\url{{https://inet.omnetpp.org/docs/showcases/tsn/trafficshaping/timeawareshaper/doc/index.html}}

\bibitem{PatentKishore2020}
K.~Kasichainula, Time sensitive networking device (Aug. 2020).

\bibitem{PatentBush2021}
S.~Bush, Time sensitive network {(TSN)} scheduler with verification (Sep. 2021).

\bibitem{2023-TSN-16}
G.~Patti, L.~L. Bello, L.~Leonardi, Deadline-aware online scheduling of {TSN} flows for automotive applications, {IEEE} Transactions on Industrial Informatics (2022).

\bibitem{2023_Simulation_15}
C.~Sommer, R.~German, F.~Dressler, Bidirectionally coupled network and road traffic simulation for improved {IVC} analysis, {IEEE} Transactions on Mobile Computing 10~(1) (2011) 3--15.
\newblock \href {https://doi.org/10.1109/TMC.2010.133} {\path{doi:10.1109/TMC.2010.133}}.

\bibitem{Takrouni2017}
M.~Takrouni, M.~Gdhaifi, A.~Hasnaoui, I.~Mejri, S.~Hasnaoui, Design and implementation of a simulink {{DDS}} blockset and its integration to an active frame steering blockset conformed to {{SAE}} {ElectricVehicle}, in: 2017 {IEEE} /{ACS} 14\textsuperscript{th} International Conference on Computer Systems and Applications ({AICCSA}), IEEE, 2017.
\newblock \href {https://doi.org/10.1109/aiccsa.2017.52} {\path{doi:10.1109/aiccsa.2017.52}}.

\end{thebibliography}



\end{document}